\documentclass{jfm}

\usepackage{graphicx}
\usepackage{subcaption} 
\usepackage{bm} 
\usepackage{newtxtext}
\usepackage{newtxmath}
\usepackage{natbib}
\usepackage{hyperref}
\hypersetup{
    colorlinks = true,
    urlcolor   = blue,
    citecolor  = black,
}

\newcommand{\RomanNumeralCaps}[1]
\linenumbers

\title{Interfacial deformation and energy exchange in strong free-surface turbulence}

\author{Andre Calado\aff{1}\corresp{\email{andre.calado@gwu.edu}}
 and Elias Balaras \aff{1}  }
\affiliation{\aff{1}Department of Mechanical and Aerospace Engineering, The George Washington University, Washington D.C., USA}

\begin{document}
\maketitle

\begin{abstract}
This study investigates the dynamics of strong free-surface turbulence (SFST) using direct numerical simulations (DNS). We focus on the energy exchange between the deformed free-surface and underlying turbulence, examining the influence of Reynolds ($Re$), Froude ($Fr$), and Weber ($We$) numbers. The two-fluid DNS of SFST at high $Fr$ and $We$ is able to incorporate air entrainment effects in a statistical steady-state. Results reveal that high $We$ primarily affects entrained bubble shapes (sphericity), while $Fr$ significantly alters free-surface deformation, two-dimensional compressibility, and turbulent kinetic energy (TKE) modulation. Vorticity flux across the interface occurs from viscous diffusion of surface-parallel structures. At lower $Fr$, kinetic energy is redistributed between horizontal and vertical components, aligning with rapid distortion theory (RDT), whereas higher $Fr$ preserves isotropy near the surface. Evidence of a reverse or dual energy cascade is verified through third-order structure functions, with a strong net reverse cascade near the integral length scale, and enhanced vertical kinetic energy in upwelling eddies. Discrete wavelet transforms (DWT) of TKE show weaker decay at smallest scales near the interface, suggesting contributions from gravitational energy conversion and reduced dissipation. The wavelet energy spectra also exhibits different scaling laws across the wavenumber range, with a $-3$ slope within the inertial subrange.
These findings highlight scale- and proximity-dependent effects on two-phase TKE transport, with implications for sub-grid modeling. 
This work underscores the need for advanced analytical tools that allow for localization in both physical and spectral domains to further elucidate the complex energy cascade mechanisms in SFST.
\end{abstract}



\section{Introduction}
\label{sec:intro}

Turbulence interactions with a free-surface are a main driver in the exchange of mass and momentum for a variety of situations. For example, in oceanic flows proper understanding of aeration is needed to evaluate gas transfer between the ocean and the atmosphere \citep{KrausBusinger,Chanson1995,Herlina2019,Farsoiya2023}.  Near the hull of a ship entrained air interacts with its components and propulsor to produce a bubbly wake which is critical for hydrodynamic performance \citep{Masnadi2019}.  These types of flow are very complex, with turbulence interacting with both phases, and current two-phase Reynolds Averaged Navier-Stokes (RANS) approaches that are routinely used in engineering applications have to be modified from their single-phase formulations to account for different mechanisms in turbulent kinetic energy (TKE) transport and dissipation \citep{Salvetti1997,Frederix2018,Yuan2024,Draheim2024}.

Dimensionless numbers which are relevant to characterize these flows are the Reynolds ($Re$), Froude ($Fr$), and Weber ($We$),  with defining velocity and length scales typically associated with the bulk turbulence.  A qualitative map for different regimes based on $Fr$ and $We$ numbers is proposed by \citet{Brocchini2001}. 
In cases where both $Fr$ and $We$ are low, the gravitational and surface tension forces are large enough to minimize the deformation of the free-surface. This is the case of weak free-surface turbulence (WFST), where the the eddies under the surface play a role on the diffusion of gas through the interface, with distinct cases depending on the amount of shear due to the gas velocity above the liquid \citep{Hunt1984}.  A relationship between gas-transfer velocity and free-surface divergence is established in \citet{McKenna2004}.
For strong free-surface turbulence (SFST), both $Fr$ and $We$ are large, the free-surface then becomes notably wavy, with `scars' and entraining the air above \citep{Flack1961,Brocchini2001,Calado2024-GFM}. The  case of turbulent bulk shear flow below the free-surface with air entrainment  has been studied both experimentally by \citet{Bardet} and through Direct Numerical Simulations (DNS)  by \citet{Yu2019}.

Early DNS of low $Fr$ turbulence  interactions with a free-surface by \citet{Shen1999} identified a thin viscous layer $(\delta_v \sim Re^{-1/2})$, and a thicker blockage/source layer of the order of the integral length scale $(\delta_s \sim \ell)$, where turbulent stress components are redistributed (vertical fluctuations dampened). Experiments in turbulent open-channel flow show that the submerged turbulence drives the low wavenumber footprint of the free-surface (such as the integral length scale), whereas for high wavenumbers, the capillary-gravity waves dominate \citep{Savelsberg2008}.
Similar observations were made by \cite{Guo2010} using DNS to study the interaction between a free-surface and underlying homogeneous  isotropic turbulence (HIT) at a Taylor microscale Reynolds $Re_{\lambda} \sim 30$ using non-linear boundary conditions. 
Spectral analysis of the surface deformation showed propagating waves and turbulence induced roughness, where low wavenumbers correspond to traveling waves, while turbulent roughness appears in all wavenumbers. The relative importance of gravity and surface tension is a function of the critical wavenumber $\tilde{k}_{cr} = \sqrt{We/Fr^2}$. 
By examining the TKE budgets they concluded that energy is primarily exchanged through free-surface upwelling and downwelling events.  \citet{Flores2017} performed spectral analysis of decaying turbulence close to a flat stress-free surface, where an inertial subrange is seen for horizontal velocity components, and roughly one integral length scale from the surface the flow tends toward isotropy. 

\citet{Sarpkaya1996} commented that for the case of submerged HIT below a free-surface, there is a transfer of turbulent kinetic energy from the vertical component to the horizontal directions (quasi-2D turbulence), and that the free-surface acts as an `imperfect mirror' with some vortices attaching to the interface, while others flatten with axis parallel to it. Merging of same sign vortices and cancellation of opposite sign vortices can skew the eddy size distribution towards larger ones, and produce a reverse energy cascade. This effect was termed the \emph{free-surface anisotropy dynamo}. 
In DNS of open channel flow with a flat free-surface, \citet{Lovecchio2015} observe upscale energy transfer (reverse cascade) associated with large scales.
Recent experiments by \citet{Ruth2024} of submerged HIT up to $Re_{\lambda} \approx 590$, in a gravity dominated regime (low $Fr$, quasi-flat surface) show that vertical and horizontal components of TKE follow predictions of rapid distortion theory (RDT) of \cite{Hunt1978}, with integral scale of horizontal fluctuations growing as the surface is approached. Along horizontal separations the direct energy cascade of horizontal fluctuations is reduced, with a reverse cascade for vertical energy. 

Overall for cases of SFST, where the free-surface is highly disturbed,  the stretched interface extracts energy from the underlying turbulence. However, there is limited understanding of the length scales at which energy can be returned back into the flow (reverse energy cascade), and how this process depends on $Fr$ and $We$ numbers.
Recent studies focusing on interfacial energy exchange have found that for disperse droplets or equal density two-phase flows in HIT, surface tension has a scale-dependent effect leading to an enhancement of TKE at high wavenumbers, and dampening at low wavenumbers \citep[see][]{Dodd2016,Vela2021,Saeedipour2023,Crialesi2023}.  Currently most free-surface DNS assume very large or infinite $We$, and limited $Re$ due to computational cost limitations \citep{Yu2019,Gaylo2024}, or simply ignore the air phase by assuming a flat slip wall or free-surface boundary conditions \citep{Guo2010,Lovecchio2015,Flores2017}.   On the other hand, experimental studies are limited by the Earth's gravitational environment and working fluid properties to very low $Fr$ and $We$ numbers \citep{Variano2008,McKenna2004,Ruth2024,Jamin2025}. Additional challenges arise from entrained air bubbles and the complexity of measuring stresses and other local variables.  

In the present work we perform high-fidelity DNS data of two-phase free-surface turbulence, including cases with air entrainment. We achieve closer matching of dimensionless numbers between DNS and available experiments, by exploiting computational efficiencies from our in-house solver. The main goal is to assess the effect of $Re$, $Fr$ and $We$ numbers on the  Reynolds stress components and interfacial energy exchange from submerged HIT. In addition, we make use of data-driven tools such as discrete wavelet transforms (DWT), which have only begun to be applied for the spectral analysis of two-phase flows \citep{Freund2019,Calado2024-PRF}.

\section{Computational Methodology}
\label{sec:methods}
\subsection{Two-phase solver}
\label{sec:solver}
We use our in-house DNS code to solve the  Navier-Stokes (N-S) equations for incompressible, isothermal flow in dimensionless form, with variable properties:
\begin{equation}
    \label{eq:mass}
    \nabla \cdot \mathbf{u} = 0,
\end{equation}
\begin{equation}
\label{eq:NS}
\frac{\partial \mathbf{u}}{\partial t} + \mathbf{u} \cdot \nabla \mathbf{u}  = - \frac{\nabla p}{\rho'}  + \frac{\mu'}{\rho'} \frac{1}{Re} \nabla \cdot \left( \nabla \mathbf{u} + \nabla \mathbf{u}^{\top} \right) + \frac{\mathbf{g}}{Fr^2}  + \mathbf{f}^T,
\end{equation}
where $\mathbf{u} = (u,v,w)^\top$ is the velocity vector, $p$ the pressure, $\mathbf{g}=(0,0,-1)^\top$ unit vector for gravity and $t$ is time. $\rho ' = \rho_i / \rho_2$ and $\mu' = \mu_i / \mu_2 $ are the ratios of density and dynamic viscosity, respectively, of the local fluid phase $(i=1,2)$, with respect to the reference denser phase (2).  An additional source term $\mathbf{f^T}$ is added to sustain turbulence, and is described in \ref{sec:setup}.
The  Reynolds, Weber and Froude numbers are hence defined:
\begin{equation}
\label{eq:numbers}
Re = \frac{\rho_2 \Tilde{U} \Tilde{L}}{\mu_2}, \qquad We = \frac{\rho_2 \Tilde{U}^2 \Tilde{L}}{\sigma}, \qquad Fr = \frac{\Tilde{U}}{\sqrt{g \Tilde{L}}},
\end{equation}
where $\Tilde{U}$, $\Tilde{L}$, $g$, $\sigma$ are the characteristic values for velocity, length, gravity, and surface tension coefficient.  
%
%
A staggered arrangement of the flow variables is adopted on a structured Cartesian grid, with all spatial derivatives approximated by second-order finite differences. A fractional step method is used to advance the N-S equations in time, where all terms are computed in an explicit fashion using a second-order Adams-Bashforth (AB2) scheme.  The pressure jump across the interface due to surface tension uses a variant of the ghost fluid method (GFM) by \cite{Fedkiw1999}. The implementation details are discussed in \cite{Dhruv2019}.  The Poisson equation for pressure is written in a constant coefficient form, based on the splitting procedure introduced by \cite{DODD2014416}.   

The interface between liquid and gas phase is tracked via conservative level-set approach. The level-set function, $\phi$, corresponds to a signed distance function from the interface such that $\phi > 0$ indicates fluid-1 (gas), and $\phi < 0$ fluid-2 (liquid). 
The unit normal vector, $\mathbf{n} = \nabla \phi / |\nabla \phi |$, and interface curvature, $\kappa=\nabla \cdot \mathbf{n}$, are computed directly from the level-set function.
The advection equation for the level-set is discretized in space using a fifth-order weighted essentially non-oscillatory (WENO-5) scheme \citep{Jiang2000} and advanced in time with a third-order total variation diminishing (TVD) Runge-Kutta scheme \citep{Shu1988}.
A dual-redistancing approach is used at every time step to ensure the level-set remains a signed distance function while maintaining global mass conservation of each phase \citep{Chang1996,Zhang1998,Calado-CAF}. 

\subsection{Problem formulation}
\label{sec:setup}
\begin{figure}
    \begin{subfigure}{0.3\textwidth}
        \centering
        \includegraphics[width=\textwidth]{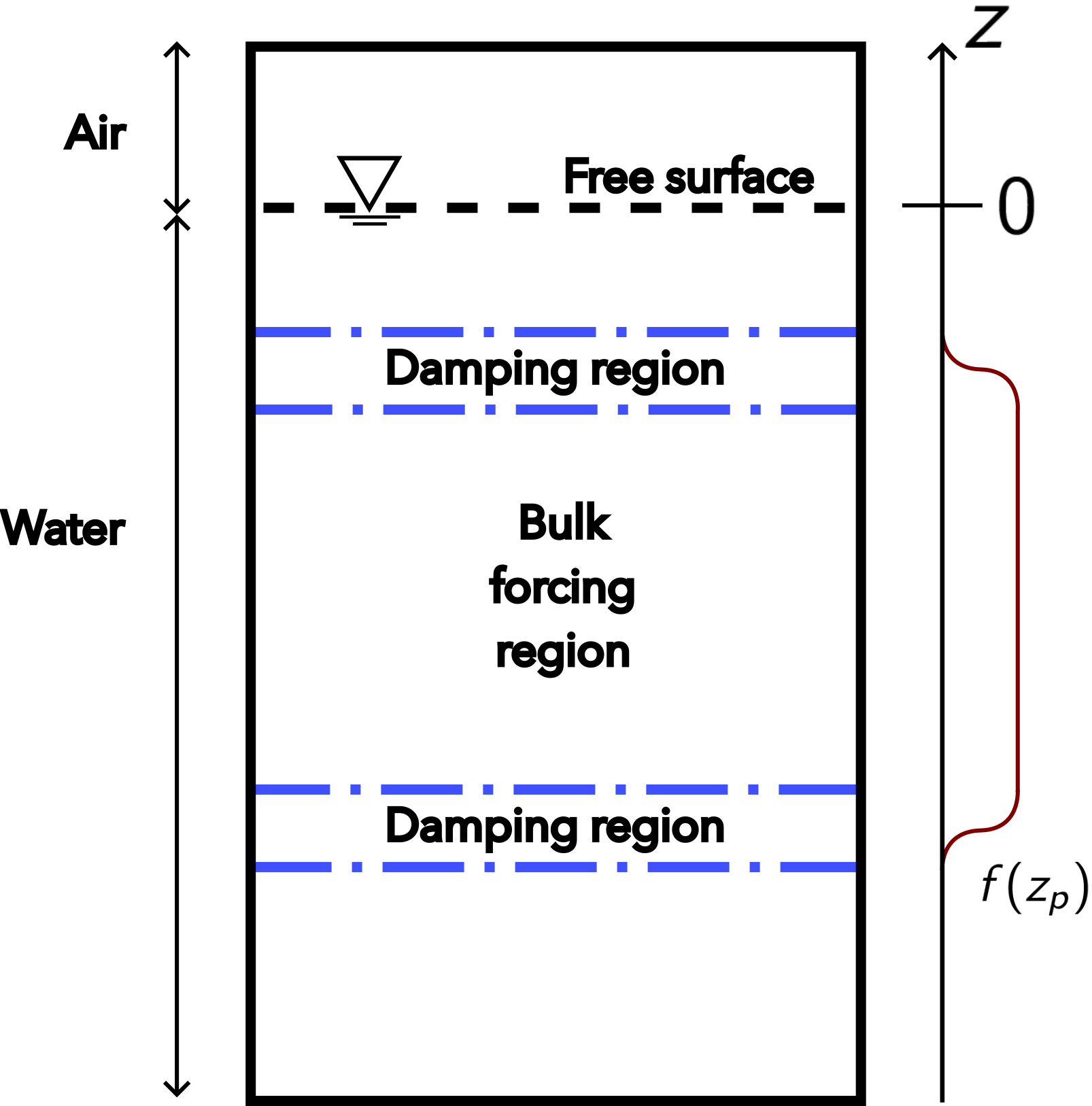}
        \caption{}
        \label{fig:turb_box}
    \end{subfigure}
    \qquad 
    \begin{subfigure}{0.6\textwidth}
        \centering
        \includegraphics[width=\textwidth]{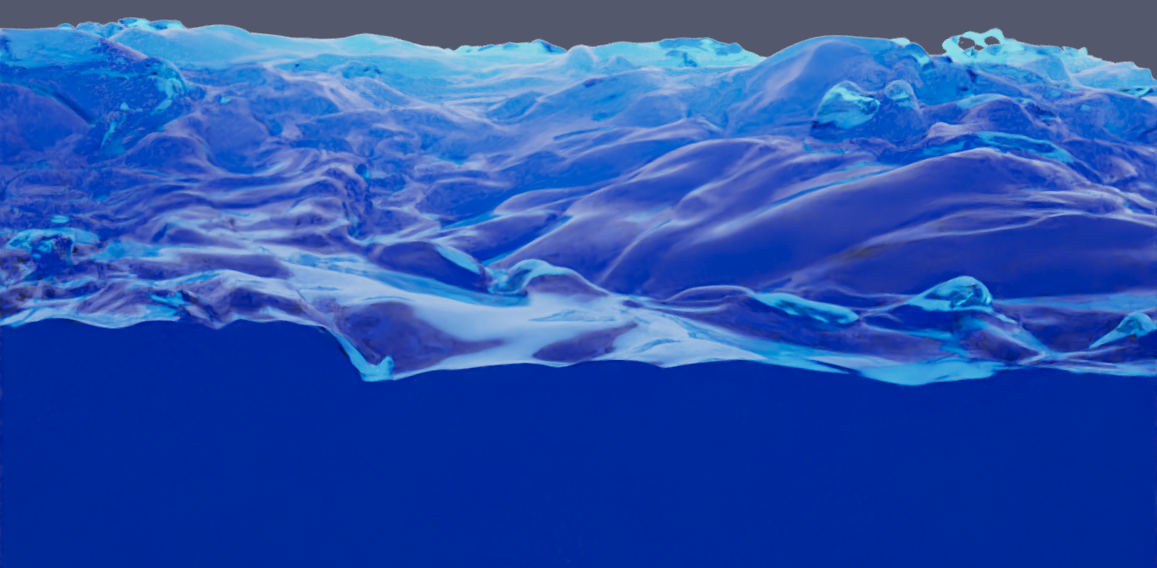}
        \caption{}
        \label{fig:render}
    \end{subfigure}
\caption{(a) Schematic view of the domain for the subsurface turbulence generation showing damping and bulk forcing regions. Mean free-surface is at $z=0$. Profile on right represents the spatial forcing function $f(z_p)$; (b) DNS rendering of turbulent free-surface.}
    \label{fig:problem}
\end{figure}

Statistically stationary HIT is sustained at a certain depth below the free-surface. Following previous studies by \cite{Guo2009,Guo2010} a bulk HIT forcing region with total height $2l_b$ is in between two symmetric damping regions of size $l_d$. We similarly  define the ratios $l_b / L = 0.75$ and $l_d / L = 0.25$, with $L=5$ being the horizontal domain size imposed by the forcing scheme by \cite{Rosales2005}. In order to preserve the isotropic nature of the turbulence, there must be sufficient length for both the bulk and damping regions \citep[see ][]{Guo2009}.   From the top damping region the distance between mean free-surface is also $0.25 L$. As such, no active forcing is present surrounding the intermittent free-surface.  The boundary conditions are periodic on the horizontal ($xy$) plane, with slip boundaries on the top and bottom ($z$).
The spatial forcing function, $f(z_p)$, depends on the vertical distance to the center of the bulk region, $z_p=|z-z_c|$, and is given by:
\begin{equation}
\label{eq:forcing_z}
  f(z_p)=\left\{   
  \begin{array}{@{}ll@{}}
    1, &  z_p \leq l_b \\
     \frac{1}{2} \left(1 - \text{cos}\left[ \frac{\pi}{l_d} \left(z_p-(l_b+l_d) \right) \right] \right), &  z_p \leq l_b+l_d \\
    0, & z_p > l_b+l_d
  \end{array}\right.
\end{equation} 
A schematic view of the domain and $f(z_p)$ is shown in Fig. \ref{fig:turb_box}. 
The instantaneous spatial average of the TKE, $k_b(t) $, within the bulk region is used to modulate the forcing coefficient in the RHS of the momentum equation:
\begin{equation}
\mathbf{f}^T = f(z_p) a_0  \frac{k_0}{ k_b(t) } \mathbf{u}, 
\end{equation}
where $a_0=1/3$ and $k_0 =3/2$ \citep{Lundgren2003,Carroll2013}.   The reference length scale is the integral scale of the generated turbulence, $\ell = u_0^3 / \varepsilon$, where the reference velocity is the root-mean-square (RMS) velocity, $u_0$, and $\varepsilon$ is turbulent dissipation rate. The total vertical domain height is $L_z = 3L = 15\ell$, with a height of air above the free-surface equal to $2.5 \ell$. 
The dimensionless parameters of the simulations relative to the integral and Taylor microscale ($\lambda$) within the bulk forcing region are given in Table~\ref{tab:parameters}. For all cases the the density and viscosity ratios of air-water are considered: $\rho_2 / \rho_1 = 833.3$, $\mu_2 / \mu_1 = 55.2$.  The grid is uniform and is of the same order of magnitude as the Kolmogorov scale $\eta = (\nu^3 / \varepsilon)^{1/4}$. The baseline grid has $256^2 \times 768$ points, with a refined grid in case F where $Re$ is increased. The grid sizes are sufficient to capture all the relevant scales of turbulence that reach the free-surface.  A snapshot rendering from the DNS showing the turbulent free-surface can be seen in Fig. \ref{fig:render}.

\begin{table}
  \begin{center}
\def~{\hphantom{0}}
  \begin{tabular}{lccccccc}
      Case  & \qquad   $Re_{\ell}$ & \qquad $Re_{\lambda}$ &  \qquad$Fr^2_{\ell}$   &   \qquad  $Fr^2_{\lambda}$ & \qquad  $We_{\ell}$ & \qquad  $We_{\lambda}$ & \qquad Grid \\[3pt]
      A & \qquad 1050 & \qquad 125 & \qquad 0.57 &\qquad 4.7 &\qquad 945 &\qquad 113 &\qquad $256^2 \times 768$ \\
      B &\qquad 1050 &\qquad 125 &\qquad 0.57 &\qquad 4.7 &\qquad 9450 &\qquad 1130 &\qquad $256^2 \times 768$  \\
      C &\qquad 1050 &\qquad 125 &\qquad 0.57 &\qquad 4.7 &\qquad $\infty$ &\qquad $\infty$ &\qquad $256^2 \times 768$ \\
     D &\qquad 1050 &\qquad 125 &\qquad 0.28 &\qquad 2.4 &\qquad 945 &\qquad 113  &\qquad $256^2 \times 768$\\
     E &\qquad 1050 &\qquad 125 &\qquad 0.14 &\qquad 1.2 &\qquad 945 &\qquad 113 &\qquad $256^2 \times 768$ \\
      F &\qquad 4228 &\qquad 250 &\qquad 0.40 &\qquad 6.7 &\qquad 5391 &\qquad 321 &\qquad $500^2 \times 1540$ \\
  \end{tabular}
  \caption{Summary of free-surface turbulence DNS cases.}
  \label{tab:parameters}
  \end{center}
\end{table}

For the analysis of the flow field data, we perform ensemble averaging over uncorrelated snapshots in time (one large eddy turnover time apart). A total of 10 snapshots are used, once the flow has reached statistical steady-state. Sensitivity of the number of snapshots used in the ensemble averaging has been verified in previous works \citep[see][]{Calado-CAF}.   We define the averaging operator $\langle \cdot \rangle$ denoting ensemble averages over the horizontal plane, and unless otherwise specified, referring to the water phase.
As proposed by \cite{Brocchini2001b} we can define an intermittency function $\gamma$ based on the average volume fraction of water. We first define the Eulerian phase function, $I(\mathbf{x},t)$:
\begin{equation}
\label{eq:phase}
  I(\mathbf{x},t)=\left\{   
  \begin{array}{@{}ll@{}}
    1 &  \textrm{in water }  \\
    0 &  \textrm{in air }  \\
  \end{array}\right.
\end{equation} 
and through the averaging operator we obtain $\gamma(z) = \langle I(z) \rangle$, as well as the intermittency layer, $\delta_i$ where $0 < \gamma < 1$.

\begin{figure}
    \begin{subfigure}{0.45\textwidth}
        \centering
        \includegraphics[width=\textwidth]{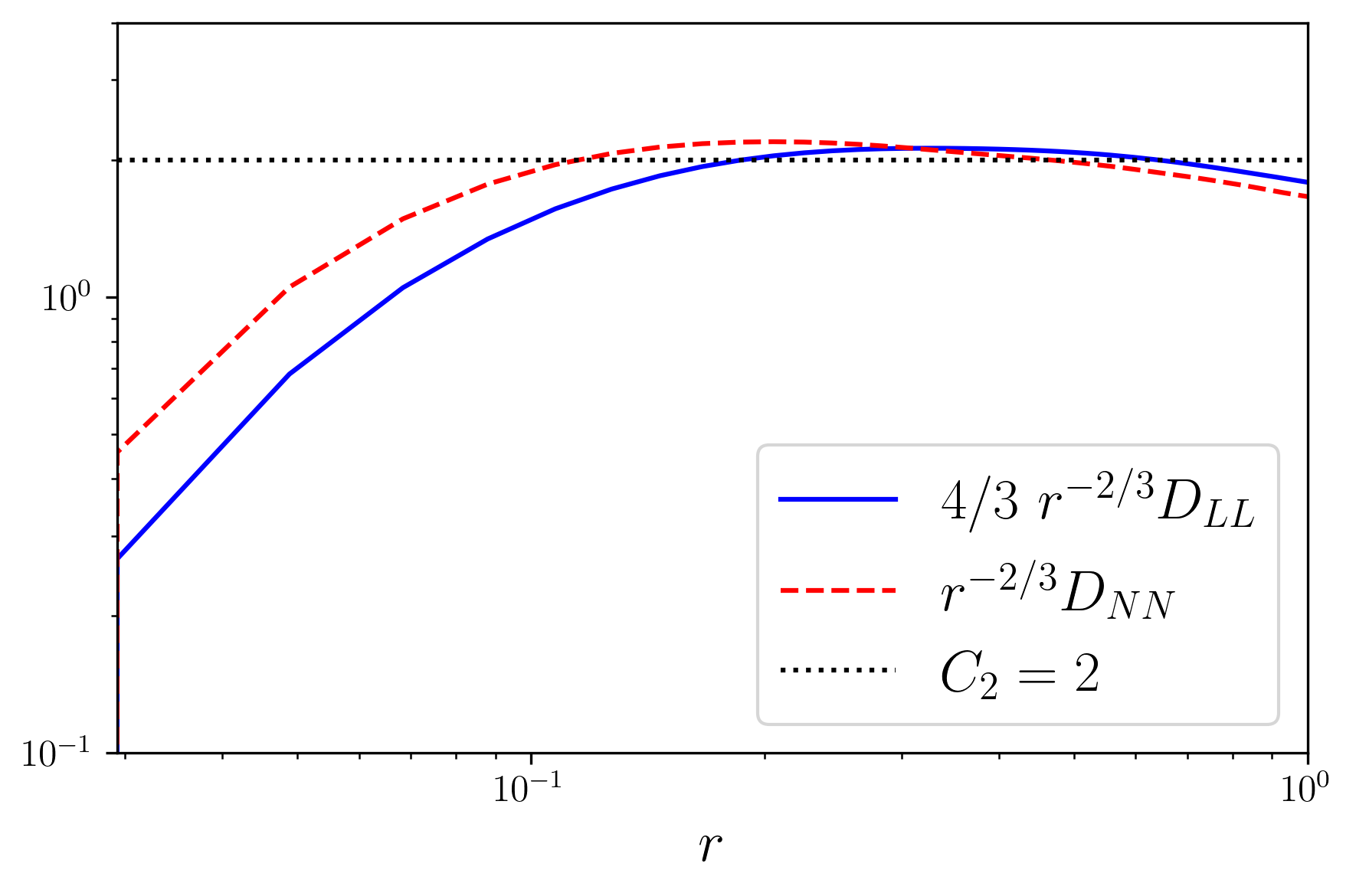}
        \caption{}
        \label{fig:structure}
    \end{subfigure}
    \qquad 
    \begin{subfigure}{0.45\textwidth}
        \centering
        \includegraphics[width=\textwidth]{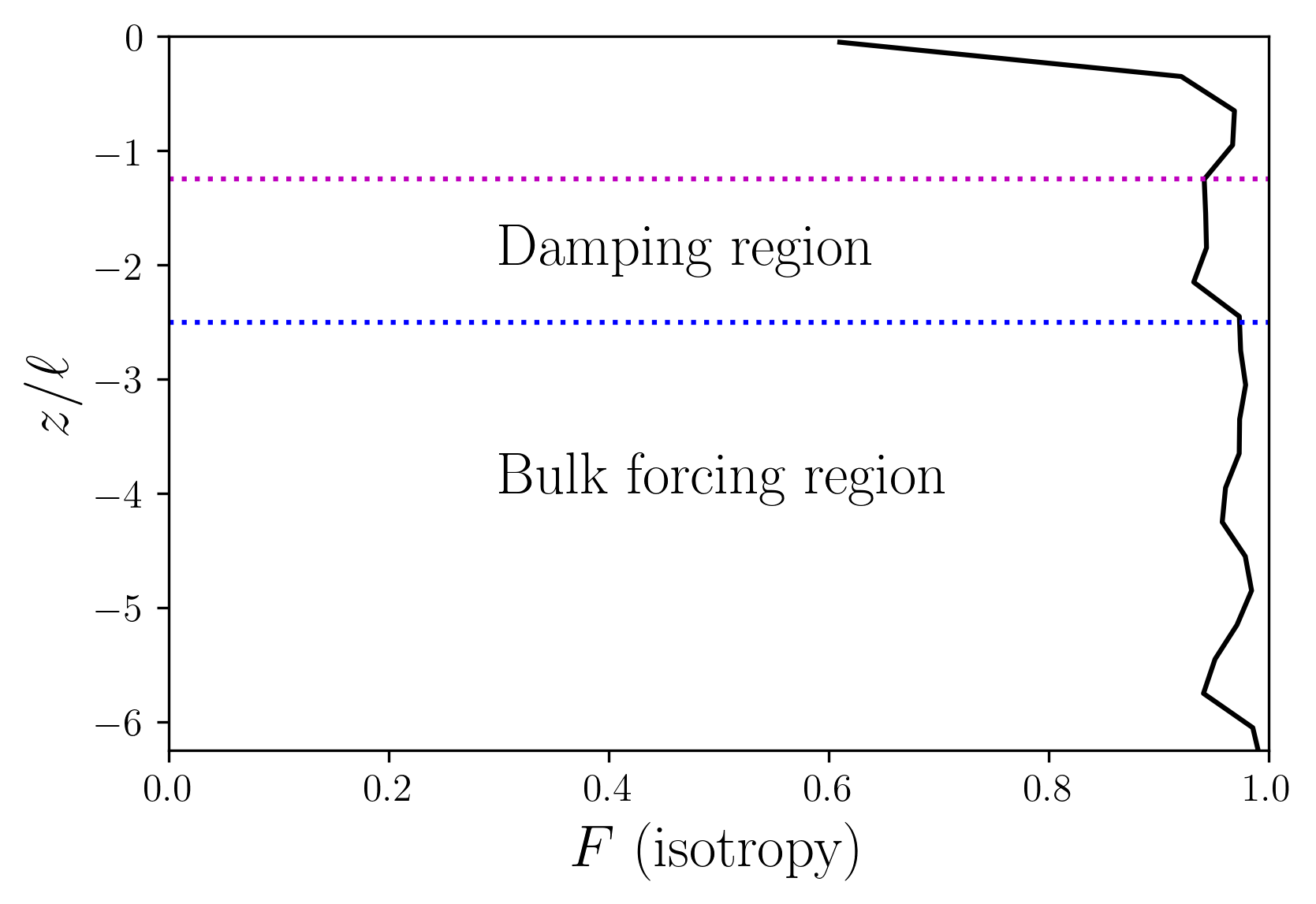}
        \caption{}
        \label{fig:isotropy}
    \end{subfigure}
\caption{(a) Pre-multiplied longitudinal and transverse second-order structure functions in the bulk turbulence region. (b) Phase-averaged isotropy scalar $F$ as a function of depth. Data is for Case A.}
    \label{fig:isoturb}
\end{figure}

The generated turbulence in the bulk is verified to be isotropic by analyzing different metrics: in Fig. \ref{fig:structure} we plot the pre-multiplied longitudinal and transverse second-order structure functions ($D_{LL}$ and $D_{NN}$), which exhibit a plateau around the constant $C_2 = 2.0$ in the inertial range \citep{Pope2000}. In addition, the level of isotropy as a function of the depth, $z$, is quantified with the isotropy scalar, $F$, which is bounded between 0 and 1 (pure isotropy):
\begin{equation}
\label{eq:F-iso}
F(z) = \frac{27  \langle u'^2 \rangle  \langle v'^2 \rangle  \langle w'^2 \rangle }{( \langle u'^2 \rangle +  \langle v'^2 \rangle + \langle w'^2 \rangle )^3}.
\end{equation}
We observe that the turbulence is indeed isotropic, with $F > 0.9$ within the bulk and damping regions, $z/\ell < -1.25$, and for the selected case A it remains isotropic until almost $z=0$. 

\section{Entrained bubble statistics}
\label{sec:bubbles}

For the higher $Fr$ number cases (A,B,C and F), the simulated conditions are found to result in air entrainment, with numerous air bubbles below the free-surface. From these simulations we intend to investigate the effects of $We$ number on the bubble sizes and their sphericity, $\alpha$, defined as the ratio between surface area of volume equivalent sphere and their actual area \citep{clift05:BDP}.  The values for $\alpha$ range between 0 and 1, where a higher value indicates closer resemblance to a spherical shape.  Note that smallest bubbles with volume below $(2 \Delta h)^3$ are considered under-resolved, and are thus excluded from the statistics.

\begin{figure}
    \centering
    \begin{subfigure}{0.45\textwidth}
        \centering
        \includegraphics[width=\textwidth]{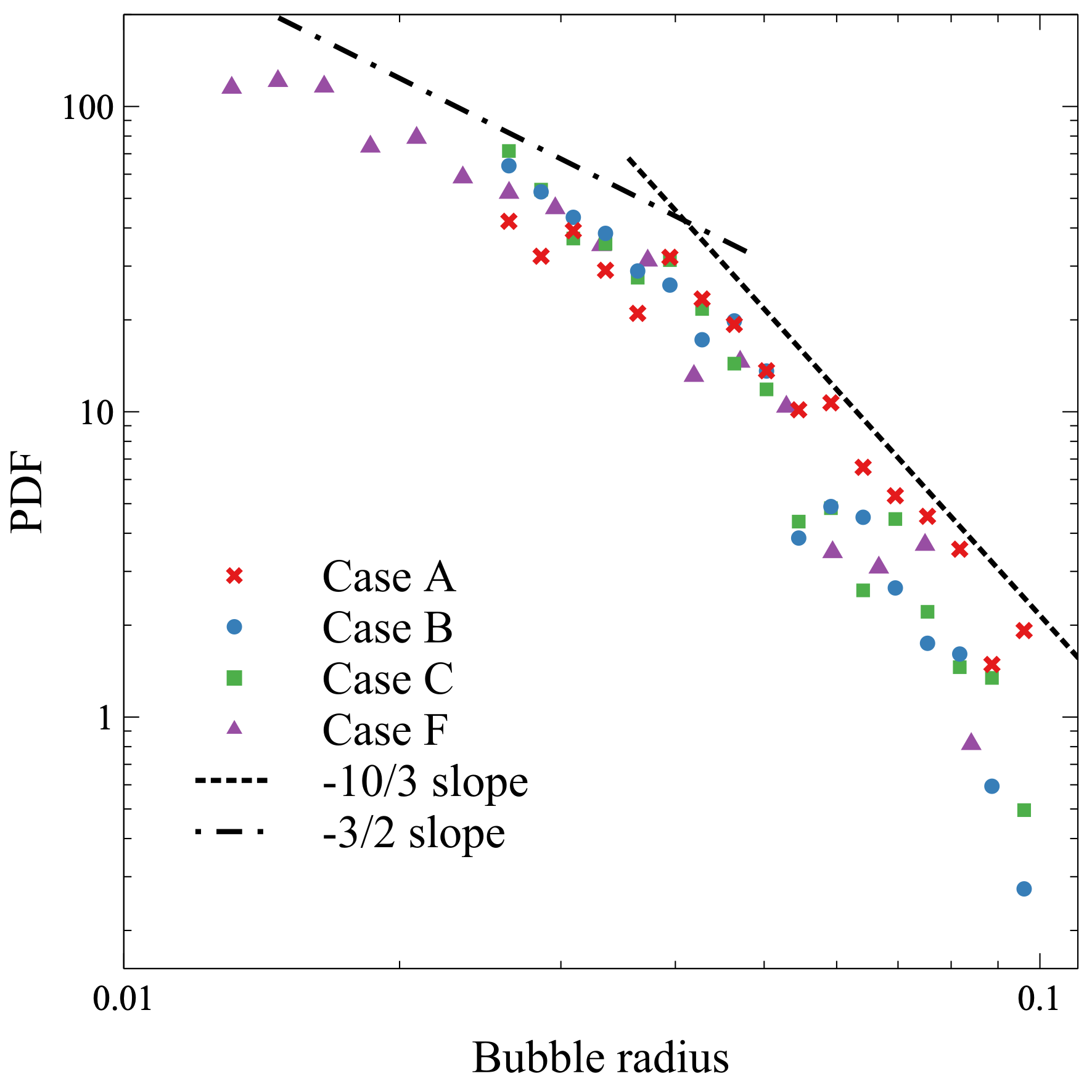}
        \caption{}
        \label{fig:bub-radius}
    \end{subfigure}
    \qquad 
    \begin{subfigure}{0.45\textwidth}
        \centering
        \includegraphics[width=\textwidth]{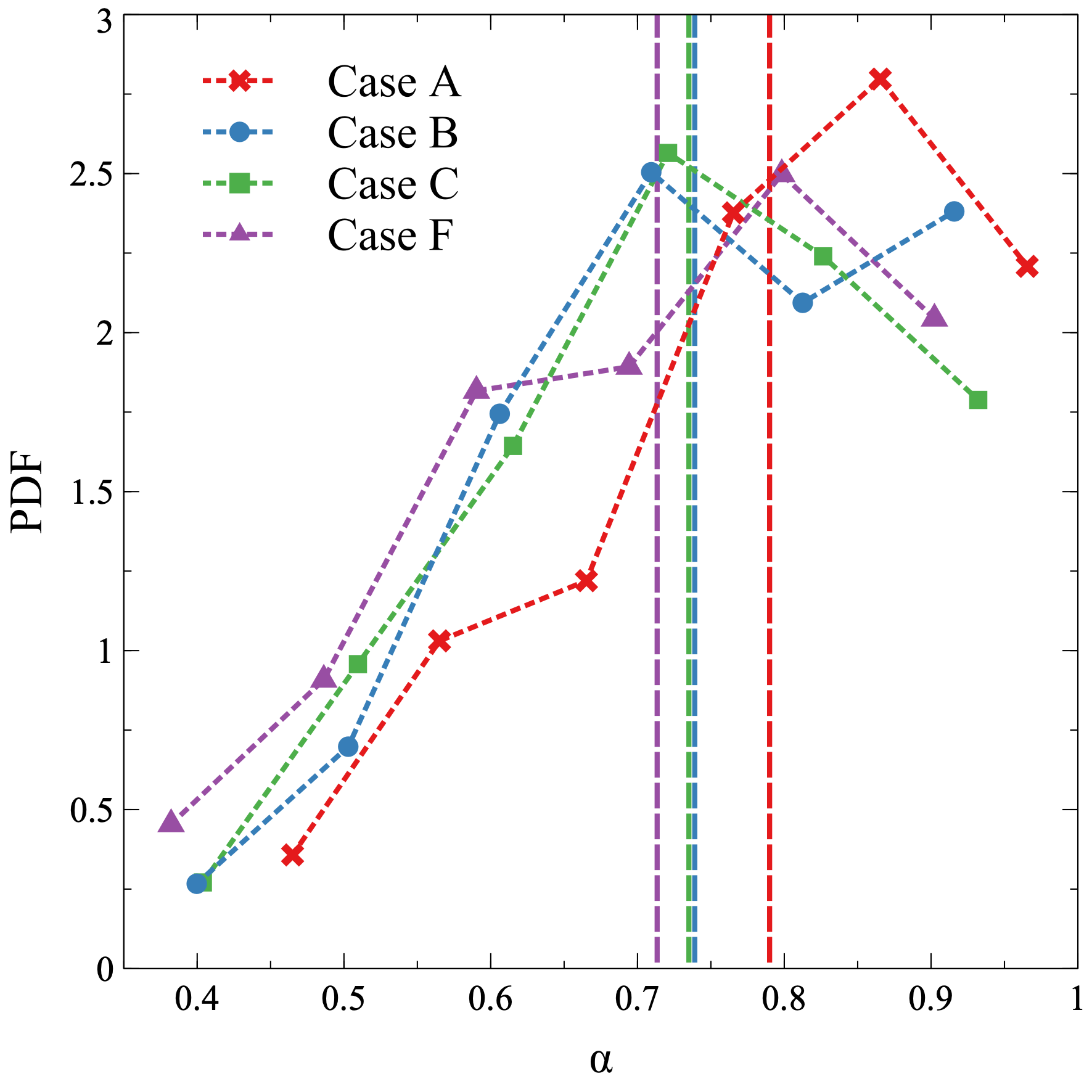}
        \caption{}
        \label{fig:bub-sphericity}
    \end{subfigure}
\caption{(a) PDF of equivalent bubble radius, and (b) PDF of bubble sphericity, $\alpha$ (mean values shown as vertical dashed lines). Note only cases with significant air entrainment are shown.}
    \label{fig:bubble-stats}
\end{figure}

The probability density function (PDF) for the equivalent bubble radius is shown in Fig.~\ref{fig:bub-radius}, where values for the different cases collapse rather well. The distinct slopes of $-10/3$ and $-3/2$ are observed for super- and sub-Hinze sizes, respectively \citep{Garrett2000,Deane2002,Yu2019}.  The influence of $We$ number on the bubble size for the tested range appears to be negligible. The distribution is also unaffected by the increase in the $Re$ number (case F), although the increased grid resolution is able to capture smaller bubble sizes which align with the $-3/2$ slope.
In Fig.~\ref{fig:bub-sphericity} we plot the PDF sphericity $\alpha$ for the same cases, along with their means (vertical dashed lines). We observe that the mean sphericity decreases as $We$ increases, with only minor differences between cases B ($We_\lambda=1130$) and C ($We_\lambda=\infty$). Case F at the higher $Re$ exhibits the lowest average sphericity. These results are somewhat intuitive, since with the increase of $We$ or $Re$ we anticipate higher deformation from the submerged turbulence and weaker restorative forces from surface tension. Note that a lower sphericity also implies greater surface area (compared to a spherical bubble), which has implications when trying to account for the total interfacial area. For example, for an average sphericity of $\alpha = 0.75$, this equates to a total surface area that is $ \sim 33 \%$ larger  than that of a sphere of equivalent volume.  Furthermore, the bubble shape and deformability can affect its rise velocity and modulate the surrounding turbulence \citep{Ni2024}.

\section{Free-surface characterization}
\subsection{Surface deformation}
\label{sec:surf-deform}

\begin{figure}
        \begin{subfigure}{0.3\textwidth}
        \centering
        \includegraphics[width=\textwidth]{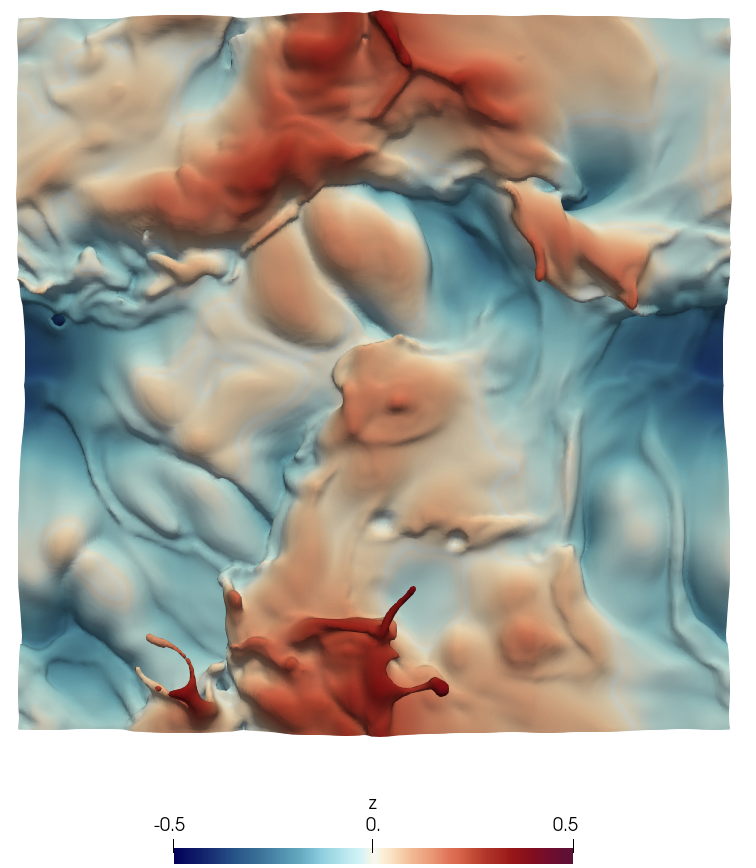}
        \caption{Case A}
        \label{fig:fr-high}
    \end{subfigure}
    \quad
    \begin{subfigure}{0.3\textwidth}
        \centering
        \includegraphics[width=\textwidth]{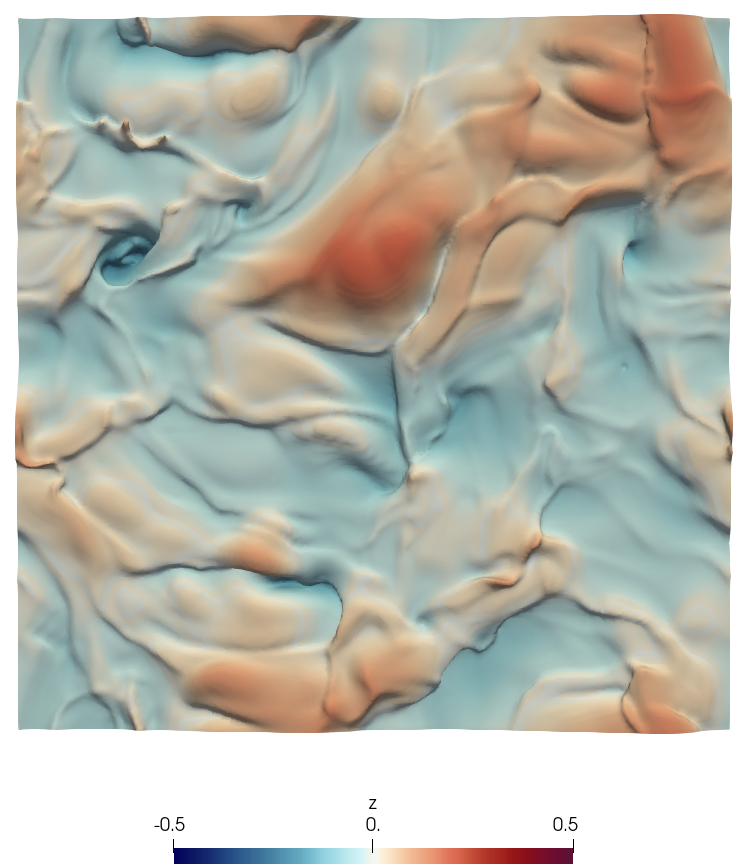}
        \caption{Case D}
        \label{fig:fr-med}
    \end{subfigure}
    \quad
        \begin{subfigure}{0.3\textwidth}
        \centering
        \includegraphics[width=\textwidth]{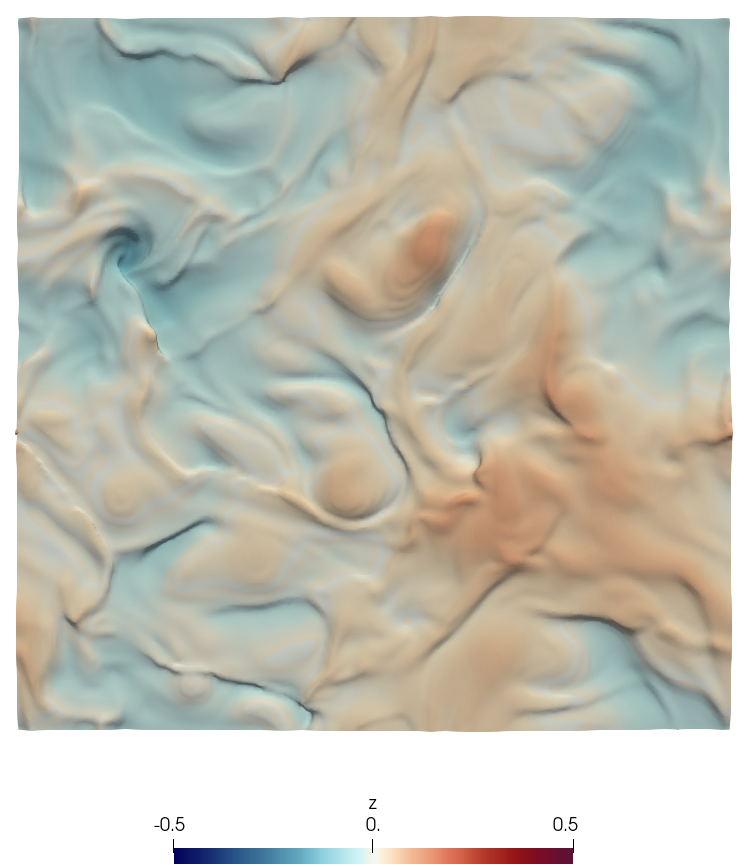}
        \caption{Case E}
        \label{fig:fr-low}
    \end{subfigure}
\caption{Snapshot visualization of interface (top view) for decreasing $Fr^2$ (from left to right:$Fr^2_{\lambda}=4.7, 2.4, 1.2$). Interface is colored by $z$ coordinate.}
    \label{fig:froude-compare}
\end{figure}

\begin{figure}
    \centering
    \begin{subfigure}{0.3\textwidth}
        \centering
        \includegraphics[width=\textwidth]{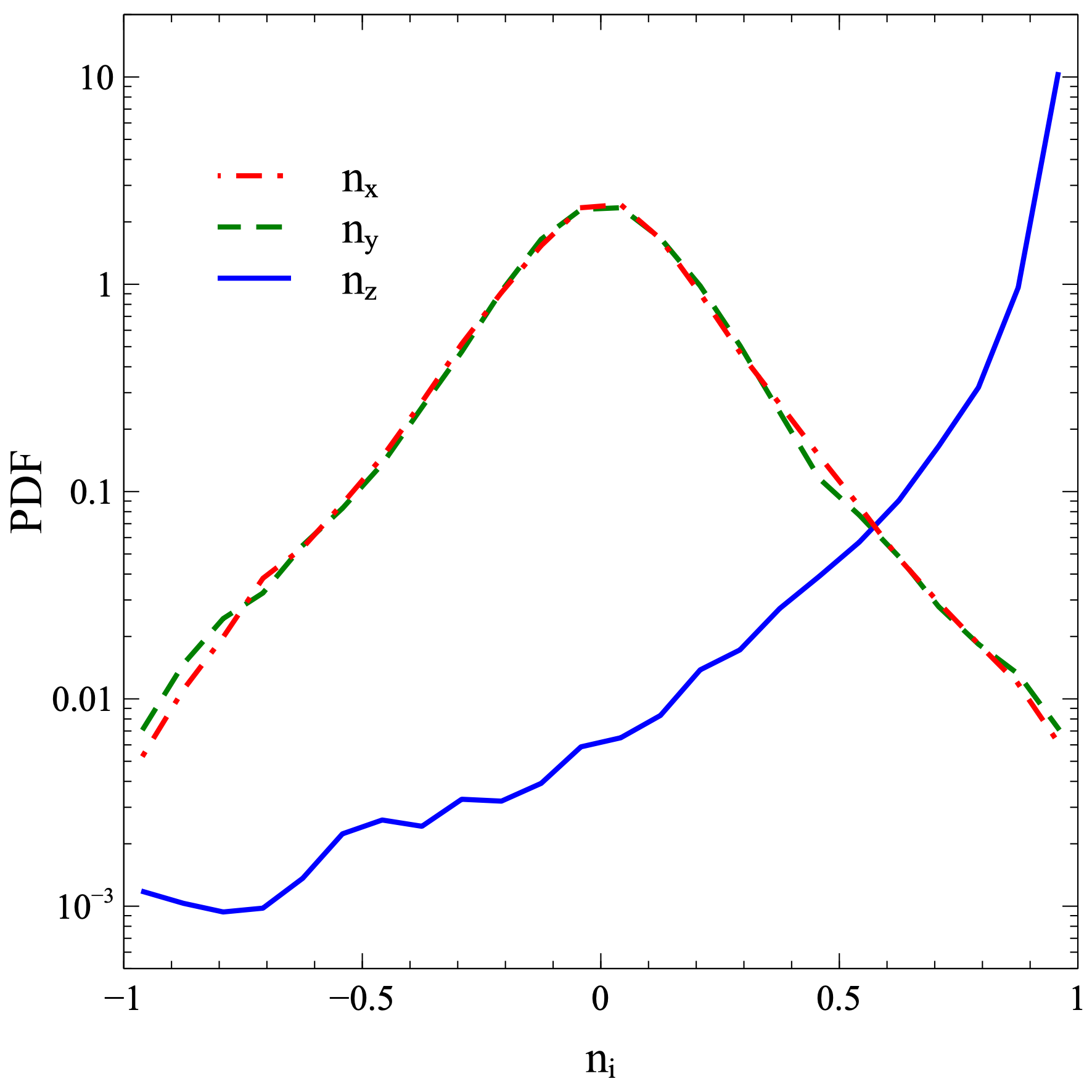}
        \caption{}
        \label{fig:pdf-normals-low}
    \end{subfigure}
    \quad 
    \begin{subfigure}{0.3\textwidth}
        \centering
        \includegraphics[width=\textwidth]{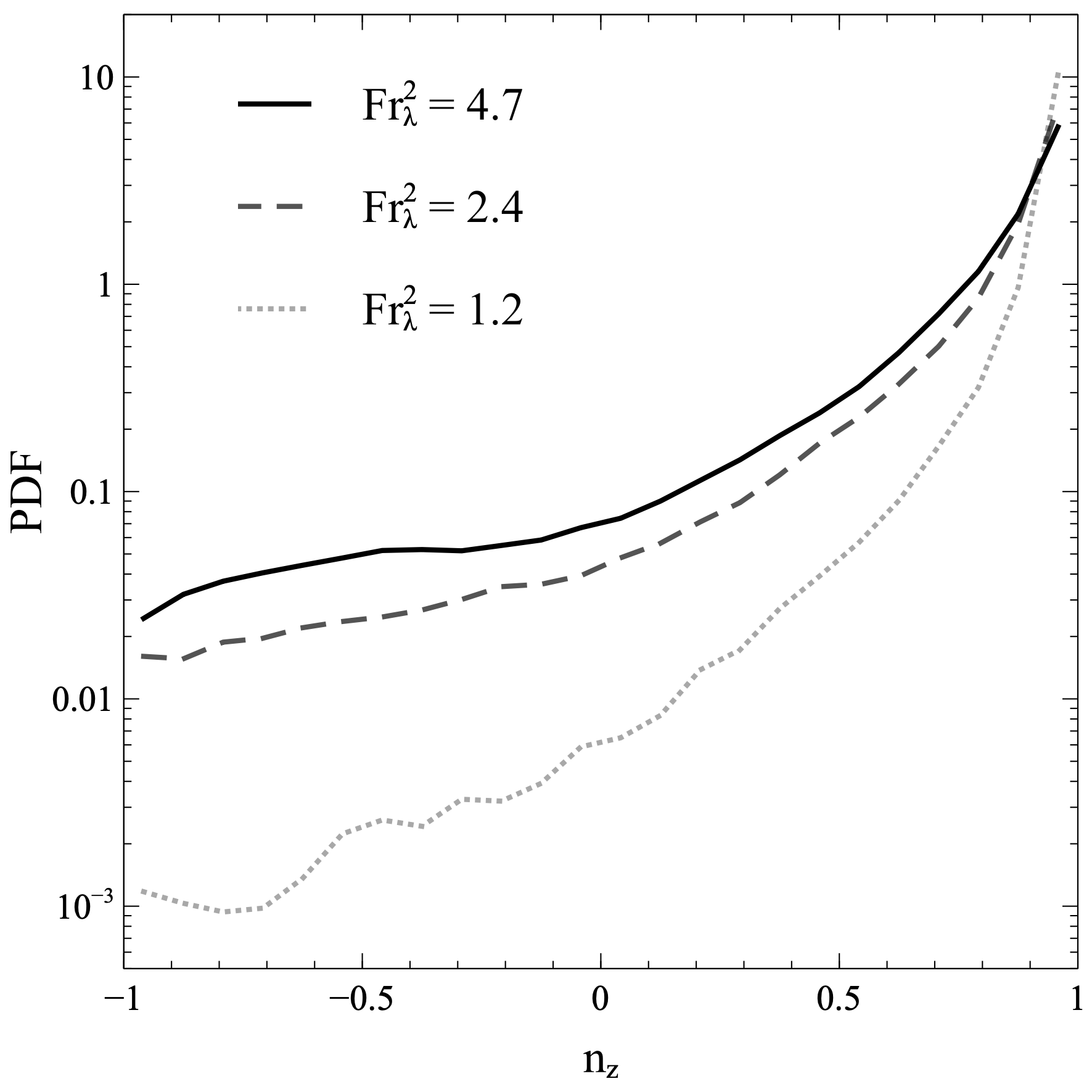}
        \caption{}
        \label{fig:pdf-nz}
    \end{subfigure}
    \quad
        \begin{subfigure}{0.3\textwidth}
        \centering
        \includegraphics[width=\textwidth]{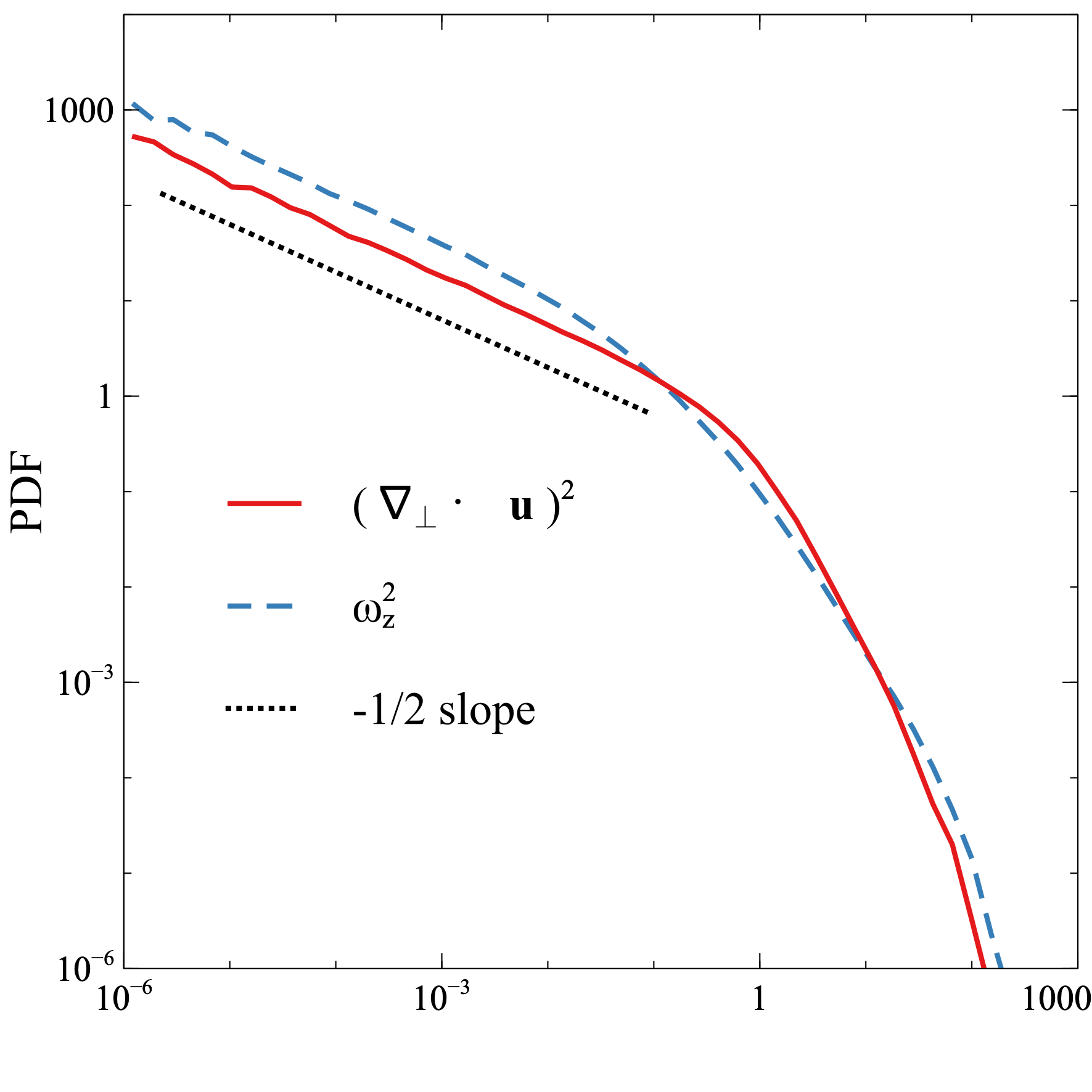}
        \caption{}
        \label{fig:pdf-surf-div-sq}
    \end{subfigure}
\caption{(a) PDF of free-surface normal components for case E, (b) PDF of $n_z$ for different $Fr$ and (c) PDF of $(\nabla_\bot \cdot \mathbf{u})^2$ and $\omega_z^2$, scaled by RMS, for case E. }
    \label{fig:pdf-normals}
\end{figure}

In the case of a quasi-flat free-surface, a linear dependency of the vertical velocity component is expected around $z=0$ such that $w(z) \sim z  \partial_z w $ \citep{McKenna2004,Lovecchio2015}.   This also implies that the surface divergence can be evaluated as $\nabla_\bot \cdot \mathbf{u} = \partial_x u  +  \partial_y v =   - \partial_z w$, and the local sign and magnitude can indicate the presence of upwelling or downwelling structures \citep{Longuet-Higgins_1996,Babiker2023}.   This definition however is not valid for larger $Fr$ and $We$ where surface deformations are appreciable (non-linear), as seen for example in Fig.~\ref{fig:froude-compare}.  The orientation of the free-surface is quantified by plotting the distribution of its normal vector components in Fig.~\ref{fig:pdf-normals-low} for the lowest $Fr$ number case (E): it is evident that the $x$ and $y$ components are symmetrically distributed around zero, and that $n_z$ is dominant.    Comparing the vertical component $n_z$ for the different $Fr$ we considered (see Fig. \ref{fig:pdf-nz}), we observe that even though $n_z \approx 1$ is most common, the distribution tends to be flatter indicating that the horizontal components become more relevant as the free-surface becomes increasingly deformed.  Hence we propose that the surface divergence to be evaluated using the 3D normal vector as,
\begin{equation}
    \label{eq:fs-div}
    \nabla_\bot \cdot \mathbf{u} = n_x \partial_x u  + n_y \partial_y v  + n_z \partial_z w.
\end{equation}

An alternative metric indicative of the deformation of the interface is related to the net stretching of the interface by the strain rate tensor $\mathbf{S}=(\nabla \mathbf{u}+ \nabla \mathbf{u}^{\top})/2$ introduced by \cite{Vela2021}, and written out in Cartesian coordinates as:
\begin{equation}
  \label{eq:vartheta}
  \begin{split}
    {\vartheta} &= -\mathbf{n}^\top \cdot \mathbf{S} \cdot \mathbf{n}  \\
     &= n_x^2 \partial_x u + n_y^2 \partial_v + n_z^2 \partial_z w +  \\
     &\;\;\;\;n_x n_y (\partial_y u + \partial_x v)  + n_x n_z (\partial_x w + \partial_z u) + n_y n_z (\partial_y w + \partial_z v)
     \end{split}
\end{equation}
This surface energy increase from net stretching $\vartheta > 0$ (or $\vartheta < 0$ in the case of compression) is scaled by the surface tension coefficient, or $We$ number as $\hat{\vartheta} = \vartheta / We$. The total energy being absorbed or released from the interface is thus proportional to $\Psi_\sigma \propto - \int_\Gamma \hat{\vartheta} d\Gamma$, and can also be related to the rate of change of total interfacial area \citep[see][]{Dodd2016,Calado2024-PRF}.
The resemblance between Eq. \ref{eq:fs-div} and \ref{eq:vartheta} is clear.
In Fig. \ref{fig:pdf-surf-div-sq} we also observe that $(\nabla_\bot \cdot \mathbf{u})^2$ and $\omega_z^2$ at the free-surface exhibit a power law tail for small values, attributed to a chi-squared type distribution where small-velocity-gradients acts as random variables, highlighted in the experiments of \cite{Qi2025}. These variables involve a single squared term, resulting in $\beta = 1$ degrees of freedom, and a $\beta/2-1=-1/2$ scaling.

\begin{figure}
       \begin{subfigure}{0.3\textwidth}
        \centering
        \includegraphics[width=\textwidth]{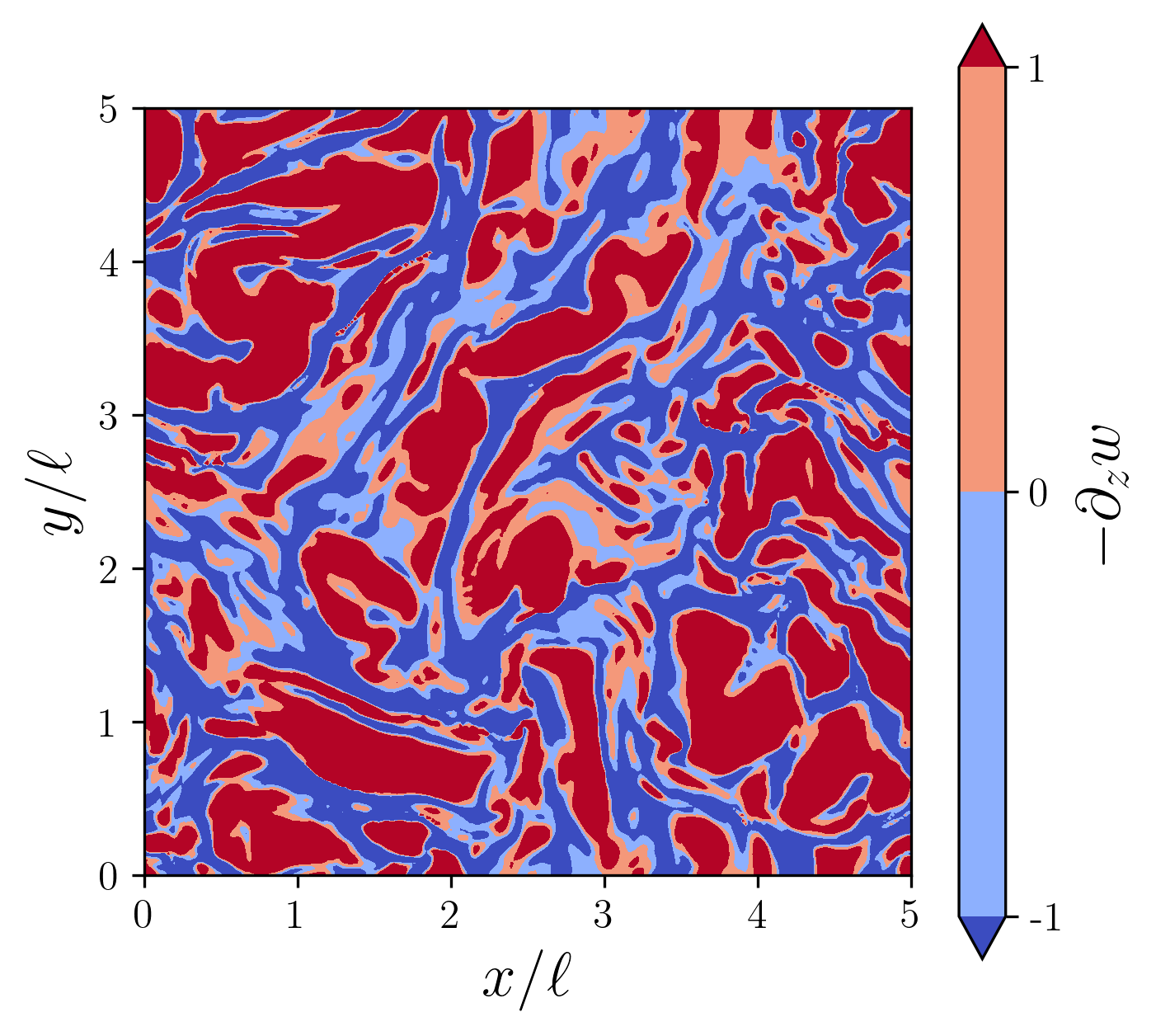}
        \caption{$- \partial_z w$}
        \label{fig:dwdz-map}
    \end{subfigure}
   \quad
    \begin{subfigure}{0.3\textwidth}
        \centering
        \includegraphics[width=\textwidth]{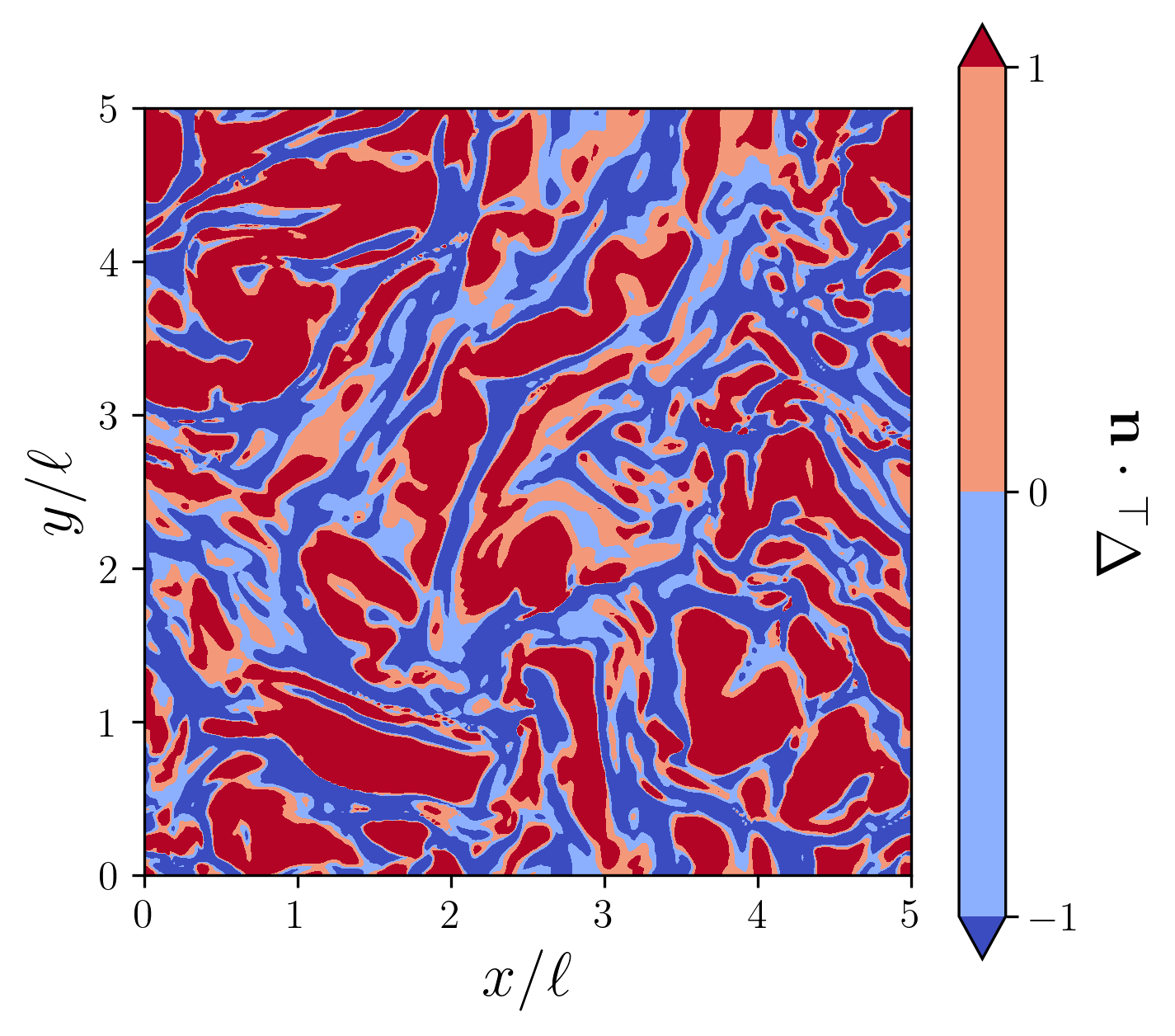}
        \caption{$\nabla_\bot \cdot \mathbf{u}$}
        \label{fig:surf-div-map}
    \end{subfigure}
        \quad
    \begin{subfigure}{0.3\textwidth}
        \centering
        \includegraphics[width=\textwidth]{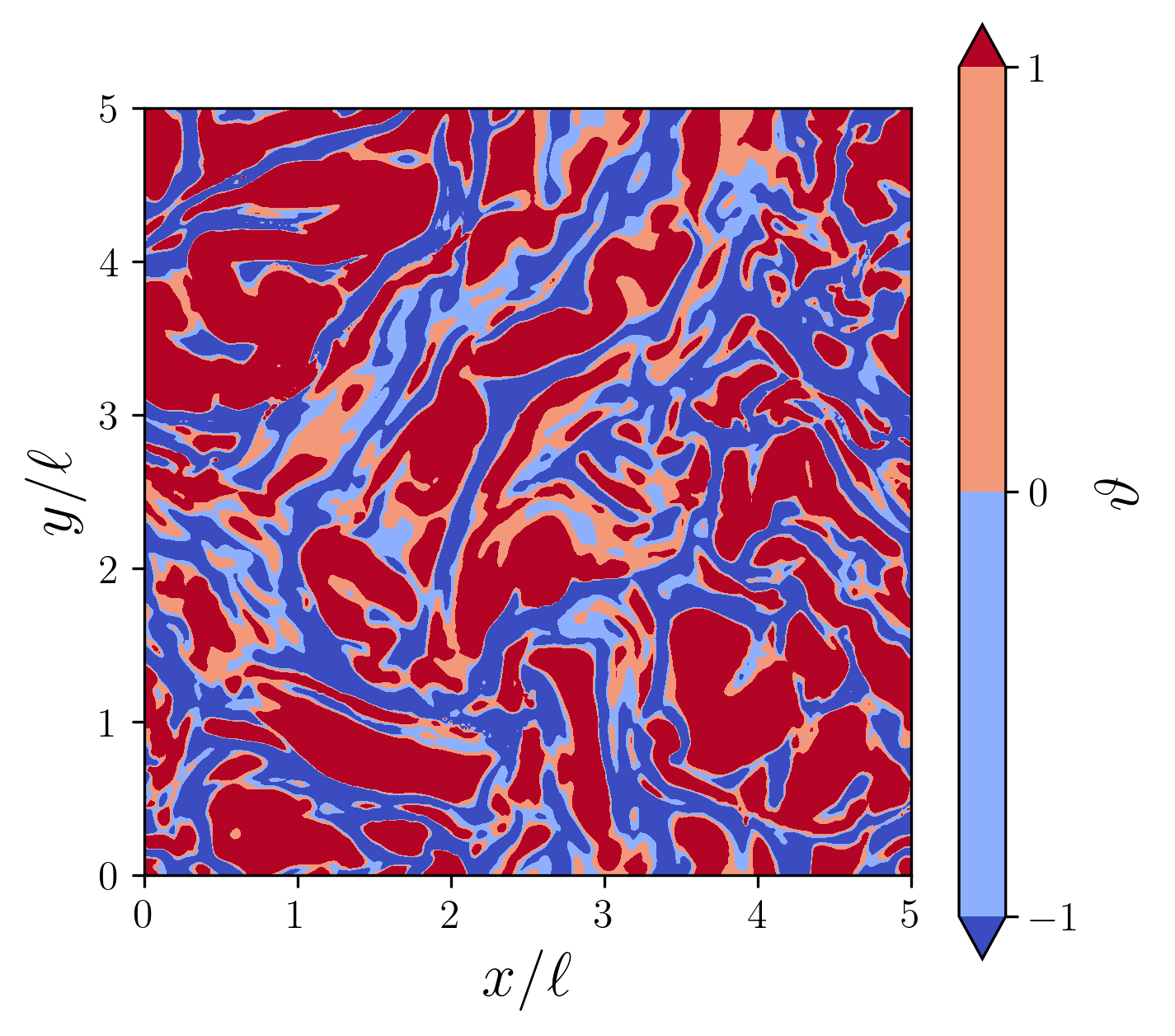}
        \caption{$\vartheta$}
        \label{fig:surf-en-map}
    \end{subfigure}
\caption{Comparison of free-surface contours for different surface deformation metrics for a snapshot of case E.}
    \label{fig:surface-maps}
\end{figure}

Ensemble averaged statistics for the free-surface are presented in Table \ref{tab:FS-stats}. We compute different quantities which reflect the extent of deformation experienced by the free-surface for the different cases.  These include the intermittency layer $\delta_i$, the ratio of total free-surface area $A_{FS}$ with initial flat surface area $A_0 = L^2$, and the surface integral $\int_{\Gamma} \vartheta d\Gamma$ indicating the net stretching effect. 
The PDF for $\vartheta$ is characterized in terms of the mean, standard deviation, skewness and kurtosis.

We observe that the free-surface statistics are comparable between cases A, B and C, with a relative increase in surface area around $\sim 20 \%$ and weakly dependent on $We$. 
Note that for this problem since $Fr^2 \ll We$, it is gravity that dominates the interfacial dynamics at larger scales, even though these three cases are considered to be in the SFST regime as shown in Fig. \ref{fig:fr-high}, with air entrainment.
The intermittency layer is found to be of the order of the integral length scale $\delta_i \sim \ell$, although decreasing with $Fr$.
Decreasing $Fr$ in cases D and E shows a reduction in the total surface area and net stretching effect from $\vartheta$, since the gravity force increases which stabilizes any disruptions from the free-surface. 
Finally for the high $Re$ case F, while the total surface area is comparable to cases A-C, the statistics for the mean and variance of $\vartheta$ are much higher, indicative of higher velocity gradients. 

Finally we comment on the fact that all cases show a negative skewness and high kurtosis (super-Gaussian) for $\vartheta$, which indicate large tails, with more extreme values of $\vartheta < 0$, or compression. These are associated with the ridges seen in Fig. \ref{fig:surface-maps} which are thin and elongated, compared to regions of upwelling where $\vartheta > 0$. As also observed by \cite{Longuet-Higgins_1996}, the regions of downwelling have signature steeper waves, and they surround smoother regions of upwelling.
This can be tied to the upwelling vs. downwelling events and their energy. \citet{Ruth2024} pointed out that downwellings compress energy to small scales, and are less energetic compared to upwellings, which are associated with larger scales. Our results for the TKE exchange near the free-surface are presented in \S \ref{sec:tke-exchange}.

\begin{table}
  \begin{center}
\def~{\hphantom{0}}
  \begin{tabular}{lccccccc}
      Case & \quad $\delta_i / \ell$ & \quad  $A_{FS} / A_0$ & \quad $\int_{\Gamma} \vartheta d\Gamma$ &  \quad Avg. $\vartheta$   &   \quad  Std. Dev. $\vartheta $ & \quad  Skew. $\vartheta$ & \quad  Kurt. $\vartheta$  \\[3pt]
      A  & \quad 0.98 & \quad 1.22 & \quad 5.57 & \quad 0.22 &\quad 5.69 &\quad -1.62 &\quad 8.12 \\
      B & \quad 0.88 &\quad 1.24 &\quad 7.60 &\quad 0.30 &\quad 4.85 &\quad -1.47 &\quad 6.78   \\
      C & \quad 0.98 &\quad 1.21 &\quad 7.47 &\quad 0.30 &\quad 4.99 &\quad -1.60 &\quad 7.44  \\
     D & \quad 0.78 &\quad 1.13 &\quad 6.21 &\quad 0.25 &\quad 8.44 &\quad -1.63 &\quad 7.78  \\
     E & \quad 0.49 &\quad 1.04 &\quad 1.19 &\quad 0.10 &\quad 8.33 &\quad -1.02 &\quad 4.43  \\
      F & \quad 0.80 &\quad 1.18 &\quad 12.65 &\quad 0.51 &\quad 16.61 &\quad -1.48 &\quad 9.96  \\
  \end{tabular}
  \caption{Ensemble statistics for free-surface deformation.}
  \label{tab:FS-stats}
  \end{center}
\end{table}

\subsection{Two-dimensional compressibility}
\begin{figure}
           \begin{subfigure}{0.3\textwidth}
        \centering
        \includegraphics[width=\textwidth]{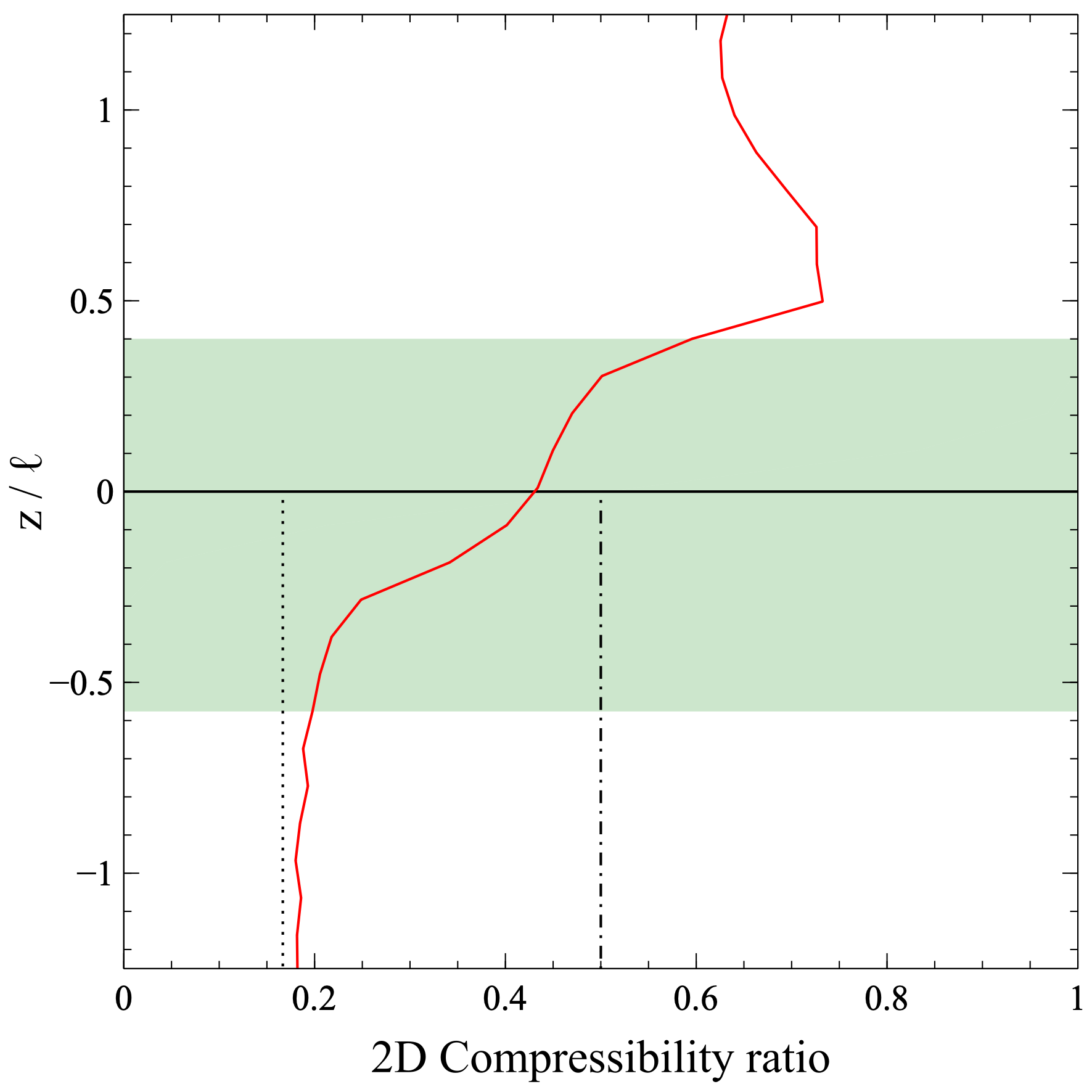}
        \caption{Case A}
    \end{subfigure}
   \quad
    \begin{subfigure}{0.3\textwidth}
        \centering
        \includegraphics[width=\textwidth]{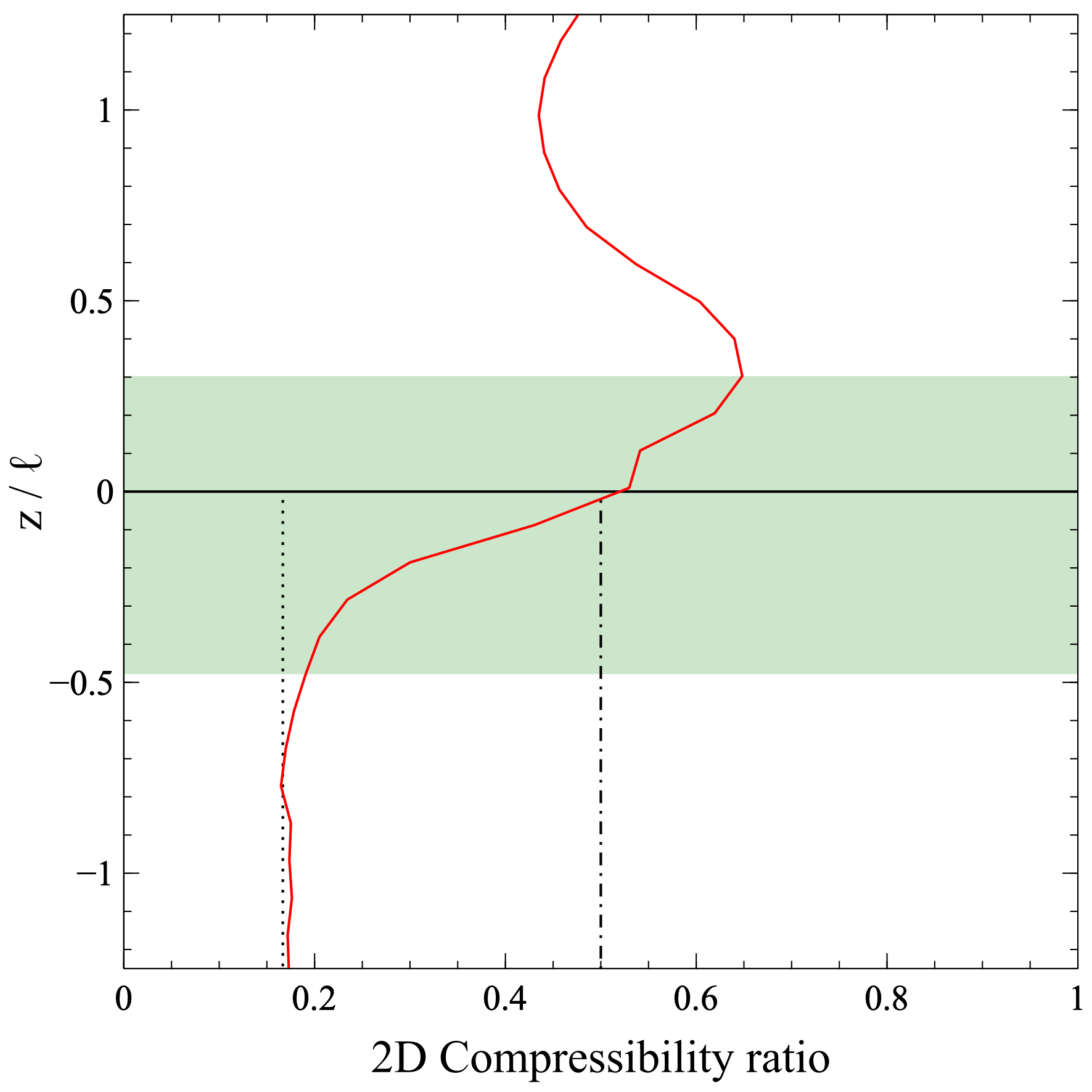}
        \caption{Case D}
    \end{subfigure}
        \quad
           \begin{subfigure}{0.3\textwidth}
        \centering
        \includegraphics[width=\textwidth]{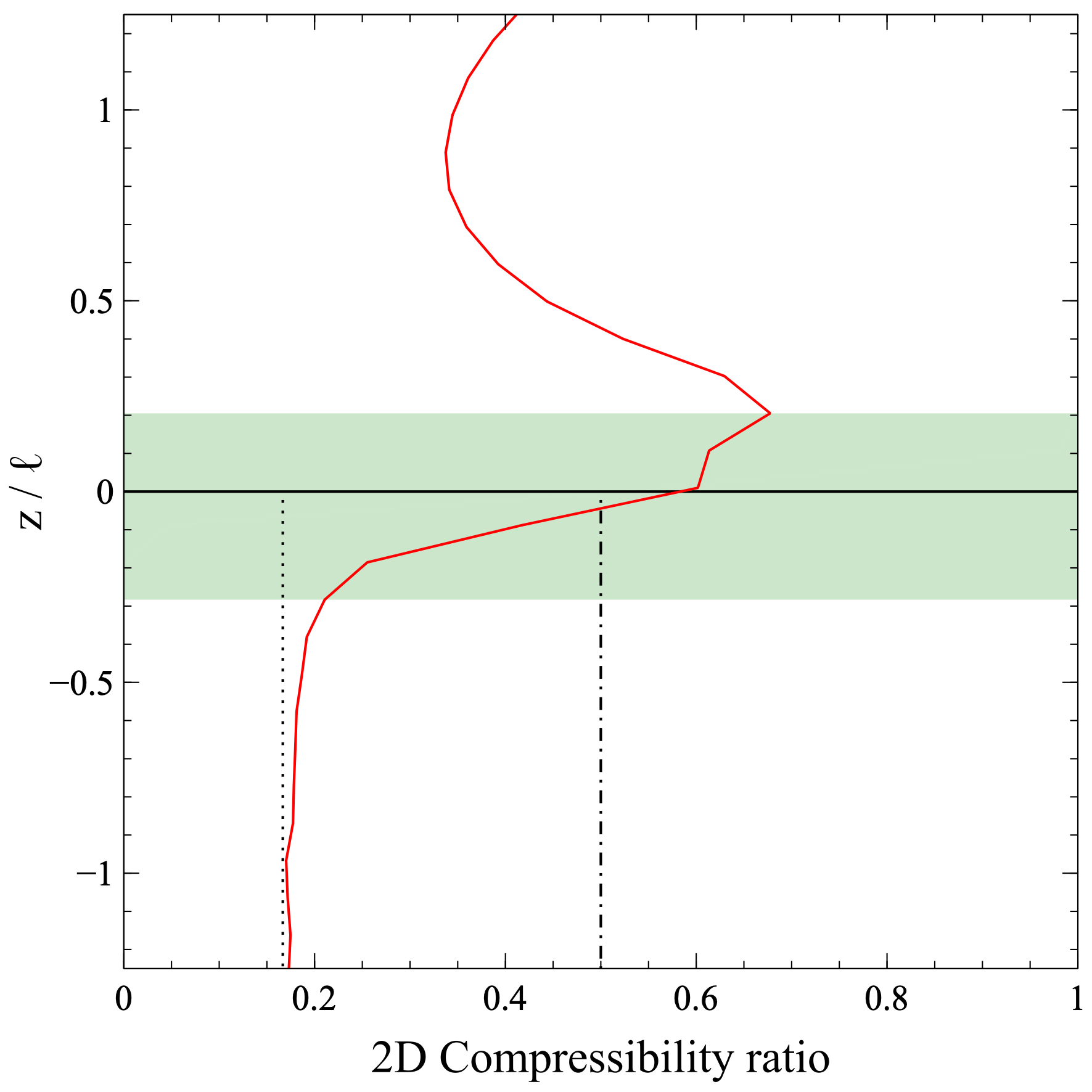}
        \caption{Case E}
    \end{subfigure}
\caption{2D compressibility ratio $\mathcal{C}$ as a function of depth. Green shaded region indicates the intermittency layer $\delta_i$; vertical lines shown for $\mathcal{C}=1/6$ (dotted) and $\mathcal{C}=1/2$ (dash-dotted). }
    \label{fig:comp-ratio}
\end{figure}

Another measure of the free-surface dynamics is the Eulerian compressibility ratio which can take values between 0 and 1.
The compressibility effect can be linked to the directionality of the energy cascade \citep{Chertkov1998,Lovecchio2015} and in free-surface flows results in Lagrangian particle clustering \citep{Boffetta2004}. 
By definition this factor is zero in three-dimensional incompressible flow, however, in the case of homogeneous flow in the plane, this ratio can be computed in two-dimensions as:
\begin{equation}
    \label{eq:compressibility}
    \mathcal{C} = \frac{\langle (\nabla_{xy} \cdot \mathbf{u})^2 \rangle}{\langle |\nabla_{xy}  \mathbf{u}|^2 \rangle} = \frac{\langle (\partial_x u + \partial_y v)^2 \rangle}{ \langle (\partial_x u)^2 + (\partial_x v)^2 + (\partial_y u)^2 + (\partial_y v)^2 \rangle }
\end{equation}
Note that in this definition we consider a continuous velocity field and ignore phase distribution.  A 2D cut through 3D HIT results in an asymptotic value of $\mathcal{C} = 1/6$, while values close to $1/2$ are observed near the free-surface in low $Fr$ number experiments and DNS with a slip wall boundary condition \citep{Cressman2004,Lovecchio2015}. The value of $\mathcal{C}=1/2$ is also the theoretical limit for compressible 2D Kraichnan flow \citep{Chertkov1998} and also implies a null cross-correlation for $\langle \partial_xu \, \partial_y v \rangle$ \citep{Boffetta2004}. Results from our DNS are given in Fig.~\ref{fig:comp-ratio} for different $Fr$ numbers.   We observe a similar behavior to previous studies, with an increasing value of $\mathcal{C}$ from $1/6$ in the pure water depths to $\mathcal{C} \approx 1/2$ at $z=0$. Since data from the air side is also available, we plot the values for $z>0$, indicating a continued increase until the upper limit of $\delta_i$ is reached.   As the $Fr$ number is decreased, the value of $\mathcal{C}$ at $z=0$ decreases, with a stronger inflection for the highest $Fr$ case. This shows that as turbulence becomes stronger, the apparent two-dimensional compressibility drops in magnitude while $\delta_i$ grows. 

\subsection{Vorticity flux}
\label{sec:vorticity}

In an incompressible interfacial flow, the only mechanism to generate vorticity is the inviscid relative acceleration between fluids due to tangential pressure gradients or body forces \citep[see ][]{Terrington2022b}. Surface-normal vorticity can appear either from advection across the interface, or from viscous diffusion of tangential vorticity. This leads to the attachment of vortex filaments to the free-surface, such that the kinematic condition is satisfied (vortex filaments do not end inside a fluid).   The no-slip condition between two fluids imposes the normal vorticity component to be continuous, while the tangential components are related to the interface curvature \citep{Dopazo2000}.

\begin{figure}
    \centering
    \begin{subfigure}{0.4\textwidth}
        \centering
        \includegraphics[width=\textwidth]{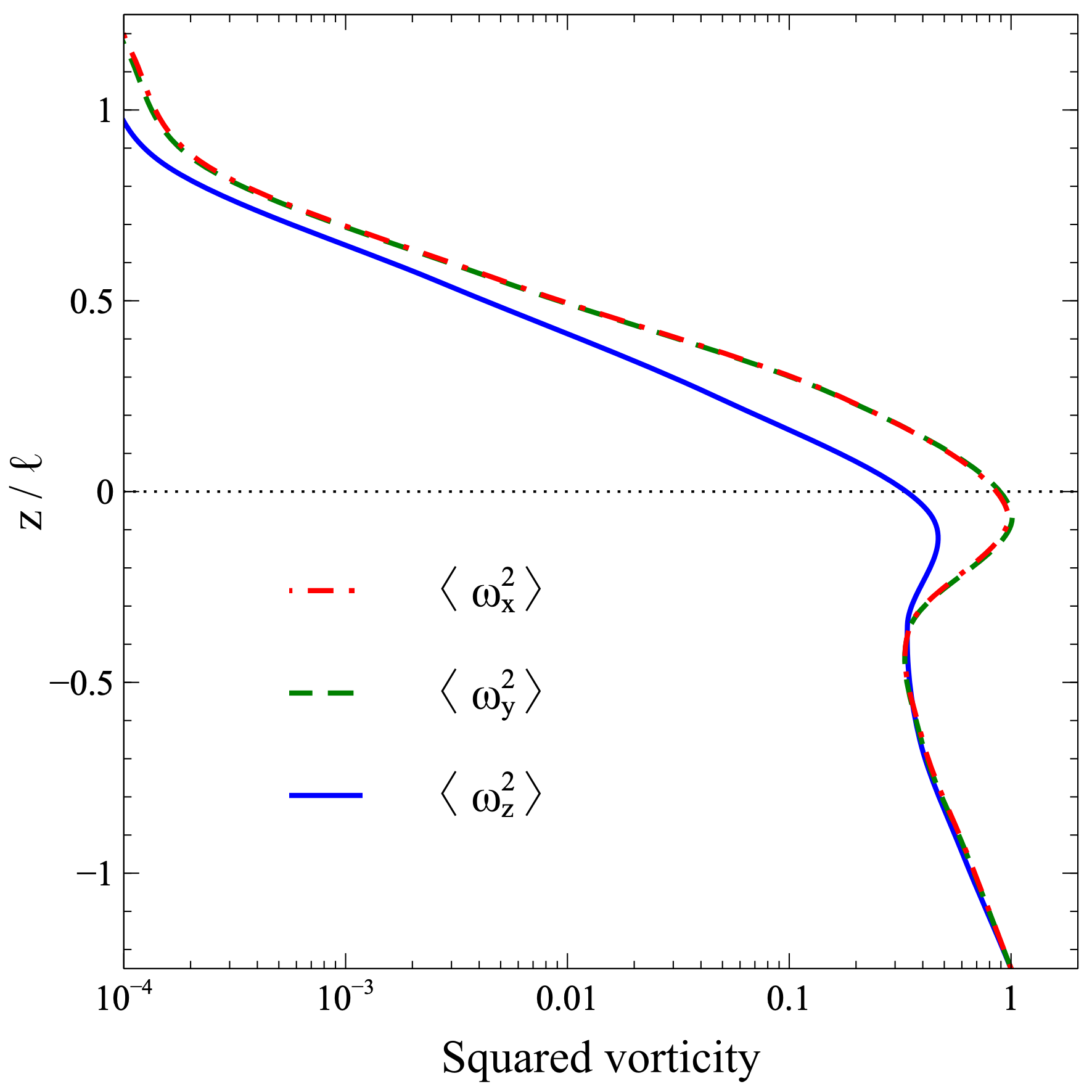}
        \caption{}
        \label{fig:vort-squared}
    \end{subfigure}
    \qquad 
    \begin{subfigure}{0.4\textwidth}
        \centering
        \includegraphics[width=\textwidth]{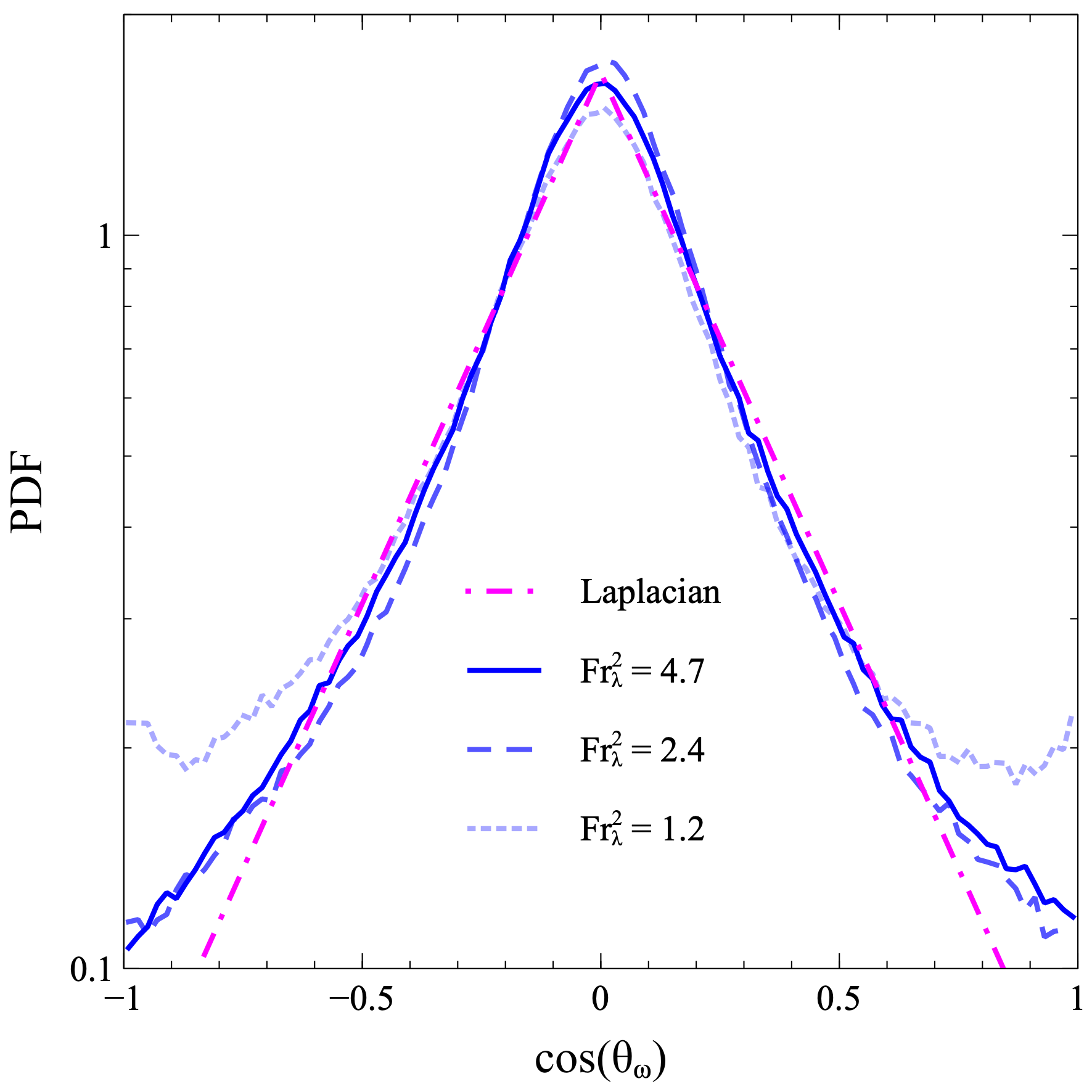}
        \caption{}
        \label{fig:vort-align-pdf}
    \end{subfigure}
\caption{(a) Squared vorticity components (non-phase averaged), normalized by values at $z/\ell=-1.25$ for case A, (b) PDF of $\textrm{cos}(\theta_\omega)$ for different $Fr$, and Laplacian distribution. }
    \label{fig:vort-align-stats}
\end{figure}

Ignoring the absolute value in the sign of vorticity, we compute the averages of squared vorticity components, $\langle \omega_i^2 \rangle$, and plot them in Fig. \ref{fig:vort-squared} for case A. Note these averages are not phase-averaged. The vorticity is seen to be isotropic below $z/\ell < -0.5$, with a change in behavior approaching $z=0$, with an inflection for all components, with a stronger increase in horizontal components. This was consistently the case for all cases, although as $Fr$ decreases, the increase in $z$ component becomes insignificant. The vorticity flux through the free-surface can be evaluated using the dot product between the vorticity $\bm{\omega} = \nabla \times \mathbf{u}$ and the normal vector:
\begin{equation}
    \label{eq:vort-align}
    \textrm{cos}(\theta_{\omega}) = \frac{\mathbf{n}\cdot \bm{\omega}}{|| \bm{\omega} ||}
\end{equation}
Plotting the PDF of this function shows a symmetric distribution around $\textrm{cos}(\theta_\omega) = 0$ that can be approximated by a Laplacian distribution $\mathcal{L} (x|s) = \textrm{exp}( - |x|/s)/(2s)$ with scale parameter $s=0.3$, as seen in Fig. \ref{fig:vort-align-pdf}. 
Using this model Laplacian distribution, equates to  $\sim 80 \%$ probability for the angle $\theta_\omega$ being greater than $60^\circ$.  This distribution was also found to be weakly dependent on $Fr$ and $We$. These results indicate that the vorticity is mainly oriented parallel to the free-surface, in agreement with other works for free-surface and disperse two-phase flows \citep{Sarpkaya1996,Masnadi2019,Vela2021,Yu2019,Bardet,Calado2024-PRF}.

\begin{figure}
\centering
    \begin{subfigure}{0.45\textwidth}
        \centering
        \includegraphics[width=\textwidth]{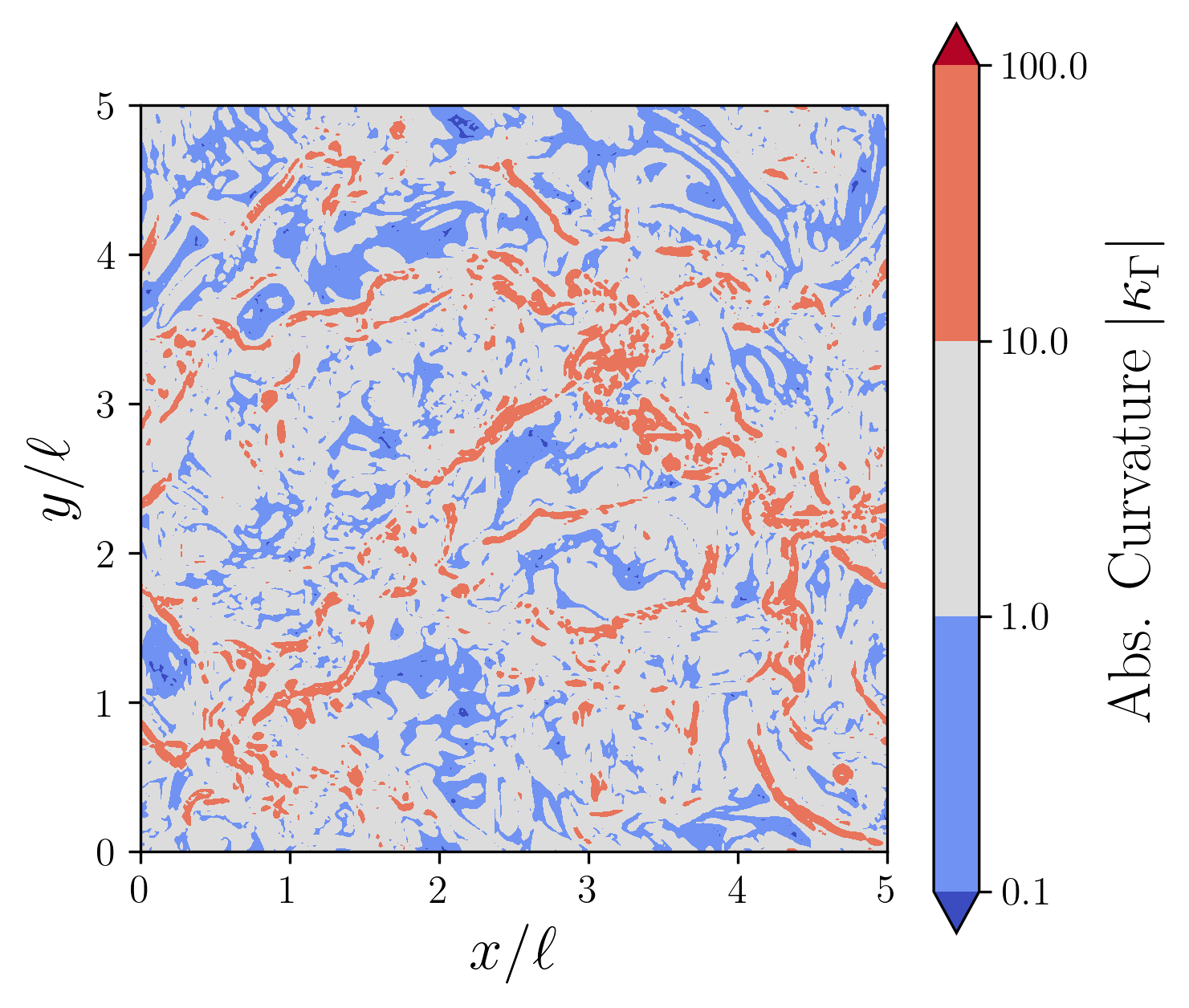}
        \caption{}
        \label{fig:curv-map}
    \end{subfigure}
      \qquad
    \begin{subfigure}{0.45\textwidth}
        \centering
        \includegraphics[width=\textwidth]{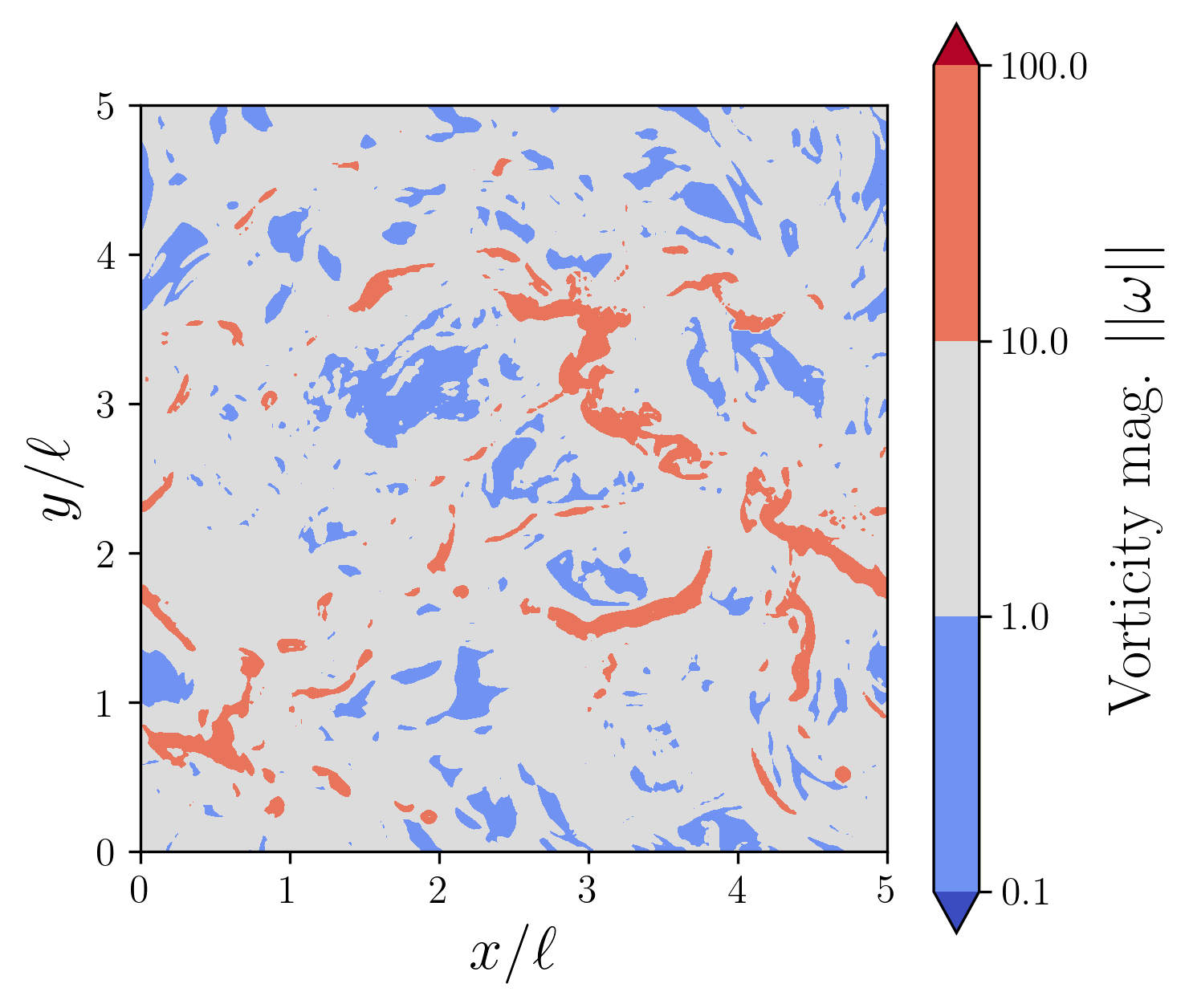}
        \caption{}
        \label{fig:vort-mag-map}
    \end{subfigure}
\caption{Snapshot contour maps of free-surface on log scale (a) absolute interface curvature, (b) vorticity magnitude. Data refers to case A.  }
    \label{fig:vort-and-curv}
\end{figure}

\begin{figure}
\centering
        \includegraphics[width=0.9\textwidth]{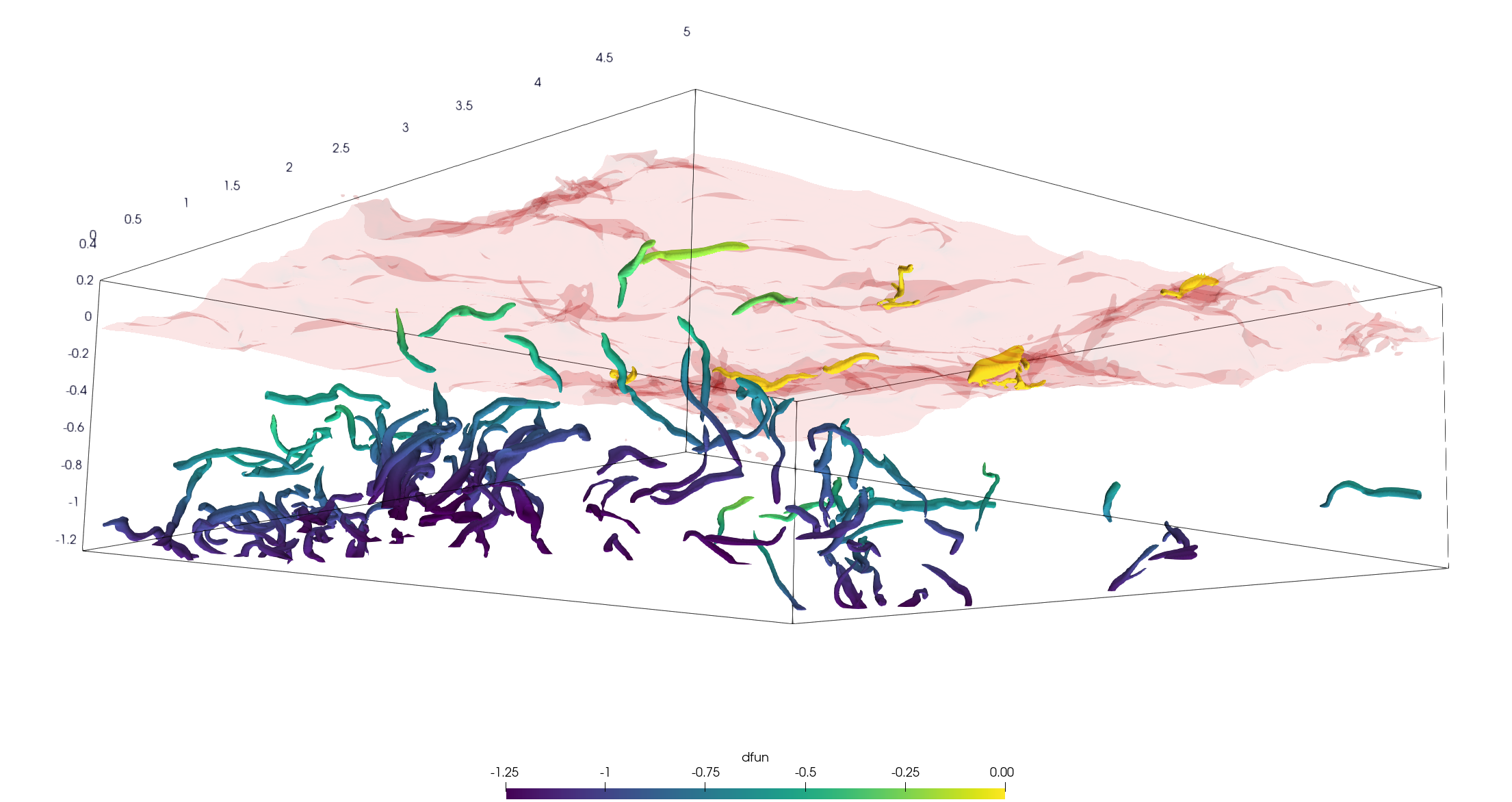}
\caption{$Q$ criterion iso-surfaces above $z/\ell > -1.25$, colored by distance to interface for case F. Only a limited number of the largest structures are shown for clarity. }
    \label{fig:q-crit}
\end{figure}

Comparing the plots of the absolute value for interface curvature $| \kappa_\Gamma|$ and the vorticity magnitude $|| \bm{\omega}||$ on a logarithmic scale in Fig. \ref{fig:vort-and-curv}, we can observe that regions of high curvature form continuous streaks or scars, as a characteristic signature of free-surface flows \citep{Brocchini2001}.   The vorticity magnitude is seen to be highly correlated with the absolute value of curvature. From these plots and the observed distributions of $\theta_\omega$ we can infer that most of the vorticity flux across the free-surface is attributed to the viscous diffusion of tangential vorticity \citep{Dopazo2000,Terrington2022,Terrington2022b}.   This is clearly visualized in Fig.~\ref{fig:q-crit} by plotting $Q = (||\mathbf{\Omega}||^2_F - ||\mathbf{S}||^2_F)/2$ criterion ($Q>0$) iso-surfaces near the free-surface, where $\mathbf{\Omega}=(\nabla \mathbf{u}- \nabla \mathbf{u}^{\top})/2$ is the rotation rate tensor, and $|| \cdot||^2_F$ is the Frobenius norm squared. Within roughly one integral length scale (blockage layer), the vortices transition from randomly oriented in the bulk, towards being parallel to the free-surface.

\section{Kinetic energy exchange}
\label{sec:tke-exchange}

\subsection{Reynolds stresses}
\begin{figure}
\centering
    \begin{subfigure}{0.35\textwidth}
        \centering
        \includegraphics[width=\textwidth]{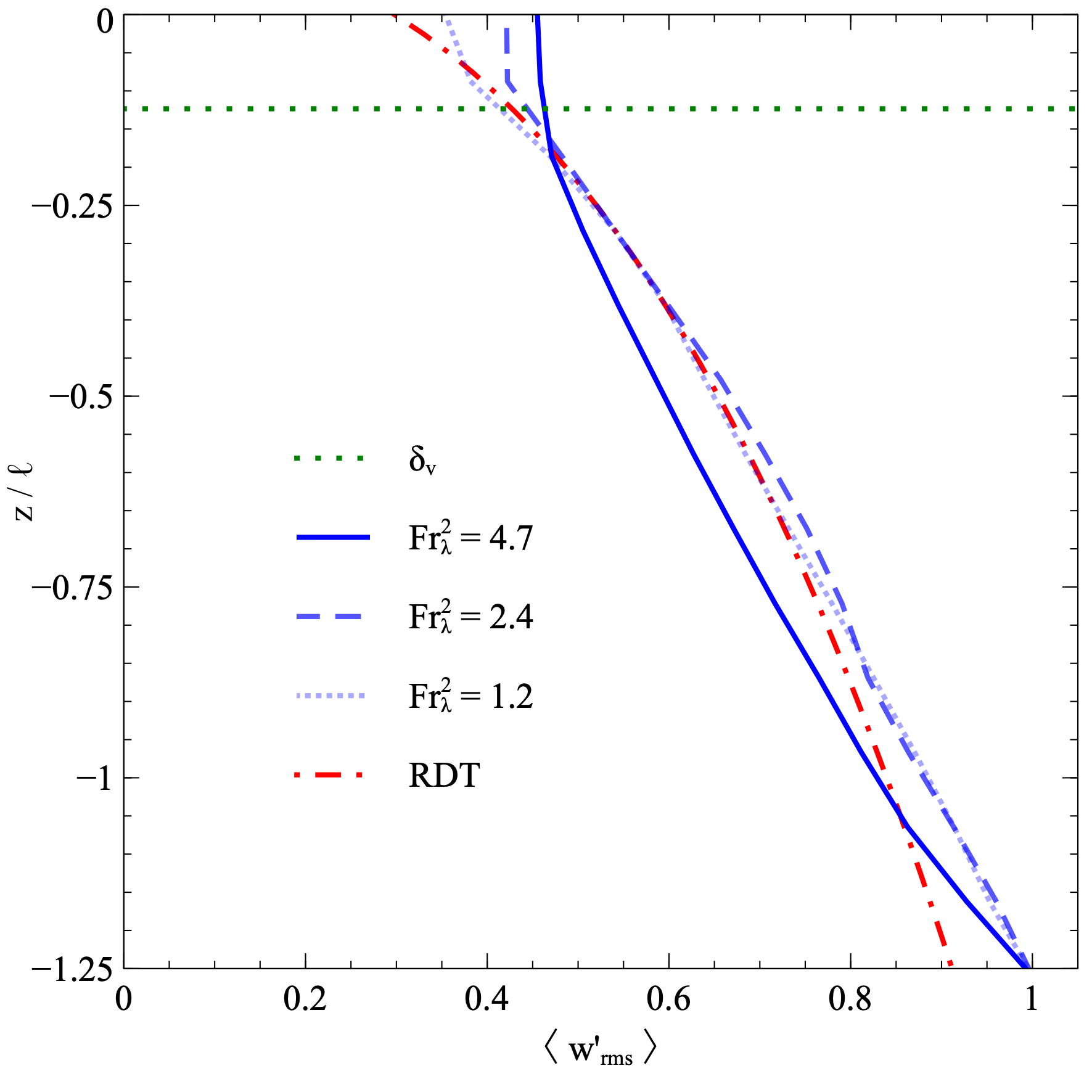}
        \caption{}
        \label{fig:w-rms}
    \end{subfigure}
      \qquad
    \begin{subfigure}{0.35\textwidth}
        \centering
        \includegraphics[width=\textwidth]{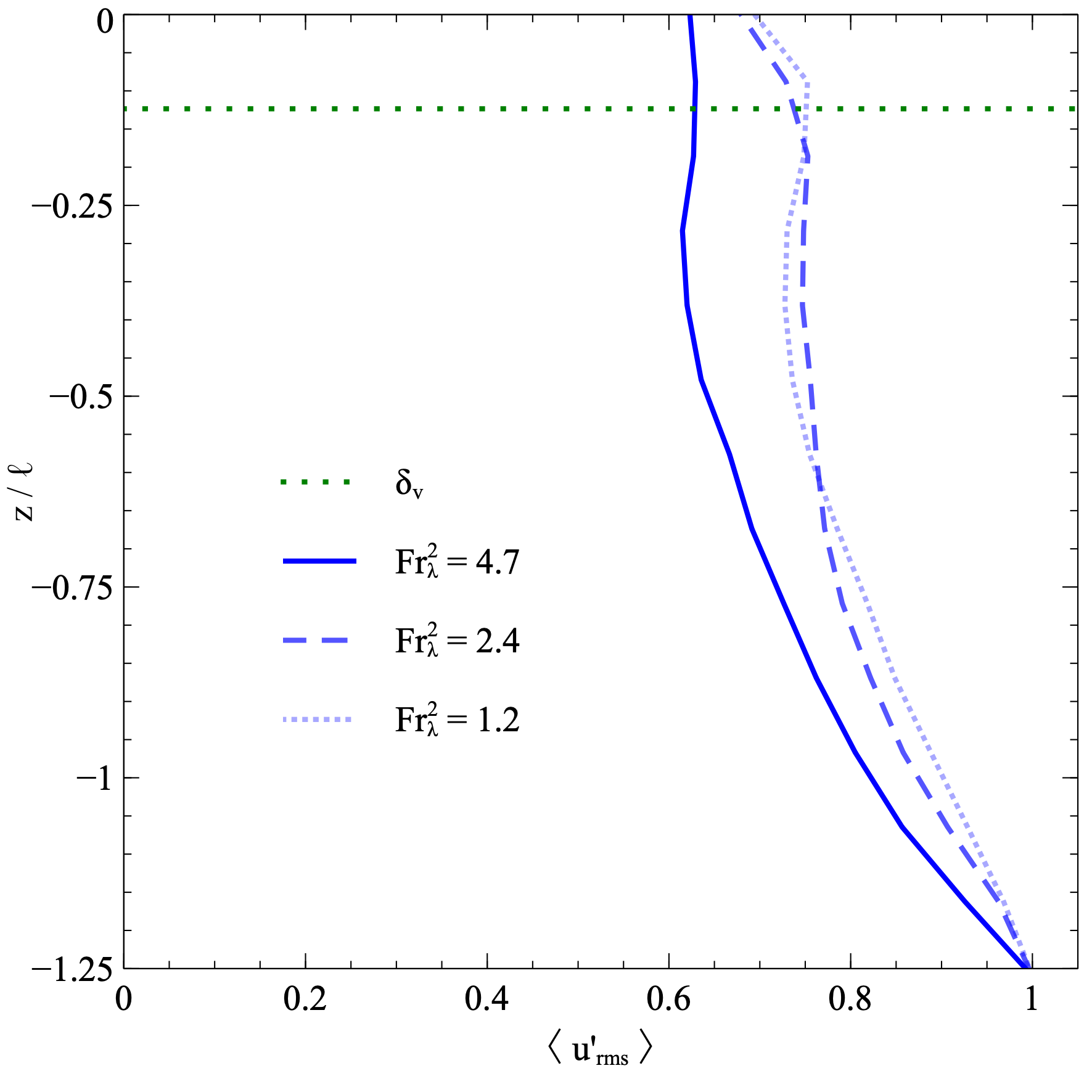}
        \caption{}
        \label{fig:u-rms}
    \end{subfigure}
\caption{RMS velocities for different $Fr$ (a) vertical component, (b) horizontal component. Data normalized by values at $z/\ell=-1.25$.  }
    \label{fig:rms-vel}
\end{figure}

First we look at the Reynolds stress components near the free-surface, and how they are modulated by the $Fr$ number.  In Fig.~\ref{fig:rms-vel} we compare the vertical fluctuating velocities for different $Fr$ numbers to RDT predictions.  Note that $\langle w'_{rms} \rangle \propto (|z| + \delta_v)^{1/3}$, with $\delta_v \approx 4 Re_{\ell}^{-1/2}$ \citep{Hunt1978,Variano2008}.  The vertical component $\langle w'_{rms}\rangle$ is seen to agree well with the trend from RDT at lower $Fr$ numbers. Potential discrepancies here can be tied to the moderate $Re$ number, since RDT is based on the inviscid assumption \citep[see][]{Ruth2024}. The horizontal components are seen to remain higher than vertical components, particularly for the lower $Fr$ numbers where blockage effects are increased.  We also note an inflection point above the viscous layer $\delta_v$. 

\begin{figure}
\centering
        \includegraphics[width=0.8\textwidth]{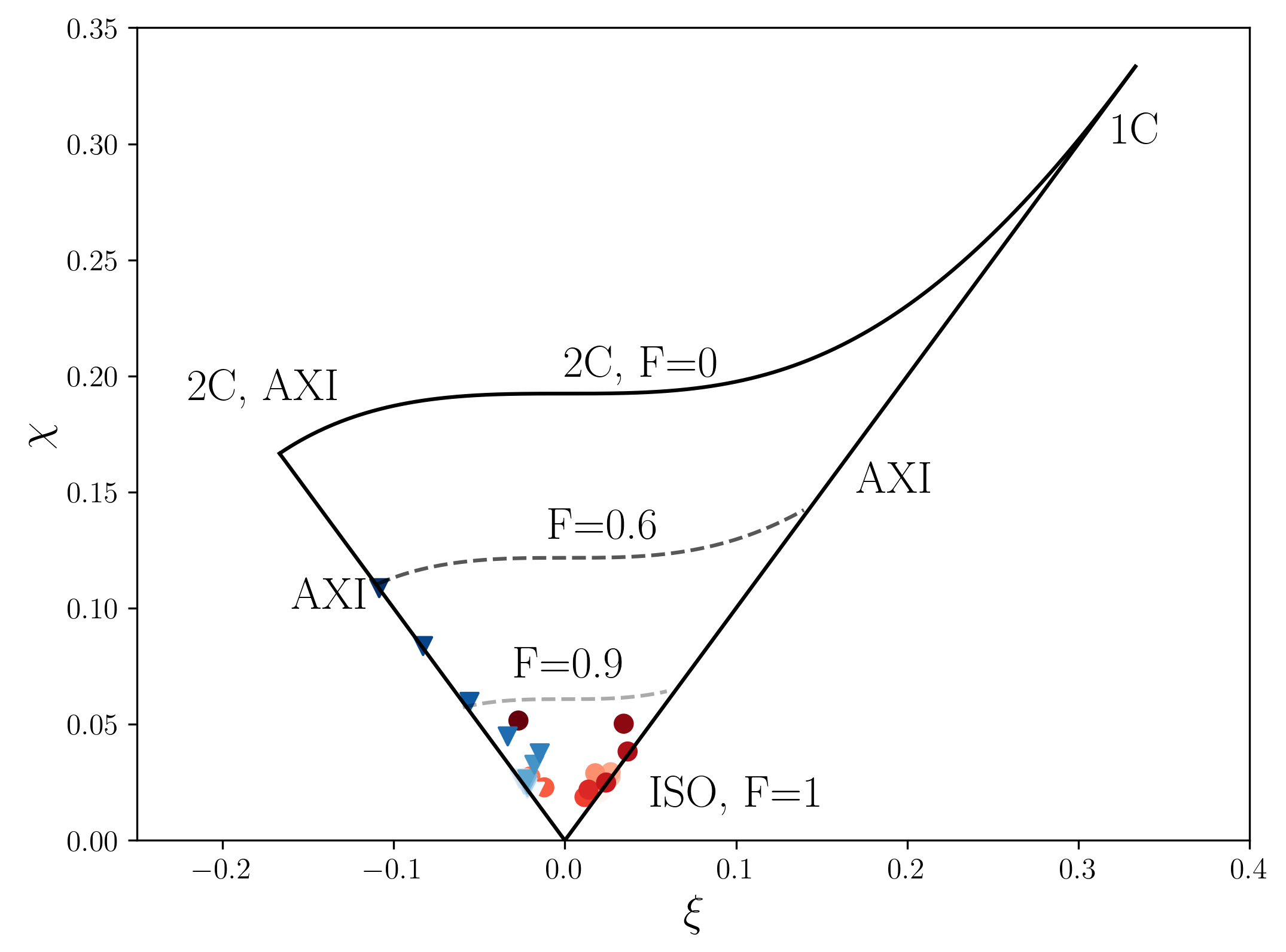}
\caption{Lumley triangle plot for case A (red circles), and case E (blue triangles) for $-1.25 \leq z \leq0$. Darker hue of the crosses correspond to depths approaching $z=0$. }
    \label{fig:lumley}
\end{figure}

The variation in Reynolds stress components can alternatively be visualized by plotting the Lumley triangle as a function of depth \citep{Pope2000,Yu2019}. Starting from the normalized anisotropy tensor $b_{ij} = {\langle u_i u_j \rangle } / {\langle u_k u_k \rangle}- 1/3 \delta_{ij}$ we compute the two independent tensor invariants $\chi$ and $\xi$, defined as $ \chi  = (b_{ij}b_{ji} / 6)^{1/2}$ and $\xi  = (b_{ij}b_{jk}b_{ki} / 6)^{1/3}$.  The cases of lowest and highest $Fr$ numbers (A and E) are plotted in Fig.~\ref{fig:lumley} for depths between between $z/\ell=-1.25$ and $z=0$. The iso-lines of $F=0.6$ and $F=0.9$ are also plotted for reference. We clearly see that in case A (higher $Fr$), the isotropy is maintained as the depth is decreased, with $F>0.9$. However, as $Fr$ is decreased in case E, the turbulence is modulated from isotropic to two-component axisymmetric, due to the increase in horizontal energy at the expense of vertical fluctuations.  These results oppose the postulation by \cite{Sarpkaya1996}, that any turbulence field approaching a free-surface becomes two-dimensional.  While this is certainly the case for lower $Fr$ numbers, the turbulence tends to remain isotropic in SFST with higher $Fr$ since there is reduced blockage effect.
 
\begin{figure}
           \begin{subfigure}{0.26\textwidth}
        \centering
        \includegraphics[width=\textwidth]{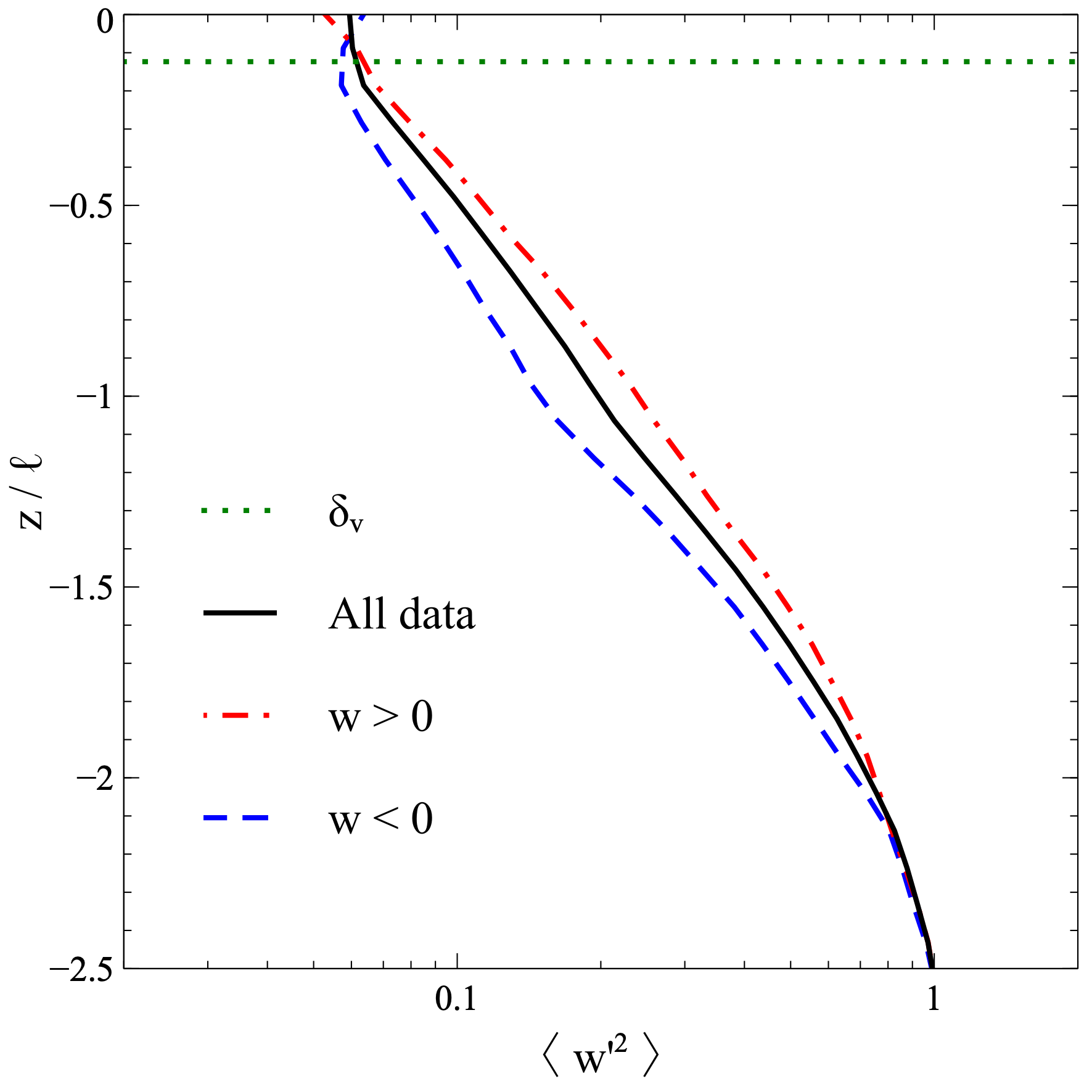}
        \caption{}
            \label{fig:cond-avg-w2}
    \end{subfigure}
   \qquad
    \begin{subfigure}{0.3\textwidth}
        \centering
        \includegraphics[width=\textwidth]{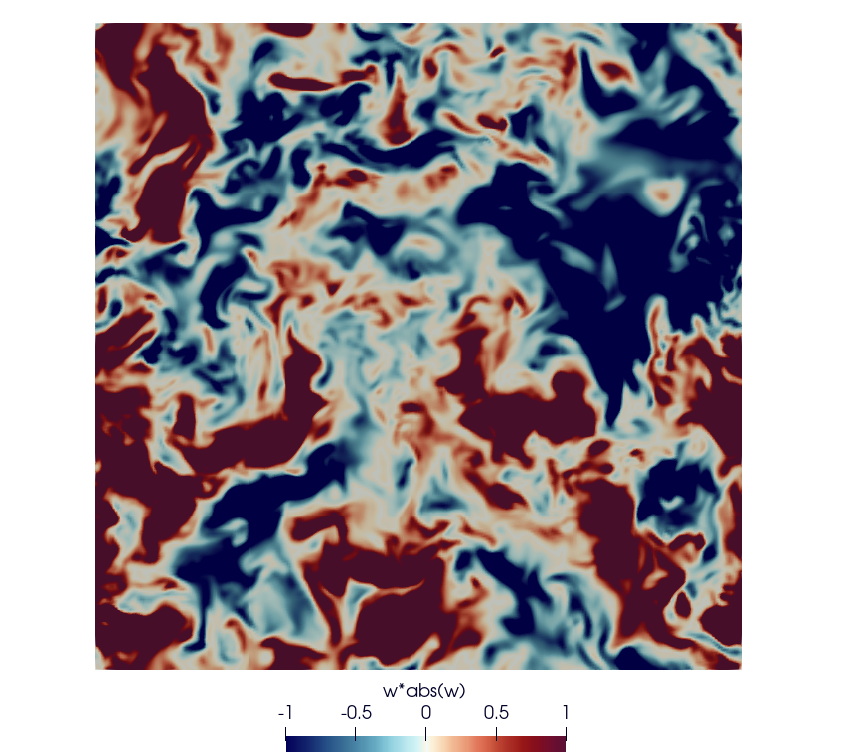}
        \caption{}
        \label{fig:w-sign-w-deep}
    \end{subfigure}
        \quad
           \begin{subfigure}{0.3\textwidth}
        \centering
        \includegraphics[width=\textwidth]{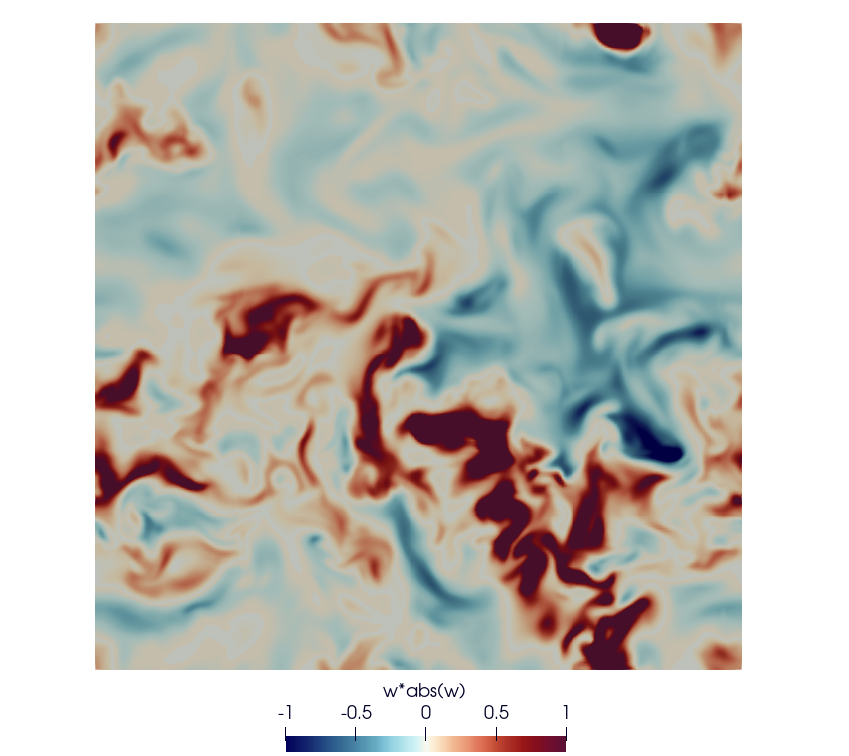}
        \caption{}
        \label{fig:w-sign-w-shallow}
    \end{subfigure}
\caption{Conditionally averaged statistics for vertical kinetic energy based on sign, normalized by values at $z/\ell=-2.5$ (a); instantaneous horizontal slice of $w'|w'|$ for depths $z/\ell=-2.5$ (b) and $z/\ell=-1$ (c). Data refers to case A. }
    \label{fig:w2-sign}
\end{figure}

Another relevant question investigated by \cite{Ruth2024} is the relative roll of upwellings vs. downwellings in transporting vertical kinetic energy. In the low $Fr$ number cases investigated in their work, they found that upwellings carry more vertical energy, while downwellings move more horizontal energy due to compressive effects from the free-surface.  We investigate these statistics for case A, by conditionally averaging the vertical energy, $\langle w'^2 \rangle$ based on its sign: $w' > 0$ for upwellings, and $w' < 0$ for downwellings. Results are given in Fig.~\ref{fig:cond-avg-w2}, on a semi-log scale and show qualitative agreement with the experiments. An instantaneous snapshot of $w'|w'|$ on horizontal slices at depths of $z/\ell=-2.5$ and $z/\ell=-1$ in Figs.~\ref{fig:w-sign-w-deep} and \ref{fig:w-sign-w-shallow} respectively, also supports this fact ($w'|w'|$ indicates both magnitude of kinetic energy and its relative motion). For greater depths, the turbulence does not sense the influence of the free-surface, and kinetic energy is transported equally in both vertical directions, while within the blockage layer, $\delta_s$, the upwellings bring the necessary energy to stretch out the free-surface. Within the viscous layer $\delta_v$ this ceases to be true, and vertical energy tends to recover isotropy.

\subsection{Presence of reverse cascade}
\label{sec:rev-cascade}
\begin{figure}
\centering
    \begin{subfigure}{0.3\textwidth}
        \centering
        \includegraphics[width=\textwidth]{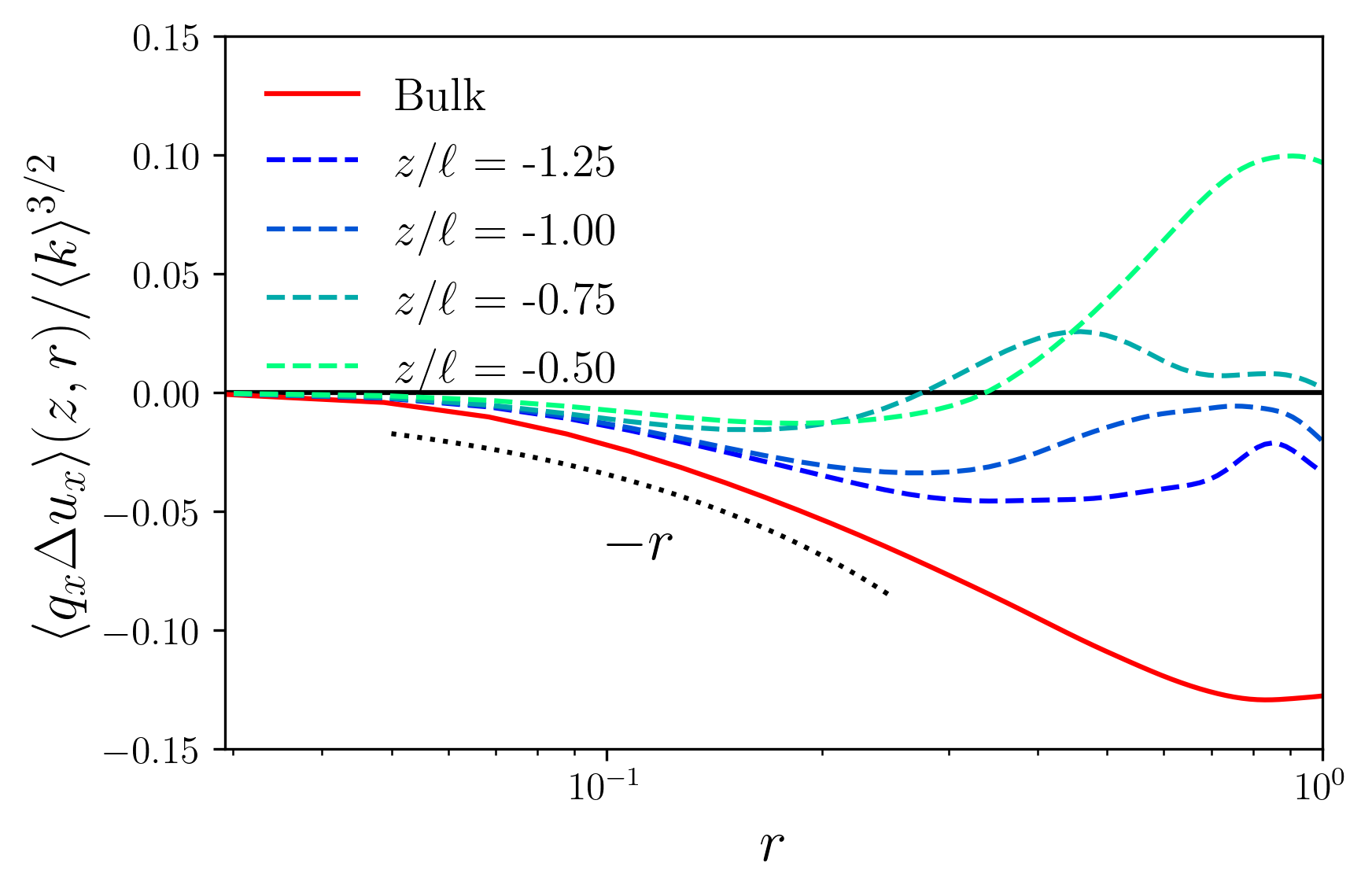}
        \caption{}
    \end{subfigure}
      \quad
    \begin{subfigure}{0.3\textwidth}
        \centering
        \includegraphics[width=\textwidth]{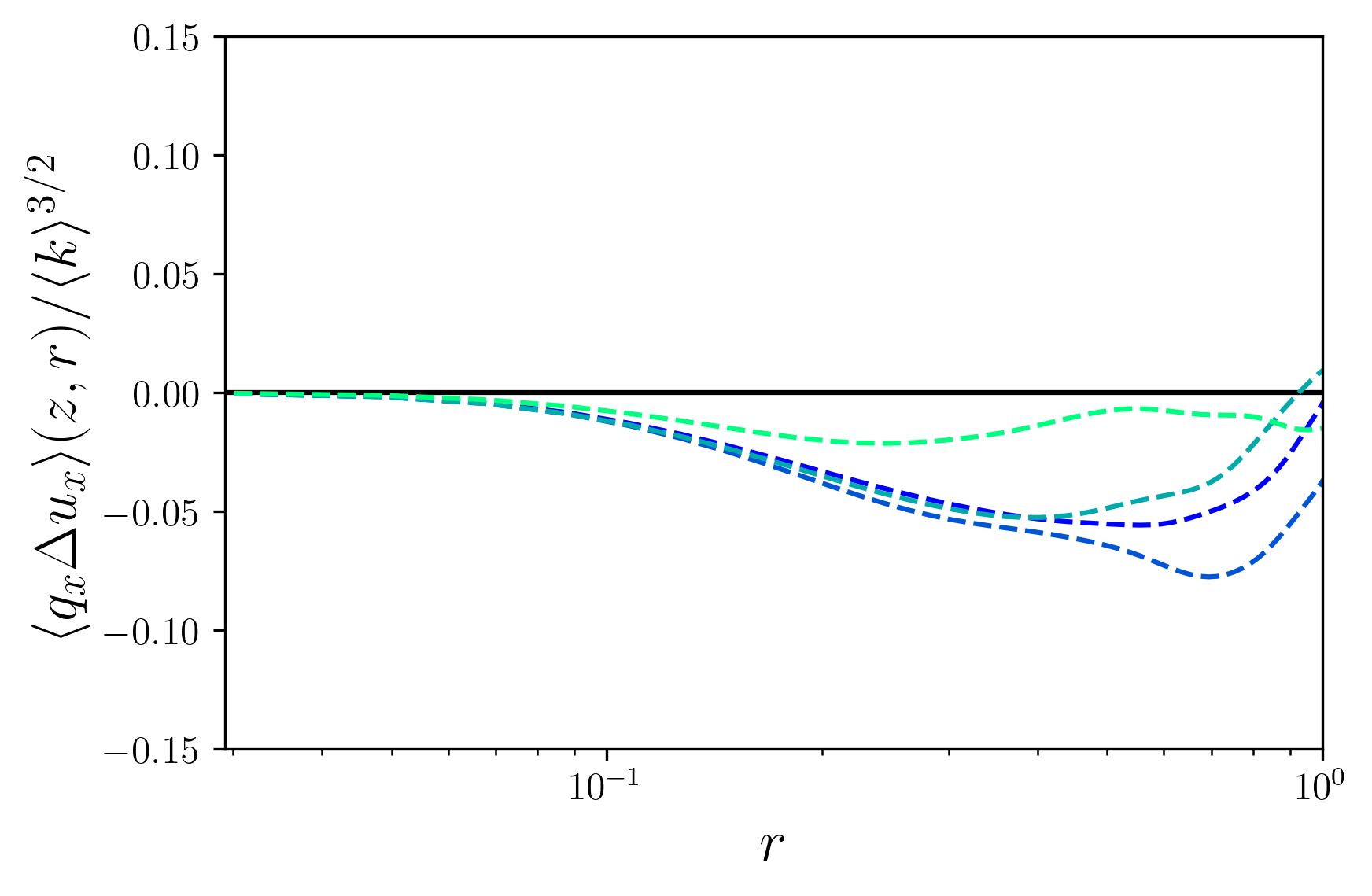}
        \caption{}
    \end{subfigure}
    \quad
        \begin{subfigure}{0.3\textwidth}
        \centering
        \includegraphics[width=\textwidth]{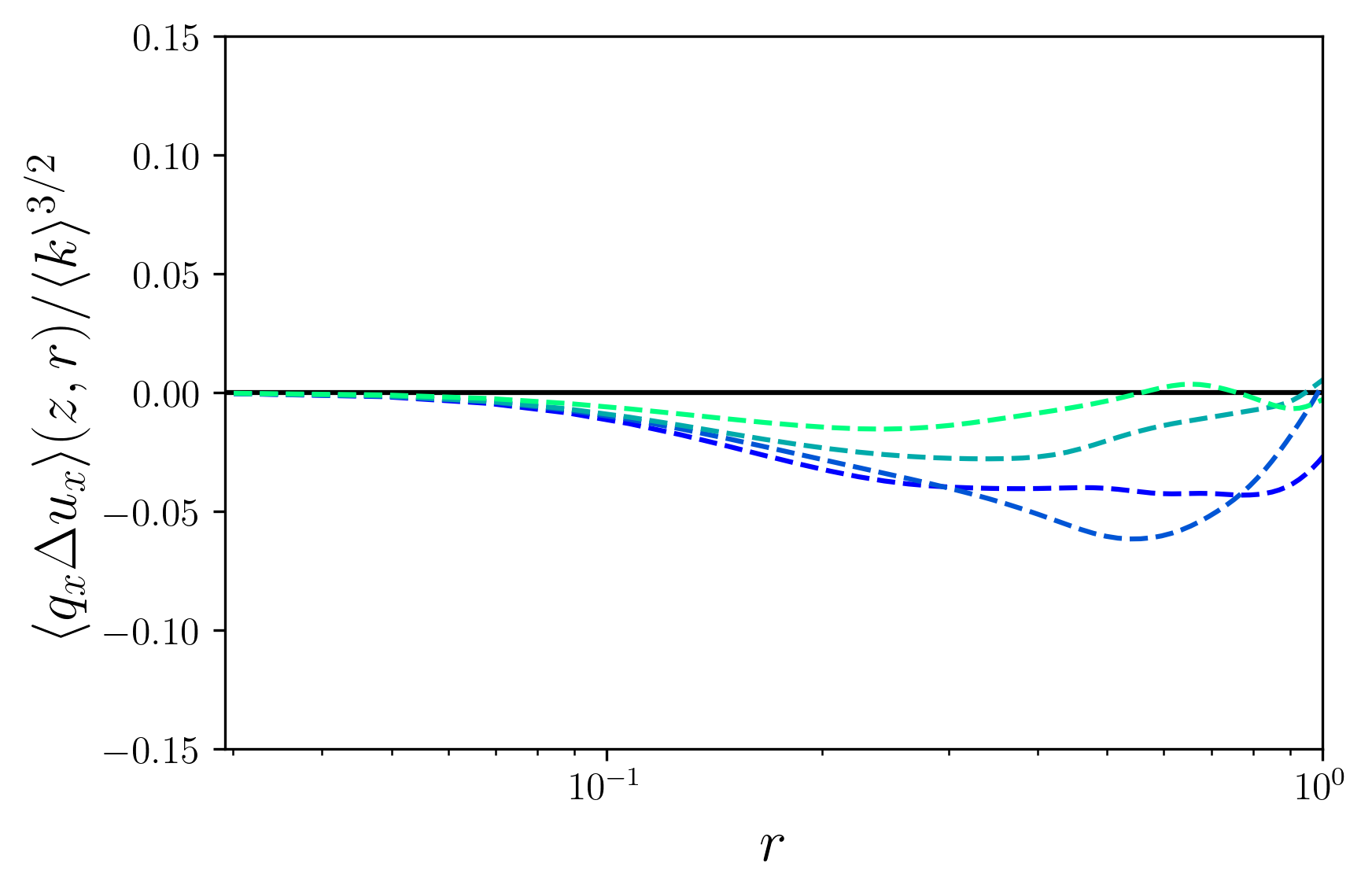}
        \caption{}
    \end{subfigure}
      \\
    \begin{subfigure}{0.3\textwidth}
        \centering
        \includegraphics[width=\textwidth]{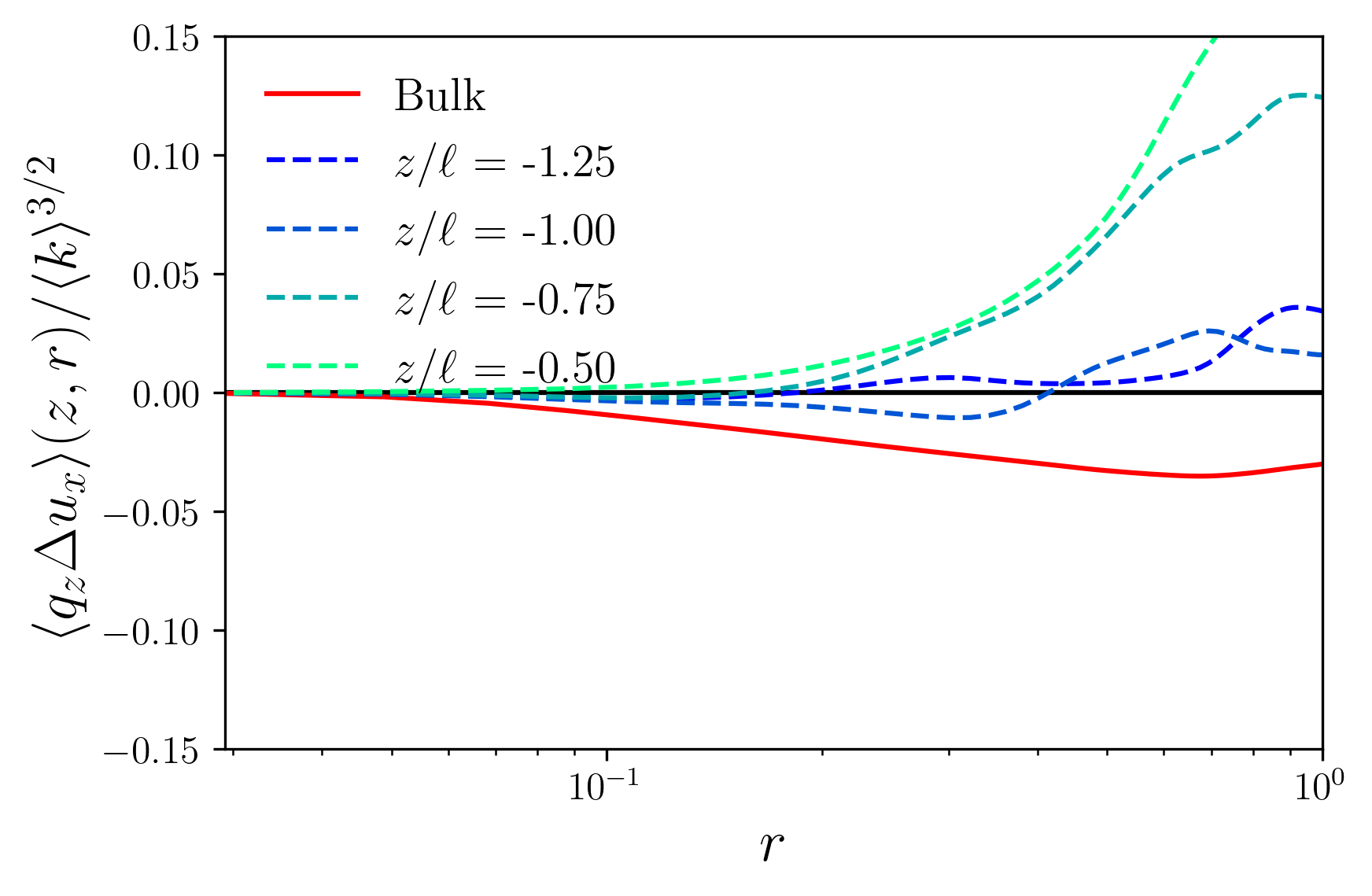}
        \caption{}
    \end{subfigure}
    \quad
        \begin{subfigure}{0.3\textwidth}
        \centering
        \includegraphics[width=\textwidth]{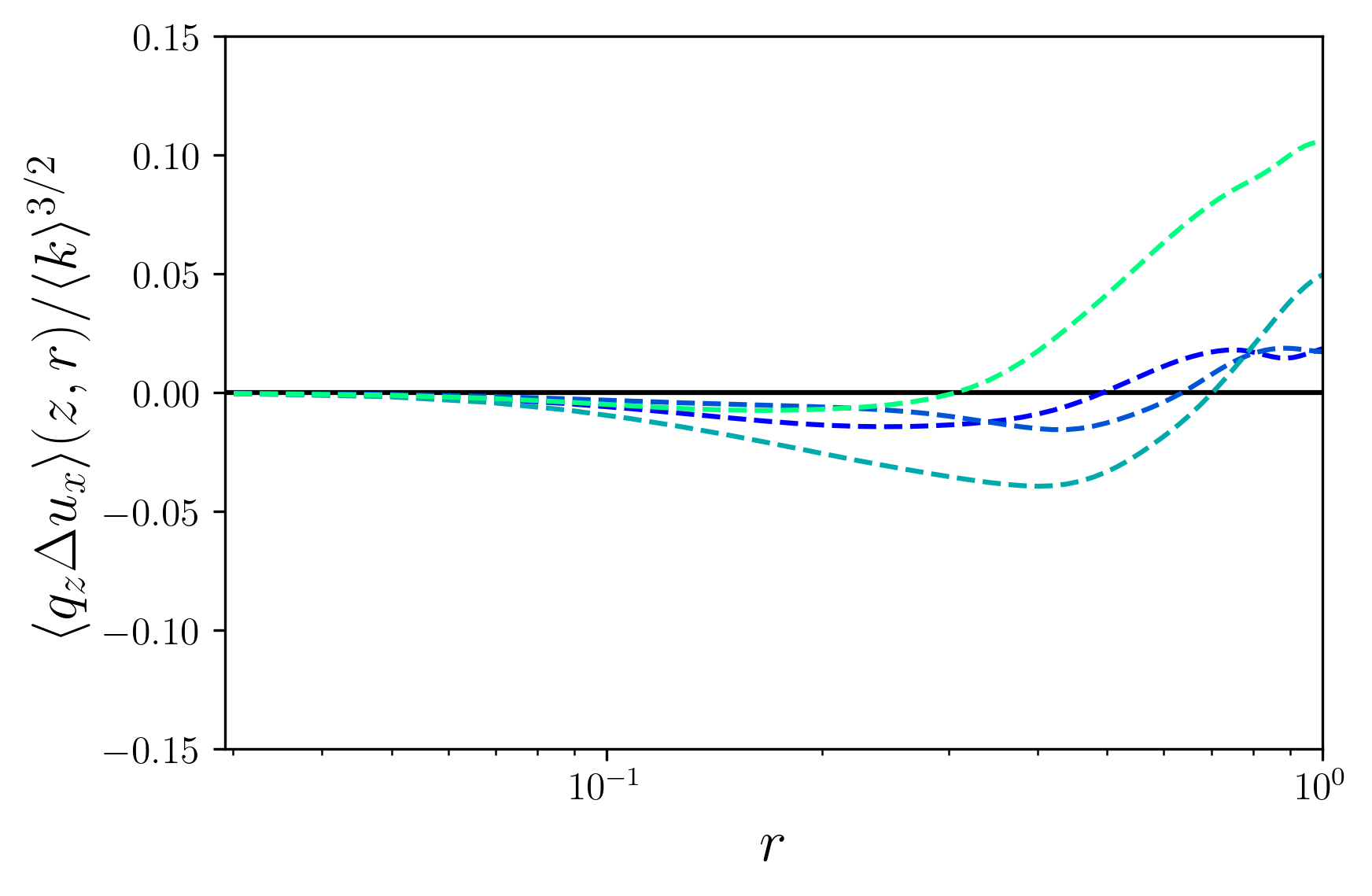}
        \caption{}
    \end{subfigure}
    \quad
    \begin{subfigure}{0.3\textwidth}
        \centering
        \includegraphics[width=\textwidth]{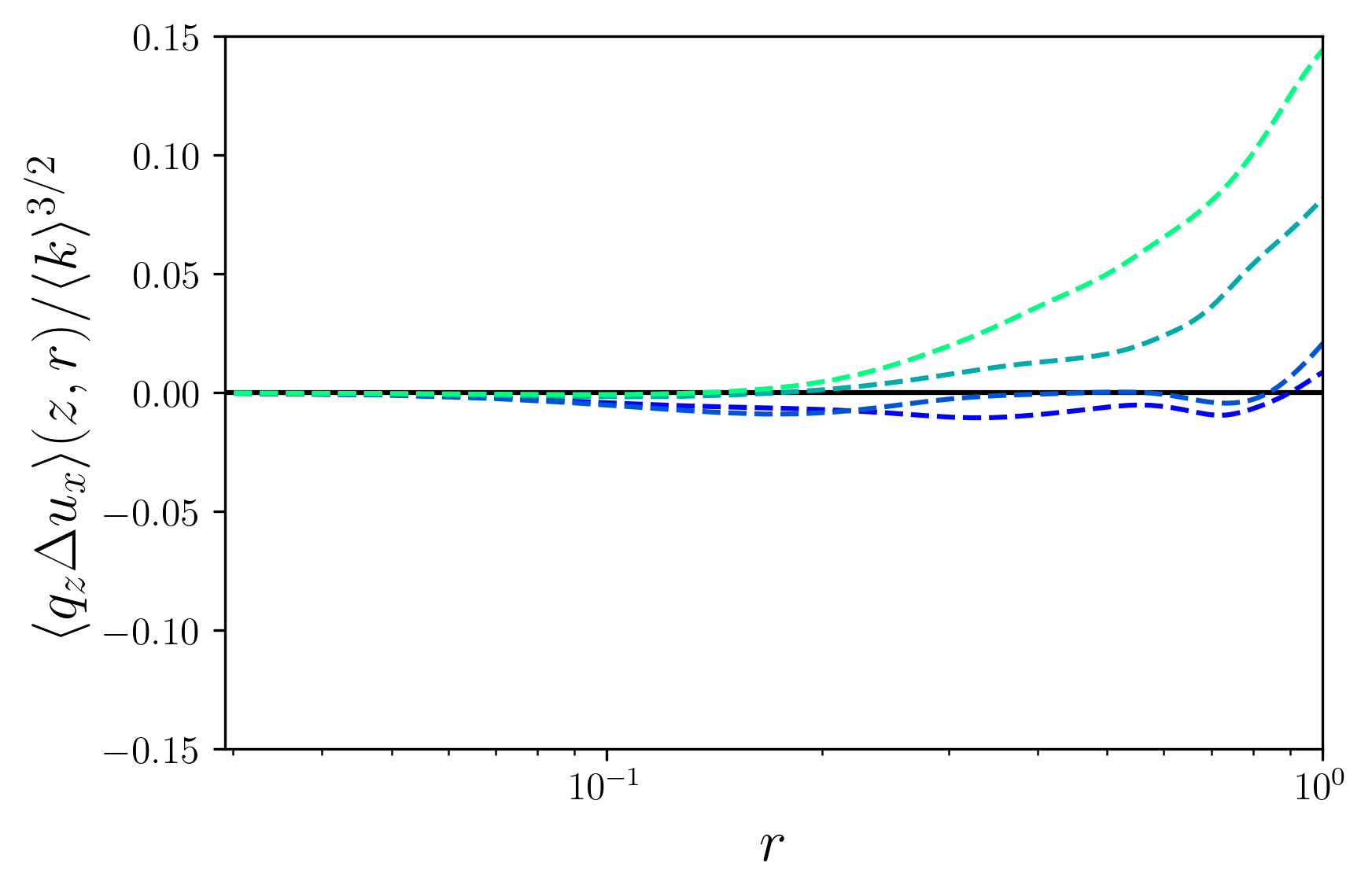}
        \caption{}
    \end{subfigure}
    \\
        \begin{subfigure}{0.3\textwidth}
        \centering
        \includegraphics[width=\textwidth]{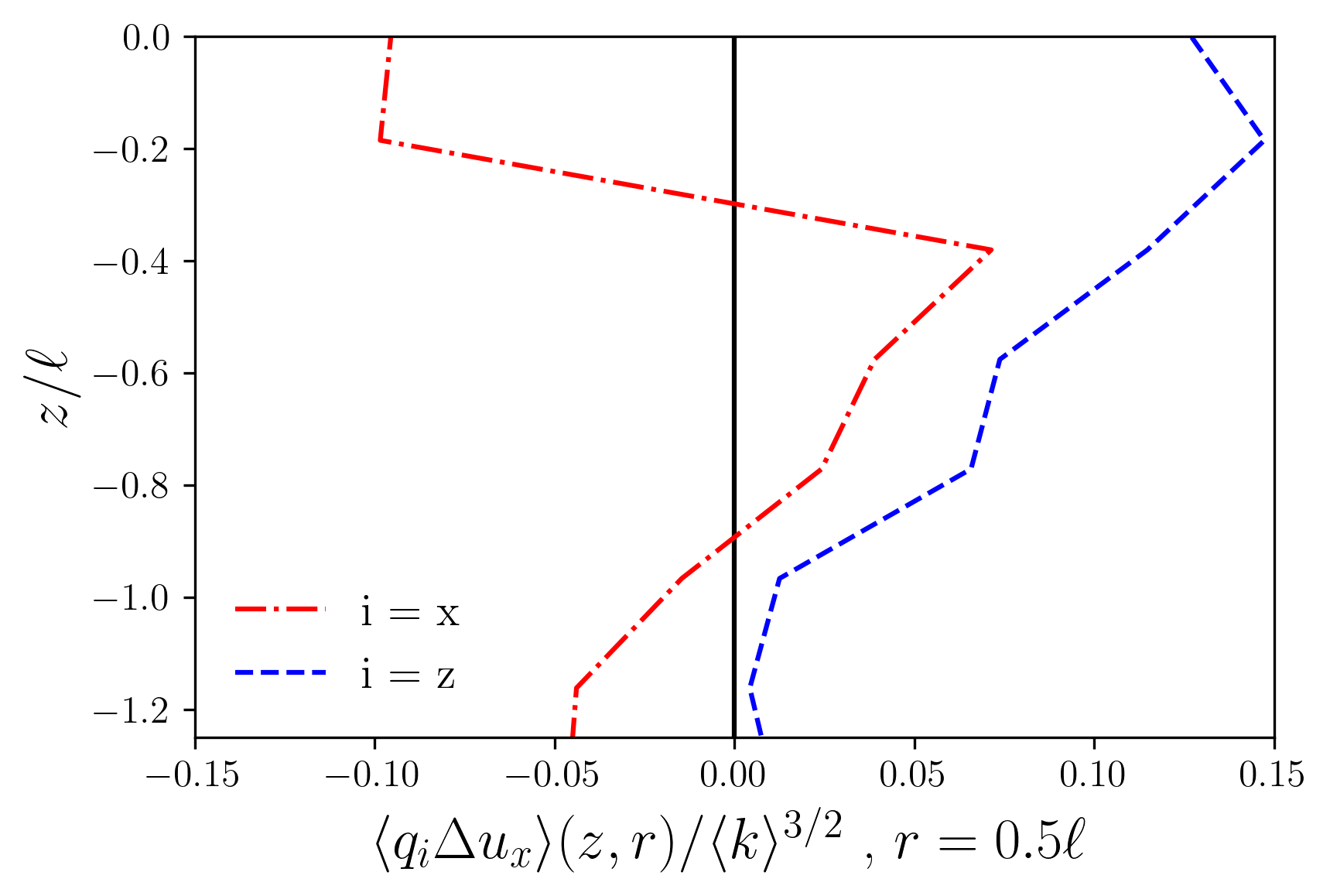}
        \caption{}
    \end{subfigure}
      \quad
    \begin{subfigure}{0.3\textwidth}
        \centering
        \includegraphics[width=\textwidth]{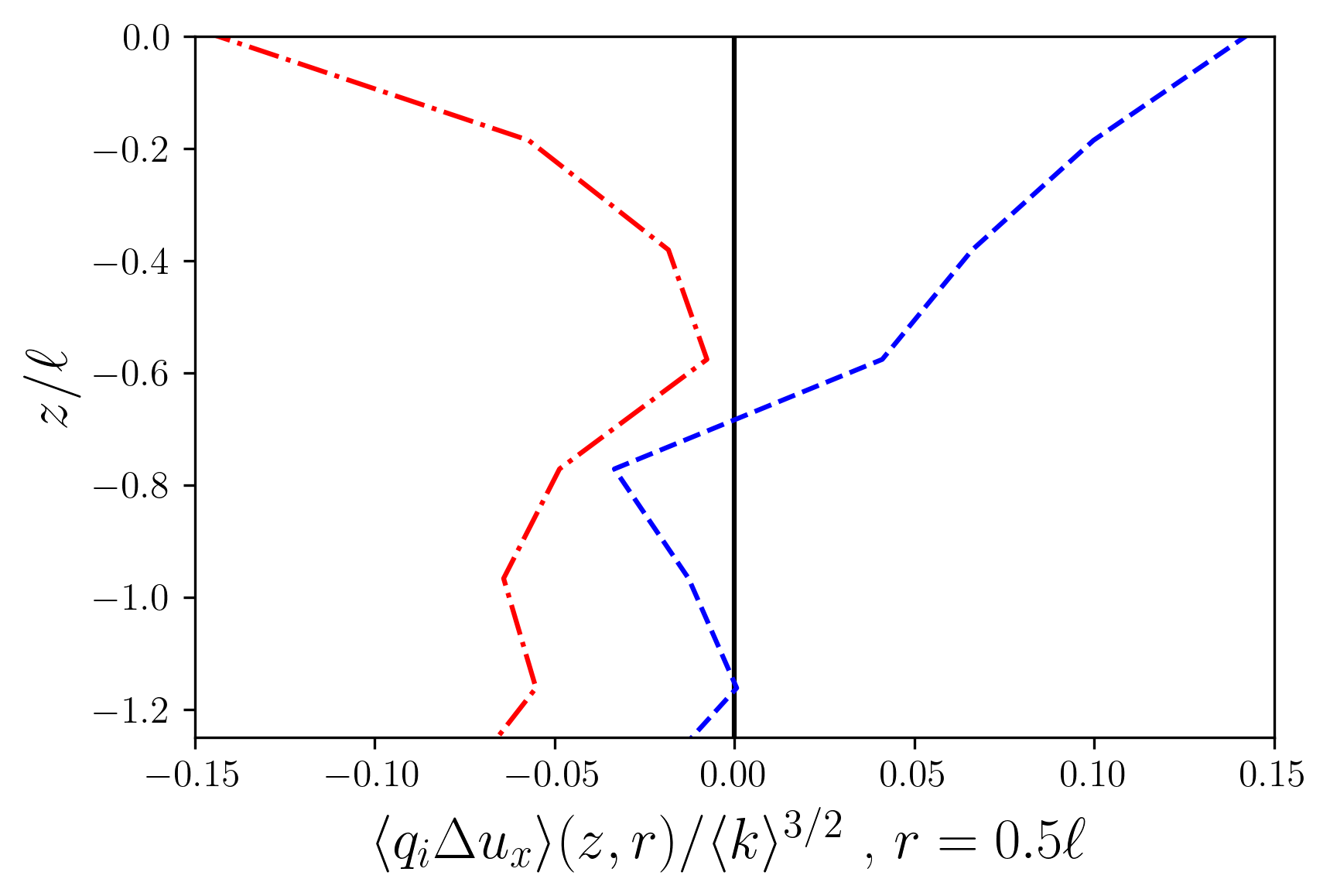}
        \caption{}
    \end{subfigure}
          \quad
    \begin{subfigure}{0.3\textwidth}
        \centering
        \includegraphics[width=\textwidth]{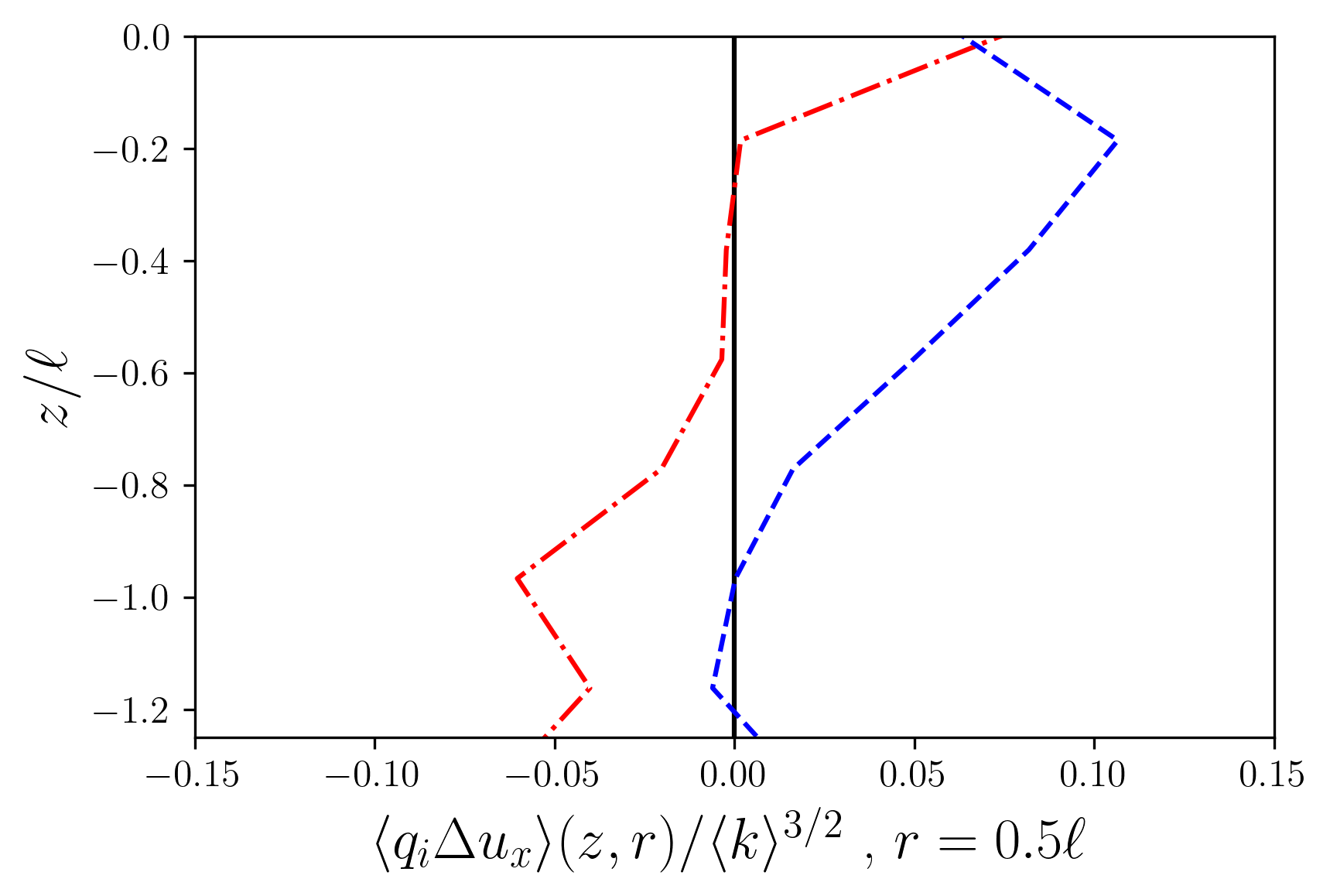}
        \caption{}
    \end{subfigure}
\caption{ Inter-scale energy transfer as function of depth and horizontal separation: (a-c) horizontal TKE and (d-f) vertical TKE; (g-i) vertical and horizontal components for $r=0.5 \ell$. Cases are A, D and E from left to right (decreasing $Fr$). Red solid line represents transfer in the bulk HIT region for reference. }
    \label{fig:third-order-struct}
\end{figure}

The third-order structure function is useful to determine the cascading direction of TKE. In particular, its sign indicates either a forward ($-$) or reverse ($+$) energy cascade at the specific length scale $r$ \citep{Lovecchio2015,Ruth2024}. According to the Kolmogorov $4/5$ law, the archetypal cascade for HIT compresses energy from large to small scales within the inertial subrange, reflected in a negative third-order structure function  \citep{Pope2000}.
Following \cite{Ruth2024} we define the TKE in the $i$-component of velocity difference between two points $q_i = \Delta u_i (\mathbf{x},\mathbf{r})^2 = [u_i(\mathbf{x}+\mathbf{r})-u_i(\mathbf{x})]^2$.  The compression or stretching of energy is thus given by $q_i (\mathbf{x},\mathbf{r}) \Delta u_j (\mathbf{x},\mathbf{r})$.  By ensemble averaging in time and along the homogeneous horizontal direction ($j=x,y $), we can define the inter-scale energy transfer rate for a separation length $r$ as a function of the vertical coordinate $\langle q_i \Delta u_j \rangle (z,r)$.  For horizontal velocity differences $\langle q_x \Delta u_x \rangle$ equates to the longitudinal third-order structure function $D_{LLL}$ \citep{Pope2000}. In Fig.~\ref{fig:third-order-struct} we plot $\langle q_x \Delta u_x \rangle$ and $\langle q_z \Delta u_x \rangle$ for depths above $z/\ell=-1.25$ for different $Fr$ numbers (cases A, D and E). Note these statistics are computed considering the water phase only, and are normalized by $\langle k \rangle^{3/2}$ at the same depth.  As a reference, we also plot the same functions for case A within the bulk HIT region.

Within the single-phase bulk region we observe the typical forward cascade (negative sign), and $\langle q_x \Delta u_x \rangle \propto -r$ within the inertial range \citep{Pope2000}.
For case A, the difference between horizontal and vertical energy fluxes are smaller compared to lower $Fr$ numbers. At greater depths the flux is still mostly negative (forward) for small scales, but much smaller in magnitude compared to the bulk. As we approach the free-surface there is a clear reverse cascade above $r/\ell \sim 0.1-0.2$, and below which the flux is close to zero, indicative of a bi-directional energy transfer (i.e. forward and reverse).
\cite{Pan1995} also investigate the inter-scale energy transfer from spectral analysis and verified that large scales have a strong net reverse cascade, while smaller scales (high wavenumbers) exhibit both forward and reverse transfer with comparable magnitudes yielding a net zero exchange.
For cases D and E, with the reduction in $Fr$ the inverse cascade is only evident for the vertical component, $\langle q_z \Delta u_x \rangle$.
Results for lowest $Fr$ case show qualitative agreement with \cite{Ruth2024}, which also showed $\langle q_z \Delta u_x \rangle > 0$ near the free-surface, as evidence of the reverse cascade effect. 
Finally we plot both $\langle q_x \Delta u_x \rangle$ and $\langle q_z \Delta u_x \rangle$ as a function of depth for $r/\ell =0.5$. 
The trends are consistent across cases in that the vertical motions are more prone to a reverse energy cascade at these larger scales, relative to horizontal ones.
The presence of the free-surface inhibits the forward cascade by limiting compression and favoring horizontal stretching due to the more energetic upwellings.

\subsection{Energy spectra}

\begin{figure}
\centering
    \begin{subfigure}{0.45\textwidth}
        \centering
        \includegraphics[width=\textwidth]{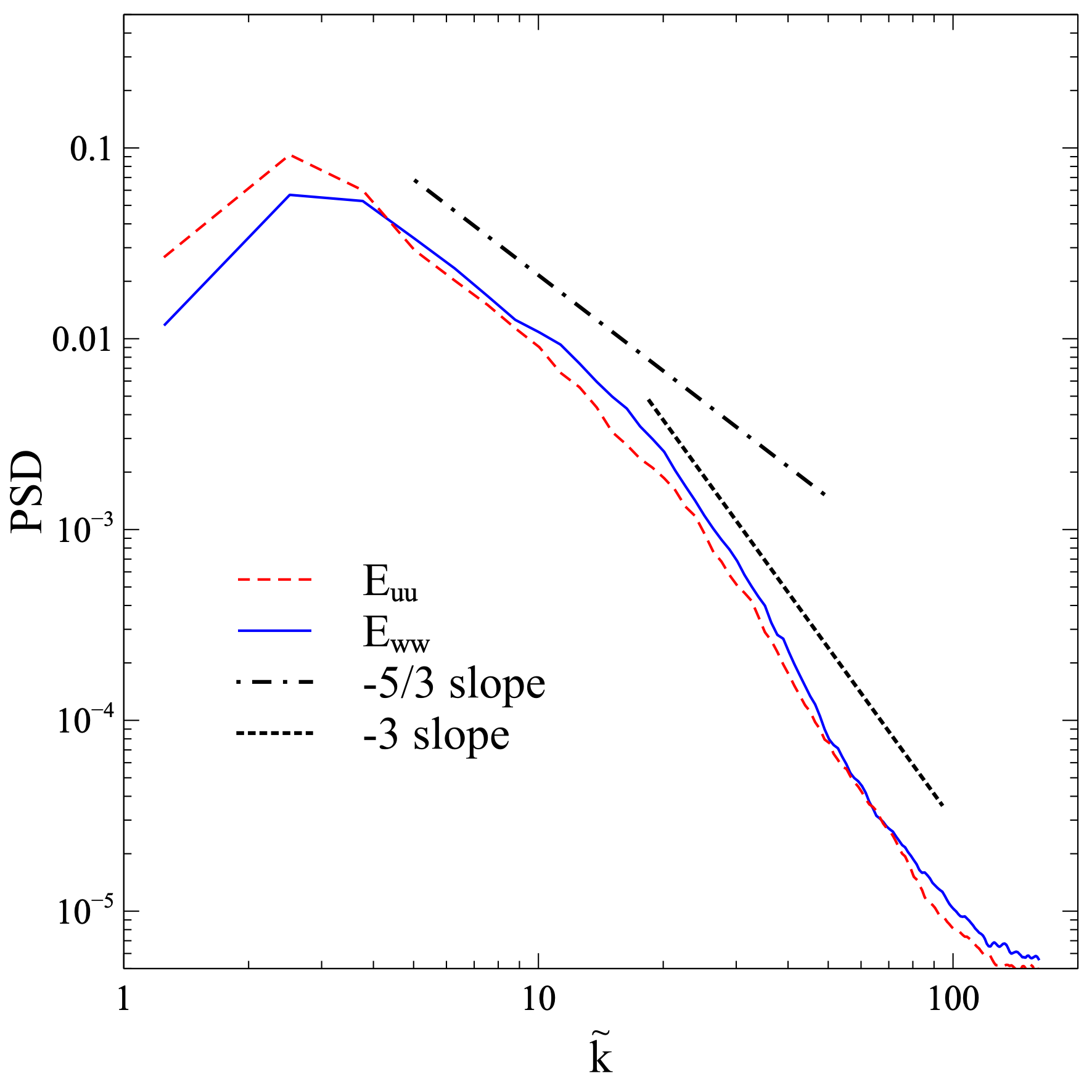}
        \caption{Case A}
        \label{fig:fourier-caseA}
    \end{subfigure}
      \qquad
    \begin{subfigure}{0.45\textwidth}
        \centering
        \includegraphics[width=\textwidth]{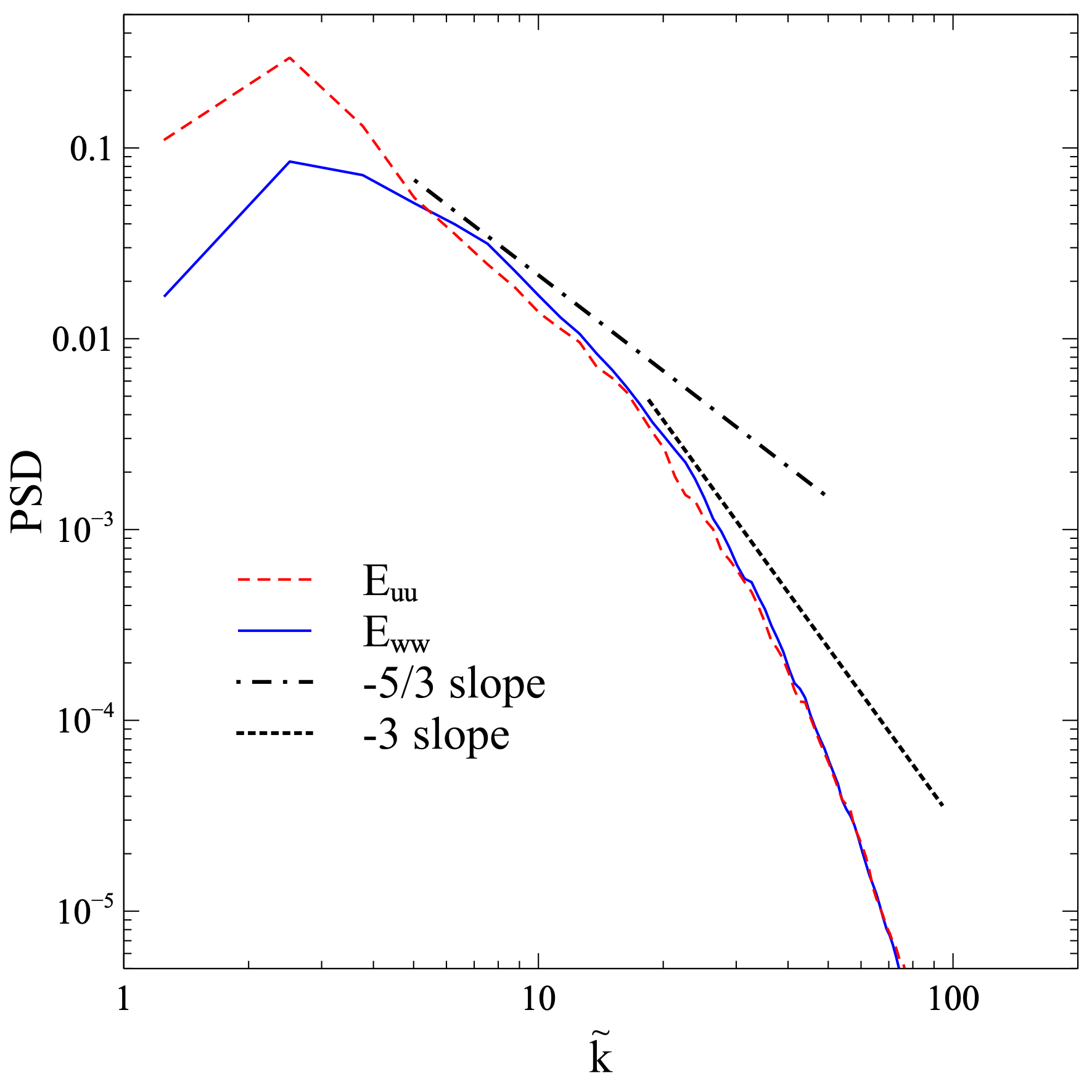}
        \caption{Case E}
        \label{fig:fourier-caseE}
    \end{subfigure}
\caption{Fourier power spectral density of kinetic energy components of water phase at $z/\ell=-0.25$ for different $Fr$.   }
    \label{fig:fourier-tke-spec}
\end{figure}

To complement the quantitative analysis of the different components of TKE and at which length scales they are present, we compute the traditional one-dimensional Fourier spectra at a fixed depth in the water phase. In Fig.~\ref{fig:fourier-tke-spec} we compare the horizontal $(E_{uu})$ and vertical $(E_{ww})$ velocity power spectral density (PSD) for cases A and E at different $Fr$ numbers at the same $z/ \ell=-0.25$.  Similar to the results from the Lumley plot (Fig. \ref{fig:lumley}), we observe that case A has higher isotropy since the PSD are comparable, and follow the $\tilde{k}^{-5/3}$ law characteristic of the inertial subrange. Case E (lower $Fr$) shows higher magnitude for $E_{uu}$ relative to $E_{ww}$ for the lowest wavenumbers which are the most energetic. For mid and high wavenumbers the PSD is comparable, indicating that anisotropy is mostly associated with the largest scales.  In addition a steeper decay in energy is noticed for the lower $Fr$ case (E), relative to case A; similar trends are also reported in \cite{Yu2019,Jamin2025}.   A transition from -5/3 to -3 slope is observed in case A, which is also reported in the DNS and experiments of \cite{Pan1995} and \cite{Kumar1998} and is attributed to a quasi-2D behavior and reverse energy cascade. 

\begin{figure}
\centering
    \begin{subfigure}{0.45\textwidth}
        \centering
        \includegraphics[width=\textwidth]{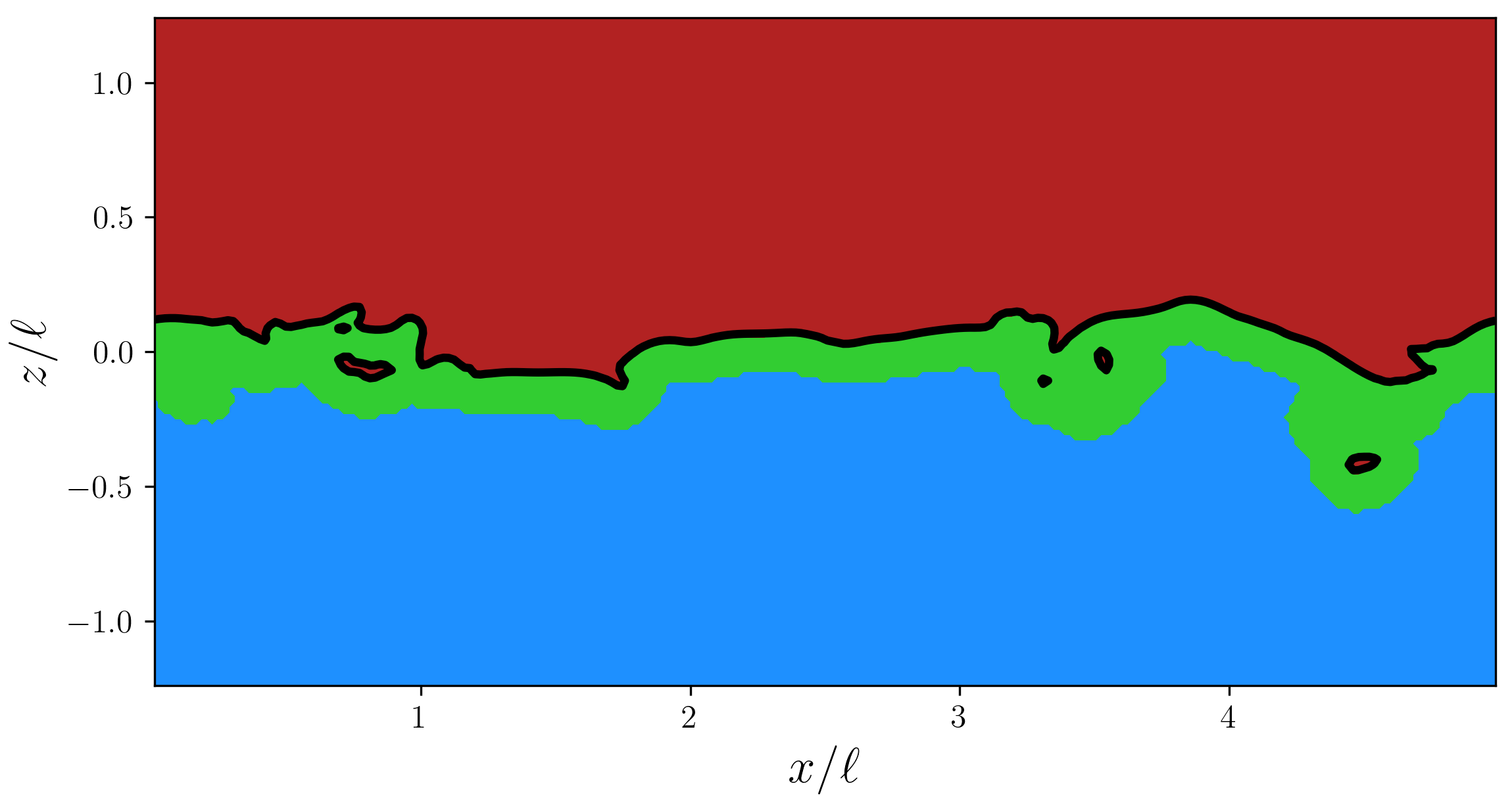}
        \caption{$m=1$}
        \label{fig:m-1}
    \end{subfigure}
      \qquad
    \begin{subfigure}{0.45\textwidth}
        \centering
        \includegraphics[width=\textwidth]{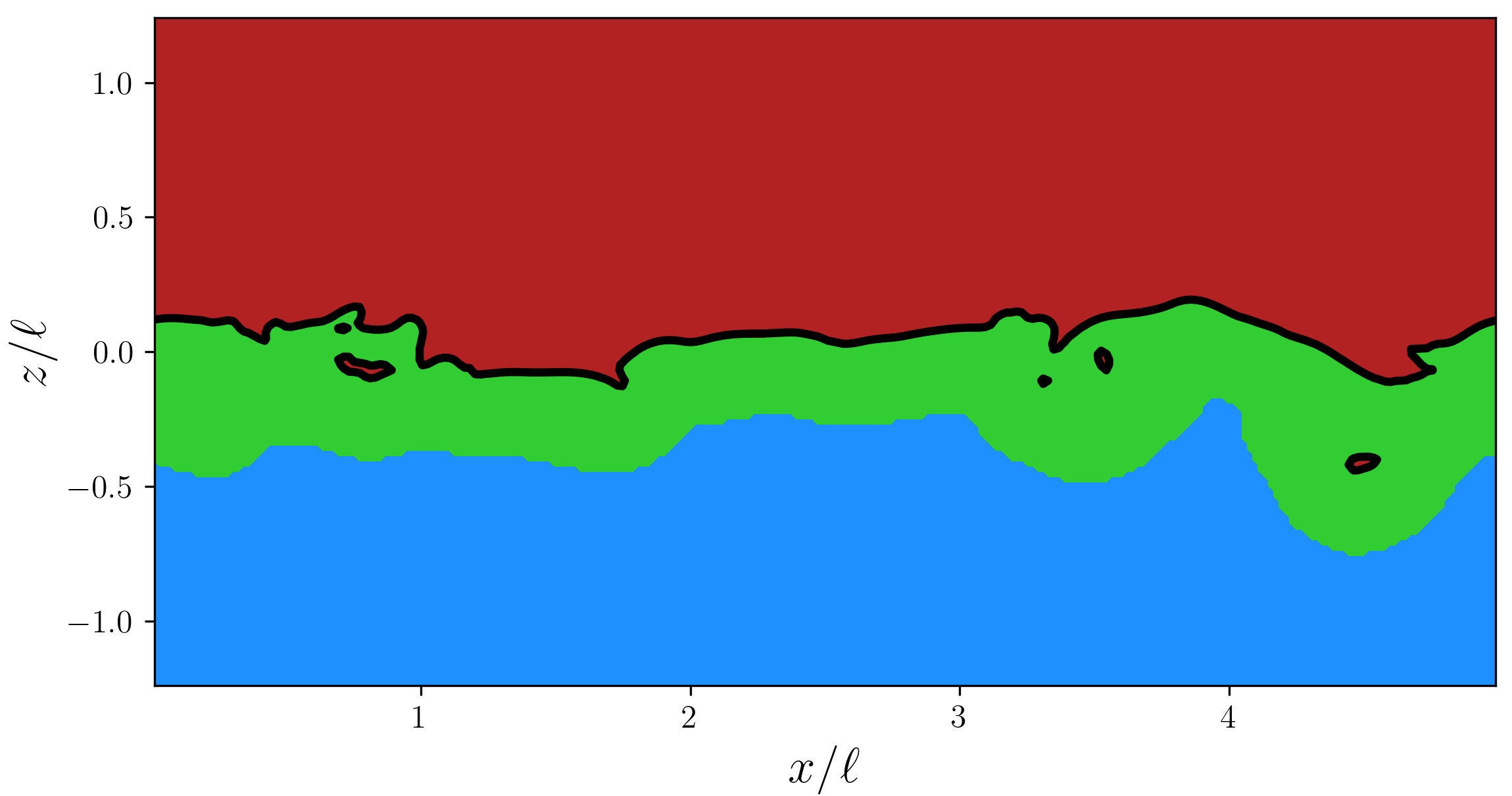}
        \caption{$m=2$}
        \label{fig:m-2}
    \end{subfigure}
\caption{Slice view of example snapshot for domain subsets used in the wavelet analysis for smallest scales. Black line represents the interface $(\phi = 0)$. Colored regions represent the air (red), interfacial (green) and water (blue) subsets. Data is for case A. }
    \label{fig:phase-subsets}
\end{figure}

A limitation of the traditionally used one-dimensional spectral analysis is that it does not differentiate between phases if used naively, and is agnostic of interface proximity. Even though the velocity field is continuous across the interface, mismatching viscosities imposes discontinuous velocity gradients which can contribute to additional energy content in the higher wavenumbers. One alternative that offers local information based on the phase or distance to the interface uses wavelets to compute the spectral energy. This is known as a phase-based wavelet analysis, and has been used in recent works by \cite{Freund2019,Calado2024-PRF}. For the sake of completeness, we include a brief description of this approach, and refer to the cited references for more details.  The method starts by decomposing the physical domain into three subsets based on the distance to the interface, $\phi$, defining liquid, gas, and interfacial subsets as $\mathcal{D}_L, \mathcal{D}_G$ and $\mathcal{D}_I$ respectively.  Due to the high density ratio, the kinetic energy associated with the air phase is negligible, and the interfacial subset only includes the neighboring region in the water side.  A multi-resolution discrete wavelet transform (DWT) of the 3D velocity field can be performed for a range of scales $1 \leq m \leq M$, from smallest $(m=1$) to largest $m=M$. The wavelet order determines the smallest scale, and doubles progressively until the largest scale, $M=\textrm{log}_2(N)$, which depends on the domain size. In our case we take the horizontal domain size $L$.  The equivalent wavenumber is given by $\tilde{k}_m= 2 \pi / (2^m \Delta h)$.  Following previous works \cite{Freund2019,Calado2024-PRF} we use the fourth-order Symlet wavelet family (orthogonal and of compact support), using eight points in each spatial direction for $m=1$.  The domain decomposition can be readily performed based on the level-set function $\phi$ and the scale-dependent threshold distance.   A discrete wavelet family of order $p$ has $n_p=2p$ points for $m=1$. The threshold distance for each scale is computed as,  $\tilde{\phi}^{(m)} = 2^{(m-1)} n_p \Delta h $. Hence the interfacial region $\mathcal{D}_I^{(m)}$ is defined for $  - \tilde{\phi}^{(m)} \leq \phi \leq 0 $.  An example of this can be seen in Fig.~\ref{fig:phase-subsets} for scales $m=1$ and $m=2$.   Note that the interfacial domain will capture the region in the vicinity of the free-surface, as well water droplets and surrounding entrained bubbles for higher the $Fr$ number cases. Nonetheless since the total interfacial area is dominated by the free-surface, the influence of disperse bubbles/droplets is negligible for the overall near surface dynamics.  The PyWavelets package is used to compute the wavelet coefficients from the DNS data with the spectra being ensemble-averaged over the different snapshots in time \citep{Lee2019}.   We note that while comparisons with traditional Fourier spectra could be made, the DWT brings additional degrees of freedom, where in 3D it considers all possible combinations of linear, planar, and volumetric dilations \citep[see for example][]{Meneveau1991}. Furthermore since the wavelet coefficients are averaged over subdomains based on the distance from the interface, it is not restricted to low $Fr$ or cases with small deformation of the free-surface.

\begin{figure}
       \begin{subfigure}{0.3\textwidth}
        \centering
        \includegraphics[width=\textwidth]{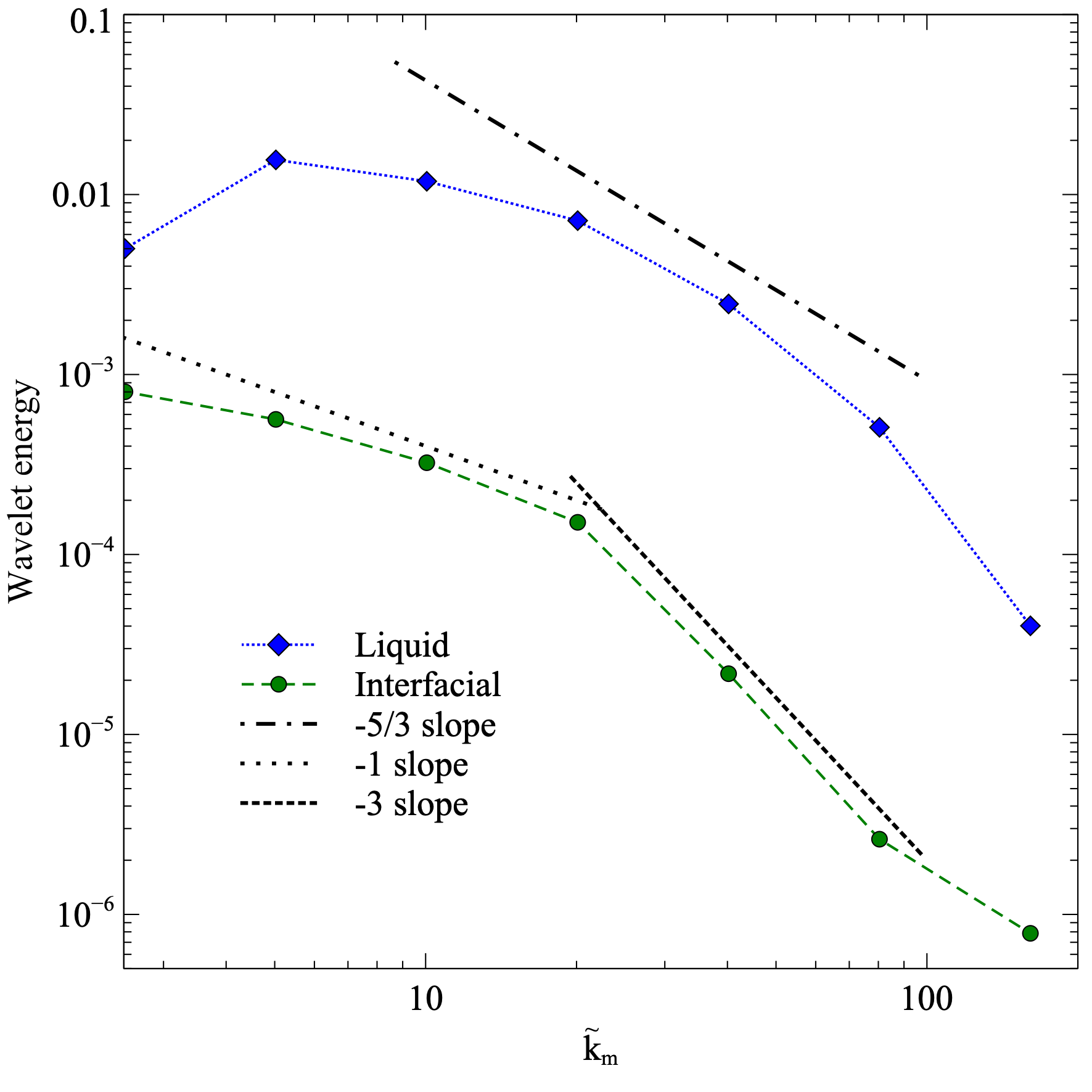}
        \caption{}
                \label{fig:wavelet-A}
    \end{subfigure}
   \quad
    \begin{subfigure}{0.3\textwidth}
        \centering
        \includegraphics[width=\textwidth]{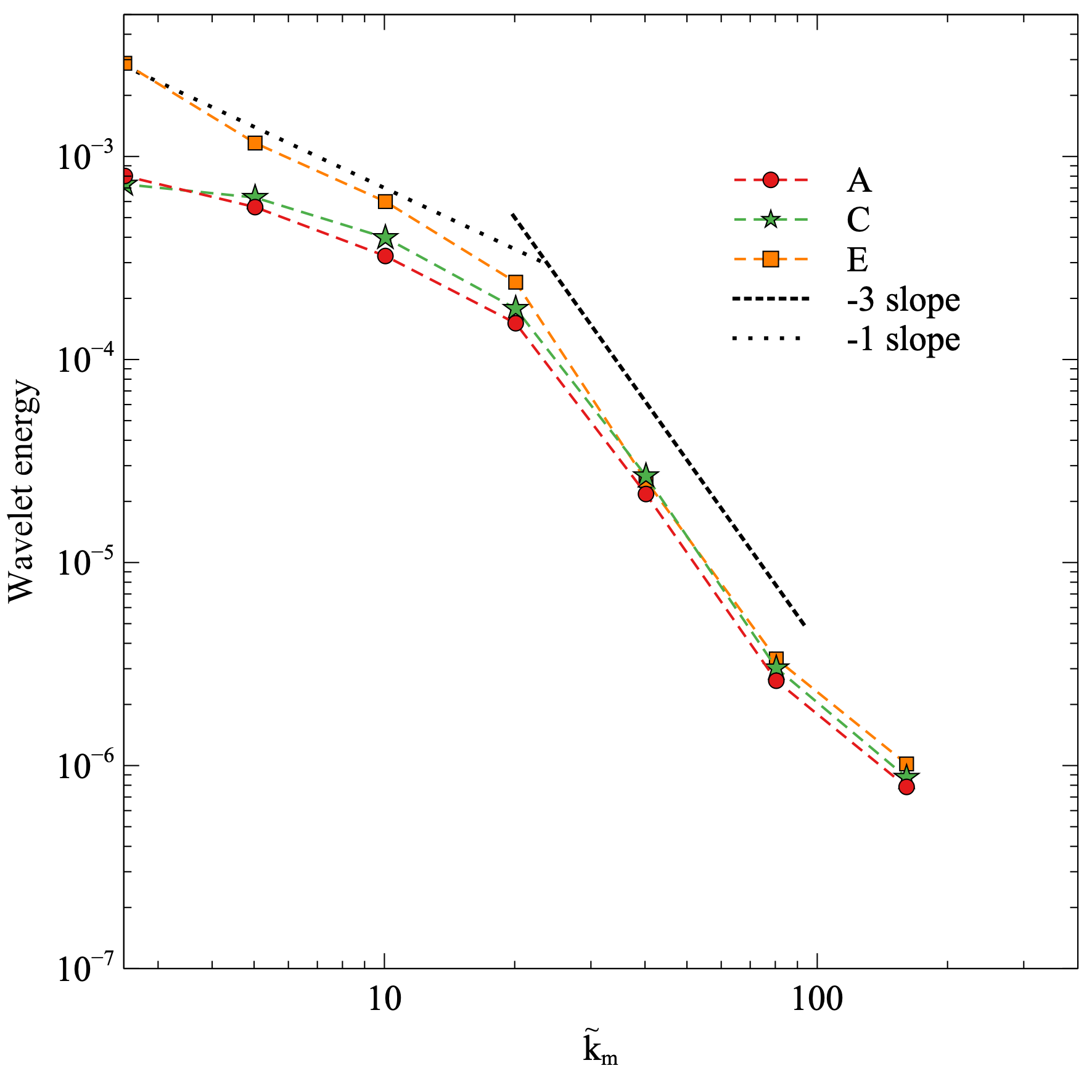}
        \caption{}
                \label{fig:wavelet-ACE}
    \end{subfigure}
        \quad
    \begin{subfigure}{0.3\textwidth}
        \centering
        \includegraphics[width=\textwidth]{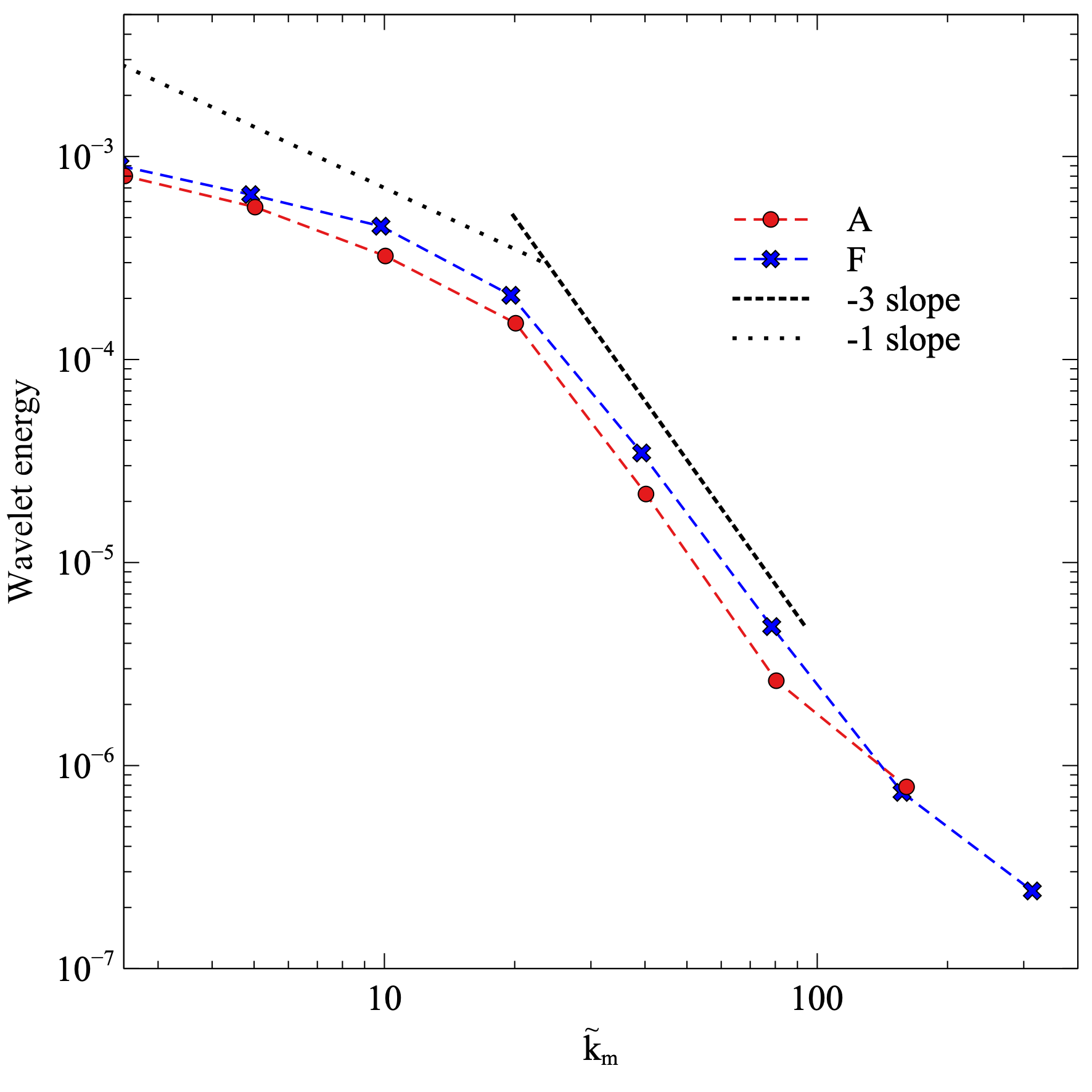}
        \caption{}
                \label{fig:wavelet-AF}
    \end{subfigure}
\caption{Wavelet spectra for case A, comparing pure water and interfacial spectra (a); interfacial spectra comparing cases A, C and E (b); cases A and F (c). }
    \label{fig:wavelet-spectra}
\end{figure}

\begin{figure}
\centering
    \begin{subfigure}{0.45\textwidth}
        \centering
        \includegraphics[width=\textwidth]{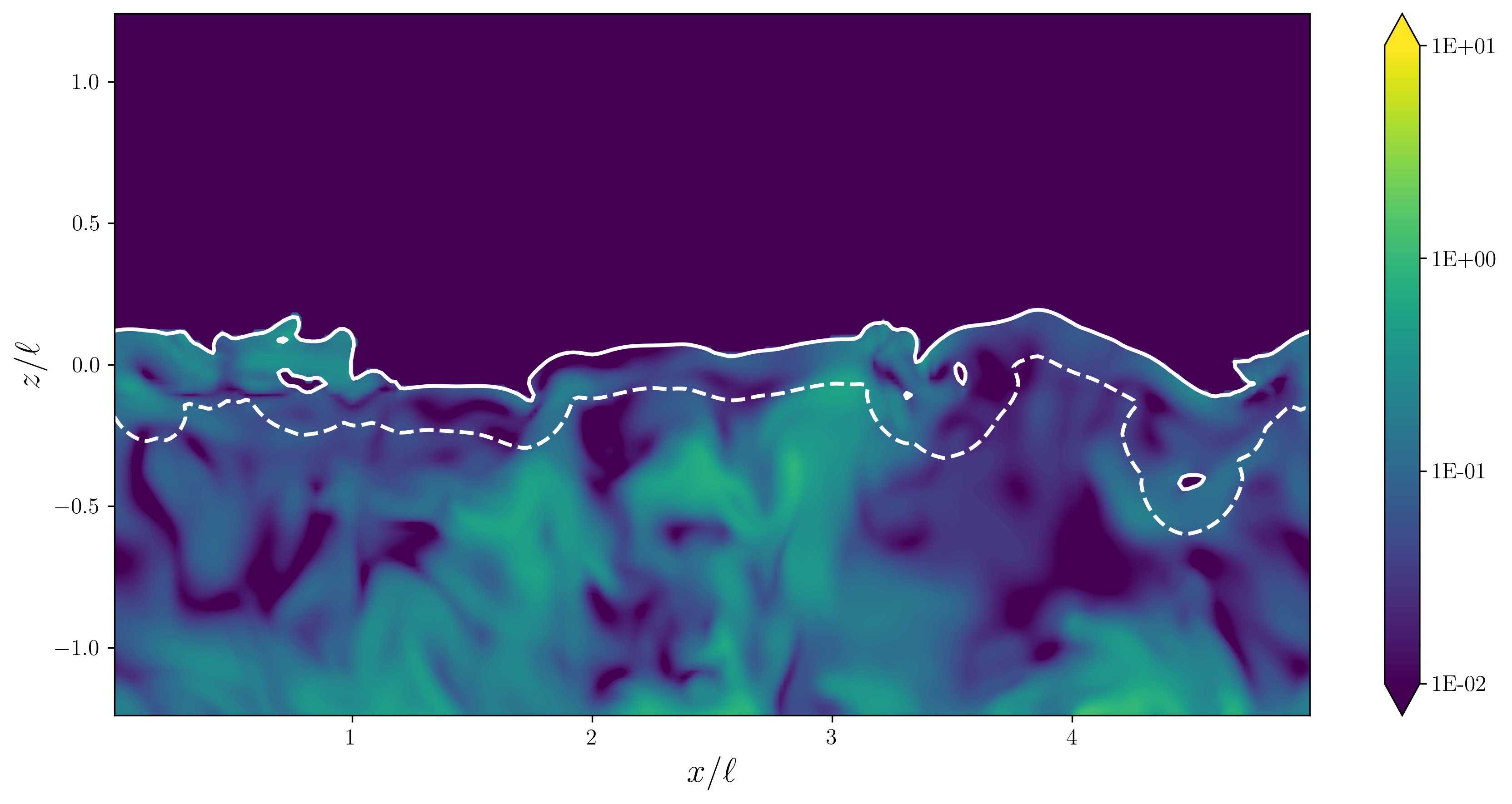}
        \caption{}
        \label{fig:wavelet-total}
    \end{subfigure}
      \quad
    \begin{subfigure}{0.45\textwidth}
        \centering
        \includegraphics[width=\textwidth]{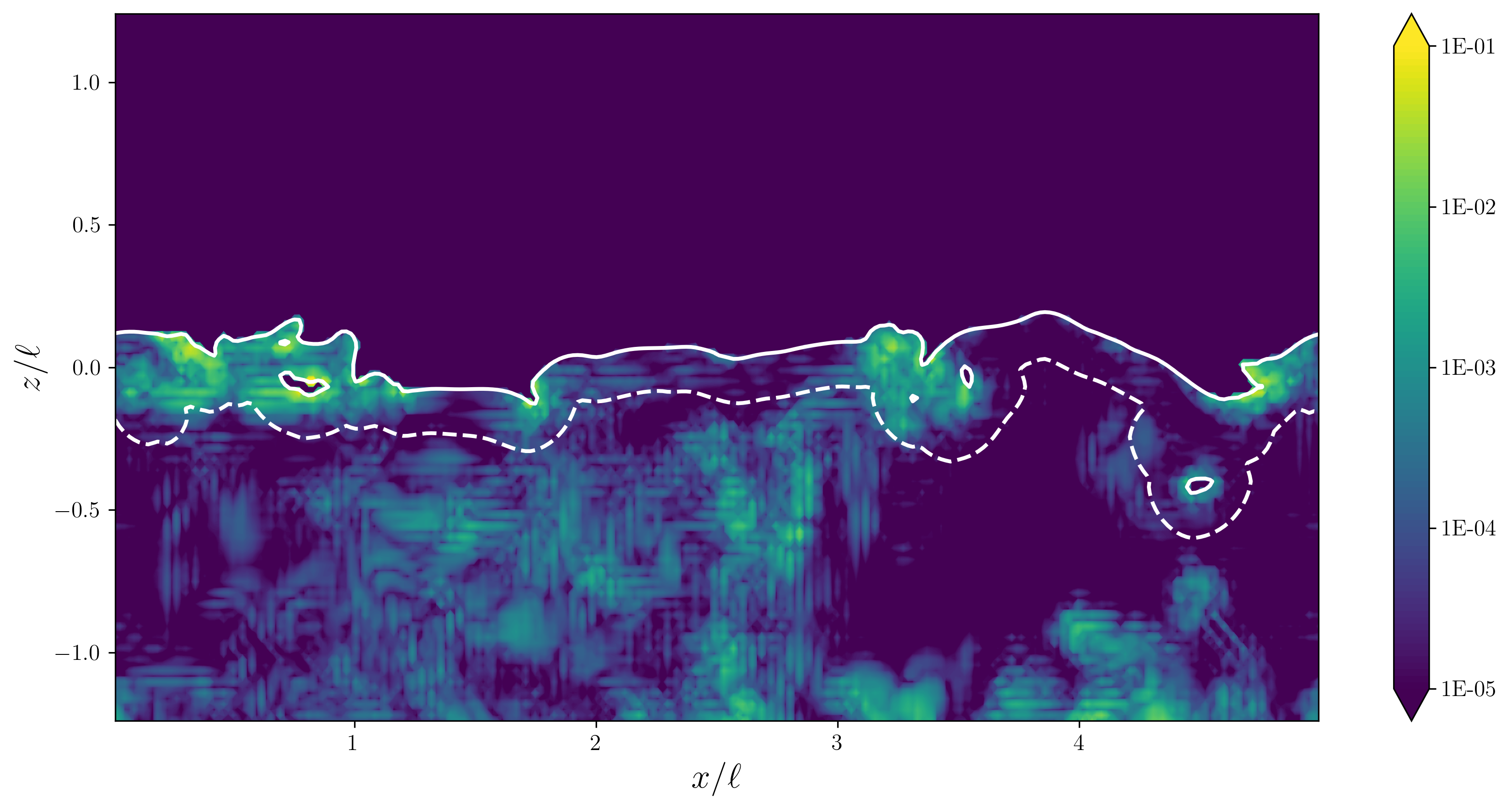}
        \caption{}
        \label{fig:wavelet-fine}
    \end{subfigure}
\caption{Contour plots for (a) total kinetic energy and (b) smallest wavelet scale $m=1$ near the free-surface. Interface is shown by the solid white line, while the interfacial region for $m=1$ is shown by the neighboring white dashed line. Data is for case A. }
    \label{fig:contour-tke}
\end{figure}

The wavelet spectra for the liquid and interfacial domains for case A are shown in Fig.~\ref{fig:wavelet-A}, and are restricted to scales smaller than $L/2$. Case A shows a higher spectral energy in the pure water phase relative to the interfacial one since it includes the bulk forcing region, and a decay approaching the $-5/3$ law. The interfacial spectra has a markedly different shape, with a -1 in the low frequency range, and a $-3$ slope for mid-range wavenumbers.   The slope changes again towards a less steep decay for the highest wavenumbers which are closest to the interface.  The interfacial spectra for cases A, C and E are given in Fig.~\ref{fig:wavelet-ACE}, exhibiting a slightly higher energy content for low wavenumbers in case E (low $Fr$), and similar trends for mid and high wavenumbers. The weak dependence on $We$ for the current parameter space is also seen here even at the smallest scales.  Finally we compare cases A and F in Fig.~\ref{fig:wavelet-AF} with small differences over the wavenumber range, although the change in slope at higher $\tilde{k}_m$ is confirmed with the improved grid resolution of case F.   This apparent pileup of energy at the smallest scales near the interface was confirmed by plotting the wavelet energy at $m=1$ in a contour slice and comparing against the total kinetic energy in Fig.~\ref{fig:contour-tke}. For all cases we consistently observe a concentration of high energy at these smallest wavelet scales next to the interface. While surface tension is known to modulate the fluid TKE if the $We$ number is low enough \citep[see e.g.][]{Dodd2016,Saeedipour2022}, we have shown in our case the $We$ is sufficiently large such that these effects would be minimized. This suggests that the higher energy content at small scales stem from the free-surface dynamics.
From the change in slope toward in the spectra we note the transition wavenumber to be around $\tilde{k}_m \approx 20$ which corresponds to a length scale $ \sim 0.3 \ell$. The dual-energy cascade discussed in \S \ref{sec:rev-cascade} likely plays a role in the -3 scaling of energy for scales smaller than this value as postulated by \cite{Pan1995}. The upper bound of the -3 slope occurs at $\tilde{k}_m \approx 80$ ($ 0.078 \ell$).

\begin{figure}
\centering
        \includegraphics[width=0.45\textwidth]{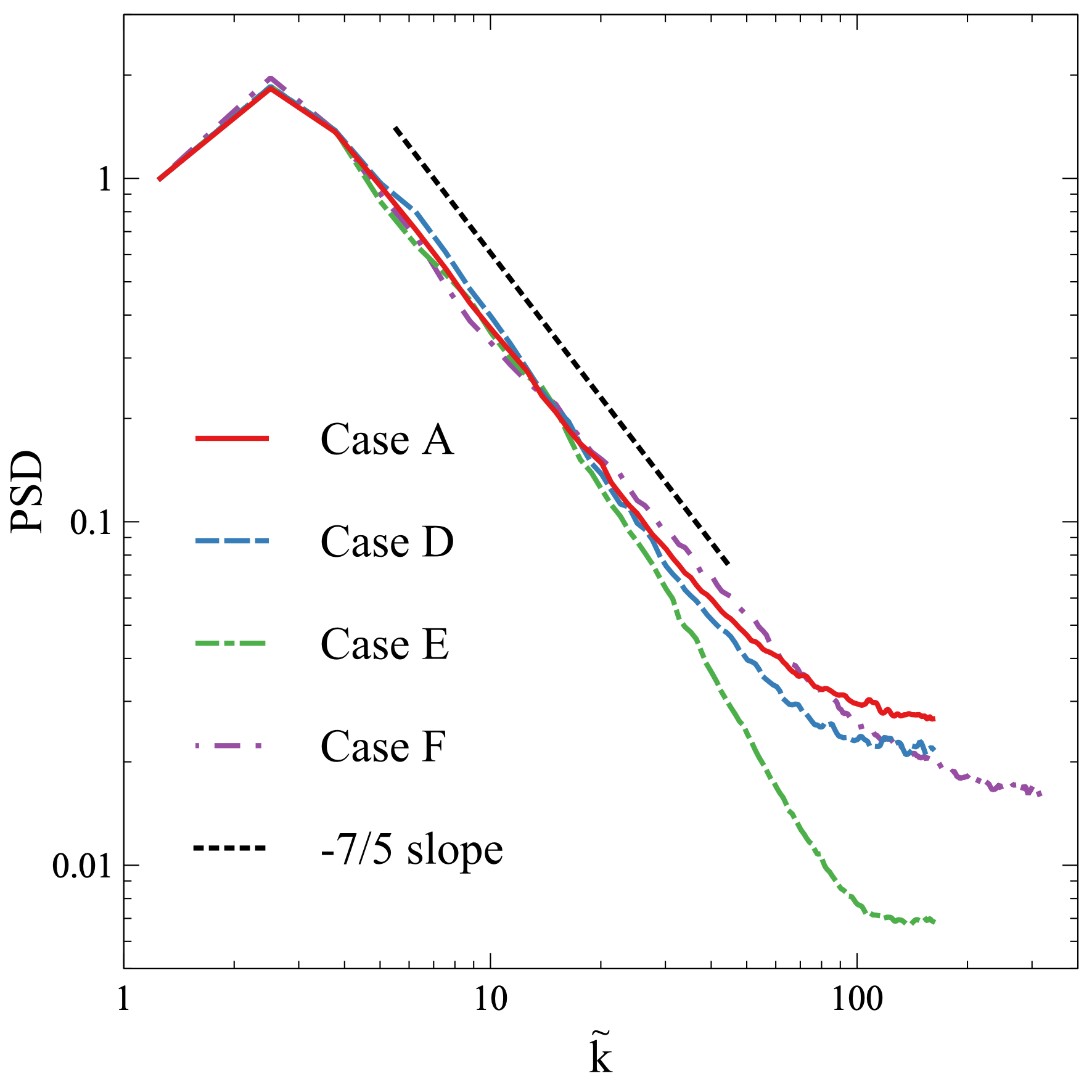}
\caption{Power spectral density of the free-surface elevation for different cases, normalized by energy of smallest wavenumber. }
    \label{fig:fs-elev-spectra}
\end{figure}

The large scale dynamics are governed by gravitational forces and dampened by the interface \citep{Frederix2018,Freund2019,Saeedipour2023}. It is well known in 2D turbulence that a change in slope from -5/3 to -3 occurs for wavenumbers larger than some forcing wavenumber, associated with a direct/forward cascade of enstrophy \citep{Kraichnan1971,Xiao2009}.  Reverse or split/dual energy cascades exist in a variety of flow configurations \citep[see][for definitions and different examples]{Alexakis2018}.  While the 2D nature is imposed in some configurations, we have demonstrated that turbulence remains three-dimensional for the higher $Fr$ cases in the present work.  In addition, under stable stratification, gravity will naturally oppose any vertical displacement of fluid, removing kinetic energy and potentially affecting the cascade mechanism and local dissipation rates, as pointed out by \cite{Bolgiano1959}.  The power spectral density of the gravitational potential energy is proportional to ${E}_g \propto z^2$, and for the values of $Fr$ and $We$ numbers in this work it makes up almost the entirety of total potential energy \citep{Guo2010}.  To quantify the spectral energy associated with the free-surface elevation, we perform a mapping of the free-surface $z$ coordinate onto a bi-dimensional grid of the same resolution as the DNS. This grid data is then used to compute 1D Fourier spectra of the elevation to measure ${E}_g$. This is given in Fig. \ref{fig:fs-elev-spectra} for cases A, D, E and F, normalizing by the energy of the first wavenumber. The data collapses and exhibits a decay with slope of $-7/5$ for a wide range of wavenumbers.  The decay however is less steep for high wavenumbers, indicating a different behavior at small scales, except for the lowest $Fr$ (case E) which has the lowest deformations.  On a relative basis, this indicates that the conversion between gravitational (potential) and kinetic energy around the free-surface to be more relevant at the smallest scales.

\begin{figure}
\centering
    \begin{subfigure}{0.33\textwidth}
        \centering
        \includegraphics[width=\textwidth]{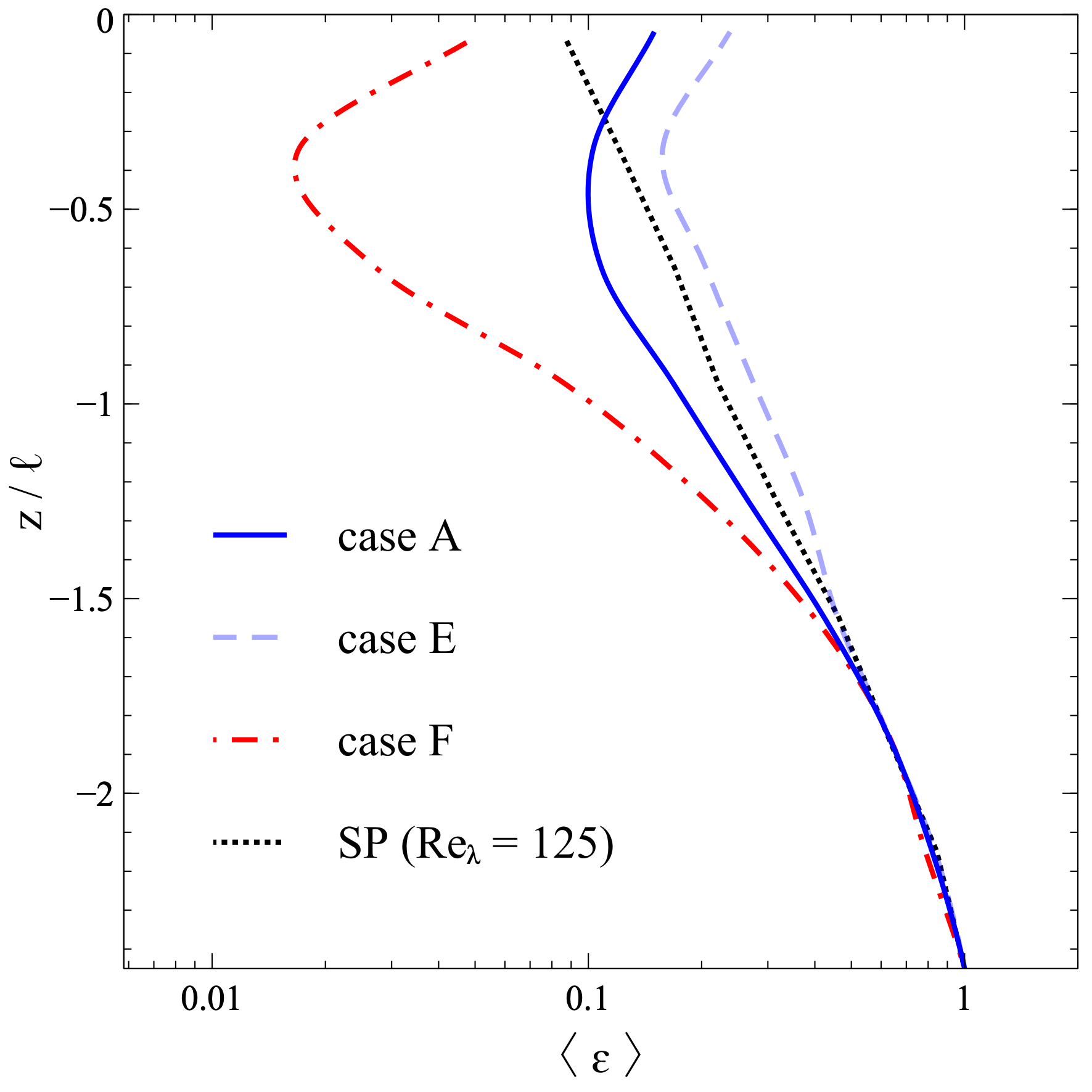}
        \caption{}
        \label{fig:diss-cases}
    \end{subfigure}
      \quad
    \begin{subfigure}{0.6\textwidth}
        \centering
        \includegraphics[width=\textwidth]{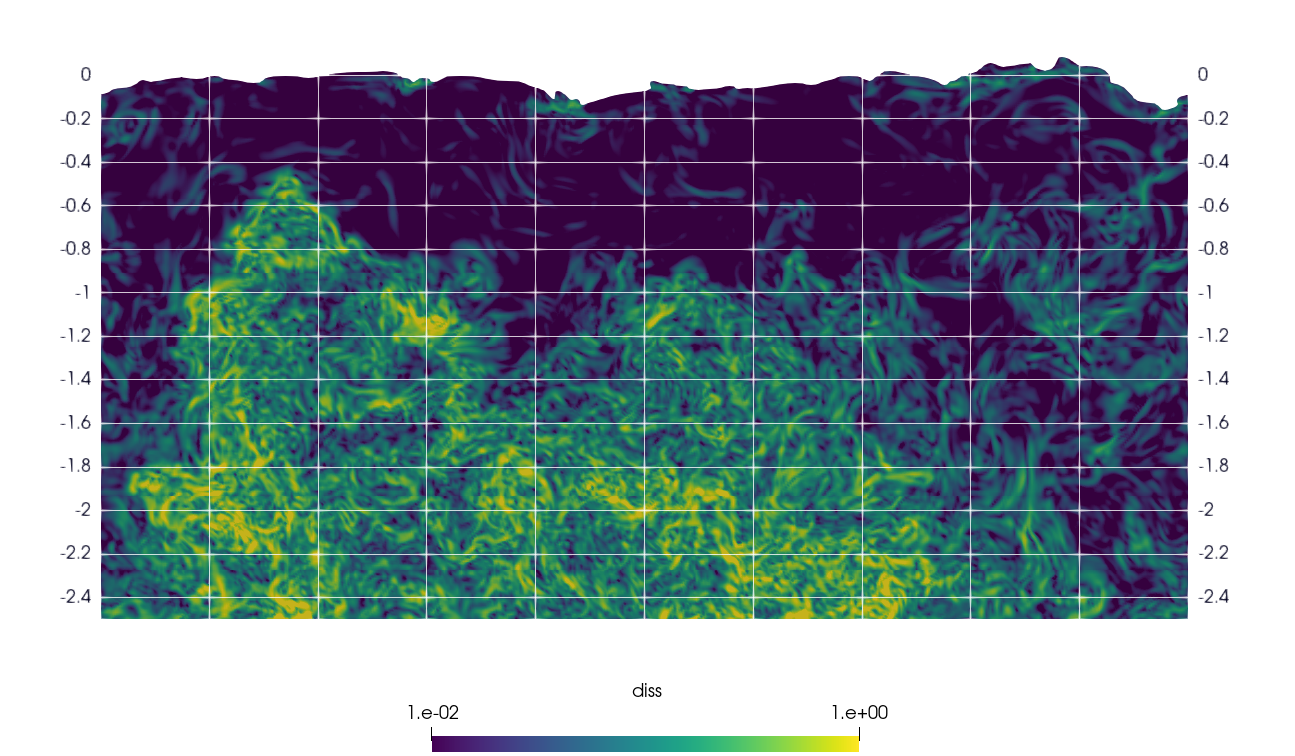}
        \caption{}
        \label{fig:diss-contour}
    \end{subfigure}
\caption{(a) Average turbulent dissipation rate $\langle \varepsilon \rangle$ in water phase for cases A, E, F, and single-phase (SP) simulation at $Re_{\lambda}=125$ for reference; data normalized by values at $z/\ell = -2.5$; (b) instantaneous contour slice (log-scale) of $\varepsilon$ for case F. }
    \label{fig:turb-dissipation}
\end{figure}

Another potential mechanism for the higher energy content in the vicinity of the interface stems from modulations in the local turbulence dissipation rate $\varepsilon$. 
In the work of \cite{Walker1996} for a flat free-surface, a decrease in dissipation results from the elimination of tangential vorticity fluctuations. While this is also the case in the local tangential directions for the deformable free-surface simulations of \cite{Guo2010}, they do not observe a substantial change in the dissipation rate near the interface.  The behavior in vorticity is markedly different between a flat free-surface (slip wall) and two-fluid interfaces (\textit{quasi-free-surface}) as shown in \S\ref{sec:vorticity} and in \cite{Terrington2022}.  We analyze the depth-averaged dissipation rate $\langle \epsilon \rangle$ in Fig.~\ref{fig:diss-cases}, along with a single-phase simulation for comparison at $Re_{\lambda} = 125$. While the decay in dissipation rate is monotonic in single-phase outside the forcing region, all two-phase cases show an inflection around $z/\ell \approx -0.5$. The higher $Fr$ in case A shows less dissipation, and similar trend is verified for case F. A snapshot of the local dissipation rate is plotted in Fig.~\ref{fig:diss-contour} for case F, showing a sudden decay in $\varepsilon$ within the blockage layer, with local areas of higher dissipation just below the interface. This analysis of $\langle \varepsilon \rangle$, $E_g$, and wavelet spectra shows that small scales within this distance from the interface are unable to effectively dissipate their energy from viscosity.

\begin{figure}
    \centering
           \begin{subfigure}{0.26\textwidth}
        \centering
        \includegraphics[width=\textwidth]{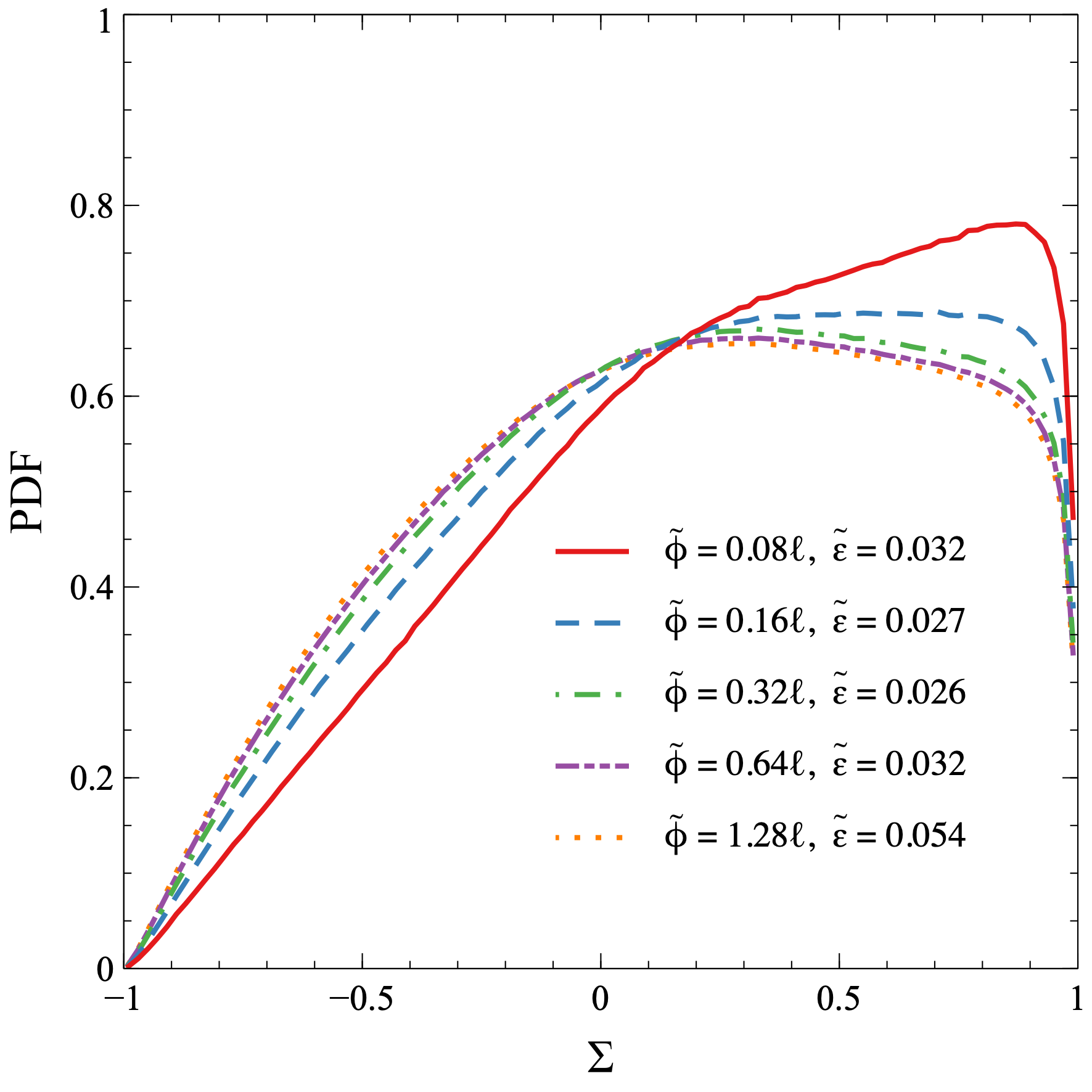}
        \caption{}
            \label{fig:pdf-topo}
    \end{subfigure}
   \qquad
    \begin{subfigure}{0.31\textwidth}
        \centering
        \includegraphics[width=\textwidth]{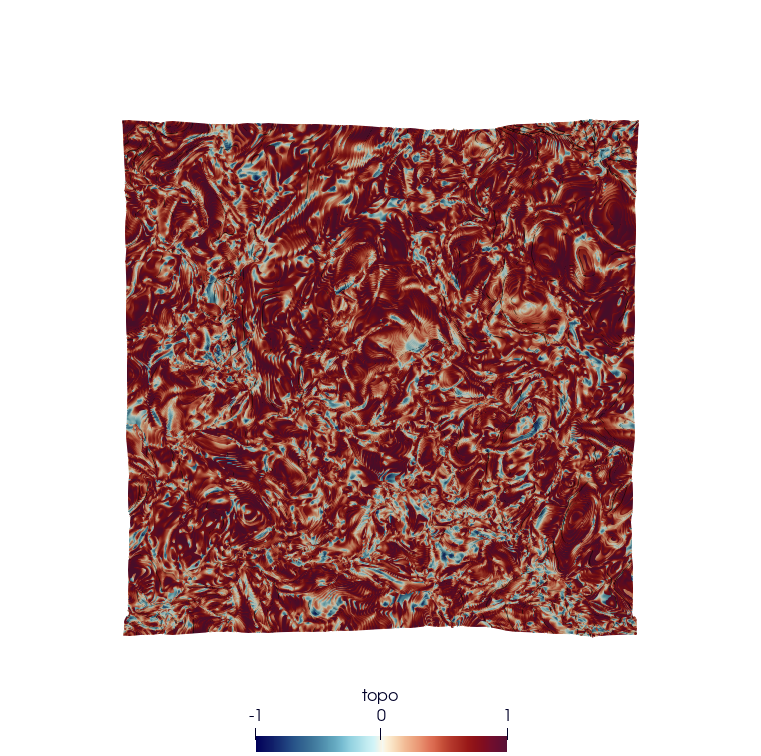}
        \caption{}
        \label{fig:topo-top}
    \end{subfigure}
        \quad
           \begin{subfigure}{0.31\textwidth}
        \centering
        \includegraphics[width=\textwidth]{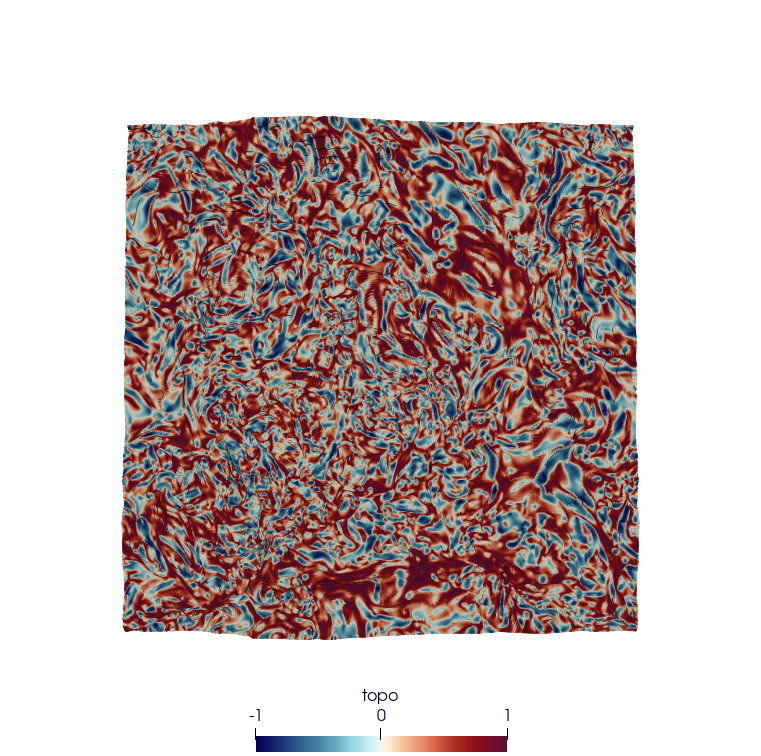}
        \caption{}
        \label{fig:topo-bottom}
    \end{subfigure} \\
       \begin{subfigure}{0.9\textwidth}
        \centering
        \includegraphics[width=\textwidth]{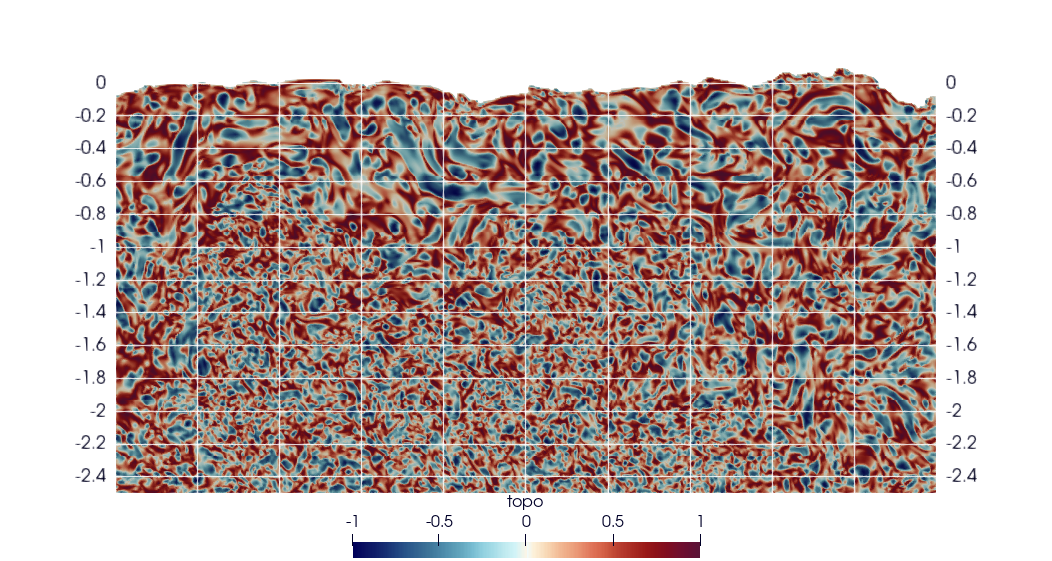}
        \caption{}
        \label{fig:topo-slice}
    \end{subfigure}
\caption{PDF of flow topology parameter $\Sigma$ for increasing distances from the interface $\tilde{\phi}$ and associated dissipation rate $\tilde{\varepsilon}$ (a); instantaneous contour plot of $\Sigma$: top view $\phi=0$ (b), bottom view $\phi=0.08\ell$ (c),  slice view (d).   }
    \label{fig:topology}
\end{figure}

To further characterize the flow near the interface and explore the dissipation regime, we introduce the topology parameter $ \Sigma = (||\mathbf{S}||^2_F - ||\mathbf{\Omega}||^2_F) / (||\mathbf{S}||^2_F + ||\mathbf{\Omega}||^2_F )$ used in \cite{Cannon2024}. This scalar parameter is bounded between $\pm 1$ and can describe states of pure rotation (-1), shear strain (0) or extensional strain (+1). We remind that only states outside of pure rotation ($\Sigma \neq -1$) indicates some level of strain in the fluid parcel and contributes toward local dissipation.   Focusing on the same scales analyzed through the multi-scale wavelet decomposition presented previously, we compute the PDF of $\Sigma$ for distances from the interface shorter than $\sim \ell$ for case F (see Fig.\ref{fig:pdf-topo}).  The averaged dissipation rates within these same regions were computed and given in the same graph as $\tilde{\varepsilon}$. Through this averaging process we provide additional confirmation of the non-monotonic behavior of the turbulent dissipation rate near the interface, and the impact on the TKE cascade.
For scales closest to the interface (within $\delta_v$) the PDF peaks close to $\Sigma \approx 1$, indicating dominance of extensional strain occurring at the interface. 
This is further confirmed by plotting $\Sigma$ and taking a top view from the free-surface ($\phi =0$) in Fig.~\ref{fig:topo-top}, compared to a view from the bottom at $\phi = 0.08 \ell$ shown in Fig.~\ref{fig:topo-bottom}: it is evident that within this short distance the flow changes to have regions of pure rotation. The extensional strain stems from the compression/stretching of the interface as observed in \S\ref{sec:surf-deform}, with large regions of net stretching and ridges with high compression.  As the distance from the interface grows, the PDFs converge and shift toward smaller values of $\Sigma$ and a more uniform distribution between shear and extensional strain.   Furthermore we can see the extent of the regions associated with pure rotation and strain to grow from Fig.~\ref{fig:topo-slice}.  

\section{Conclusions}
\label{sec:conclusions}
In this work we present two-phase DNS of SFST, investigating the effect of $Fr$, $We$ and $Re$ numbers on the surface deformation and the exchange of kinetic energy near the interface.  For the range of $We$ numbers in this work, we find that surface tension effects are only visible in the entrained air bubble sphericity, affecting the total surface area associated with the bubbles.  On the other hand, the $Fr$ number has a major influence on the free-surface area and deformation statistics, as well as the two-dimensional Eulerian compressibility factor $\mathcal{C}$.  As $Fr$ increases, the total interfacial area increases along with the net strain experienced by the free-surface. From the bulk submerged HIT, $\mathcal{C}$ increases from the asymptotic value of $1/6$ to values approaching $1/2$ as observed in the work by \cite{Cressman2004,Lovecchio2015}. Higher $Fr$ results in a slight reduction in  $\mathcal{C}$.  The vorticity flux across the free-surface is associated with viscous diffusion of vortices parallel to the interface by analyzing the dot product between the surface normal $\mathbf{n}$ and $\bm{\omega}$, and maps of interface curvature $\kappa_{\Gamma}$.

The modulation of kinetic energy between horizontal and vertical components is observed for the lower $Fr$ cases, following predictions from RDT, while higher $Fr$ cases preserve isotropy even at the surface.  The presence of a reverse or dual energy cascade is also clearly shown by analyzing the inter-scale energy transfer via third-order statistics at different depths, in agreement with the recent experiments of \cite{Ruth2024}. A strong net reverse cascade is seen for scales of the order of the integral length scale. Vertical kinetic energy is also seen to be stronger in upwelling eddies compared to downwellings by conditional averaging.

To complement the spectral analysis, we perform a phase-based DWT of the kinetic energy. We observe different power law scalings of the wavelet energy, with a stronger -3 slope for mid-range wavenumbers and a weaker decay for smallest scales closest to the interface.  The higher energy at smallest scales near the free-surface results from the conversion of gravitational energy and lower turbulent dissipation rate. The local flow topology was also investigated by quantifying the density of regions under pure rotation, shear or extension at increasing thresholds below the interface. The dissipation rate exhibits a non-monotonic behavior, with a sudden increase at the interface characterized by a dominance of pure extensional strain. Within the source layer the flow topology changes rapidly with larger regions of pure rotation, where vortices adjust from a random orientation to become parallel to the free-surface.

These findings have considerable implications for two-phase TKE transport modeling, indicating the strong dependence on scales of the modeled turbulence, as well as their proximity to the interface.  We note that studies using spectral analysis of SFST are still lacking in the literature and warranted.  One potential avenue left for future work is to use a filtering approach such as the ones in \cite{Xiao2009,Lovecchio2015,Park2025} to calculate the inter-scale kinetic energy transfer, and identify spatial structures participating in the reverse energy cascade. 
Furthermore, a decomposition of this quantity as outlined by \cite{Johnson2021} into its local/non-local vorticity stretching and strain amplification terms can help clarify the physical mechanisms involved in the free-surface turbulence energy cascade.



\backsection[Funding]{This work was funded by the U.S. Office of Naval Research grant N00014-22-1-2639 under the guidance of Dr. W.-M. Lin. The computational resources were provided through the Department of Defense High Performance Computing Modernization Program.}

\backsection[Declaration of interests]{ The authors report no conflict of interest.}


\backsection[Author ORCIDs]{A. Calado, https://orcid.org/0000-0002-9228-4746; E. Balaras, https://orcid.org/0000-0003-0132-4835}




\bibliographystyle{jfm}
\bibliography{jfm}

\end{document}